\documentclass[]{aa} 

\usepackage{graphicx}
\usepackage{txfonts}
\usepackage{xcolor}
\usepackage{amsmath}
\usepackage{amsfonts}
\usepackage{natbib}
\usepackage{longtable}
\usepackage{booktabs}
\usepackage[amsmath,thmmarks]{ntheorem}
\theoremseparator{.}
\theorembodyfont{\normalfont} 
\newtheorem{assumption}{Assumption}{\bfseries}{\itshape}
\usepackage[normalem]{ulem}
\usepackage{hyperref}
\hypersetup{
    colorlinks=true,
    linkcolor=blue,
    filecolor=magenta,      
    urlcolor=cyan,
    citecolor=teal,
    pdftitle={The cosmic DANCe of Perseus},
    }
\begin{document} 

   \title{The cosmic DANCe of Perseus}

   \subtitle{I: Membership, phase-space structure, mass, and energy distributions\thanks{Full Table \ref{table:list_of_members} is only available in electronic form at the CDS via anonymous ftp to cdsarc.cds.unistra.fr (130.79.128.5) or via https://cdsarc.cds.unistra.fr/cgi-bin/qcat?J/A+A/}}

   \author{J. Olivares\inst{1}
          \and H. Bouy\inst{2}
          \and N. Miret-Roig\inst{3}
          \and P.A.B. Galli\inst{4}
          \and L.~M. Sarro\inst{1}
          \and E. Moraux \inst{5}
          \and A. Berihuete\inst{6}
          }

    \institute{
		Departamento de Inteligencia Artificial, Universidad Nacional de Educación a Distancia (UNED), c/Juan del Rosal 16, 	
		E-28040, Madrid, Spain. \email{jolivares@dia.uned.es}
		\and
    	Laboratoire d'astrophysique de Bordeaux, Univ. Bordeaux, CNRS, B18N, allée Geoffroy Saint-Hilaire, 33615 Pessac, France.
        \and
        University of Vienna, Department of Astrophysics, Türkenschanzstraße 17, 1180 Wien, Austria.
        \and
        Núcleo de Astrofísica Teórica, Universidade Cidade de São Paulo, R. Galvão Bueno 868, Liberdade, 01506-000 São Paulo, SP, Brazil.
        \and
        Univ. Grenoble Alpes, CNRS, IPAG, 38000 Grenoble, France.
        \and
        Depto. Estad\'istica e Investigaci\'on Operativa. Universidad de C\'adiz, Avda. Rep\'ublica Saharaui s/n, 11510 Puerto Real, C\'adiz, Spain.
        }
   \date{Received; accepted}

 
  \abstract
   {Star-forming regions are excellent benchmarks for testing and validating theories of star formation and stellar evolution. The Perseus star-forming region being one of the youngest (<10 Myr), closest (280-320 pc), and most studied in the literature, is a fundamental benchmark. }
   {We aim to study the membership, phase-space structure, mass, and energy (kinetic plus potential) distribution of the Perseus star-forming region using public catalogues (\textit{Gaia}, APOGEE, 2MASS, PanSTARRS).}
   {We use Bayesian methodologies accounting for extinction to identify the Perseus physical groups in the phase-space, retrieve their candidate members, derive their properties (age, mass, 3D positions, 3D velocities, and energy), and attempt to reconstruct their origin. }
   {We identify 1052 candidate members in seven physical groups (one of them new) with ages between 3 and 10 Myr, dynamical super-virial states, and large fractions of energetically unbounded stars. Their mass distributions are broadly compatible with that of Chabrier for masses $\gtrsim0.1M_\odot$ and do not show hints of over-abundance of low-mass stars in NGC1333 with respect to IC348. These groups' ages, spatial structure, and kinematics are compatible with at least three generations of stars. Future work is still needed to clarify if the formation of the youngest was triggered by the oldest.}
   {The exquisite \textit{Gaia} data complemented with public archives and mined with comprehensive Bayesian methodologies allow us to identify 31\% more members than in previous studies, discover a new physical group (Gorgophone: 7 Myr, 191 members, and 145 $M_{\odot}$), and confirm that the spatial, kinematic, and energy distributions of these groups support the hierarchical star-formation scenario.}

   \maketitle

\section{Introduction}
\label{intro}
Young star clusters and star-forming regions are benchmarks where current theories of star formation and stellar evolution can be tested and validated. However, this validation requires statistically representative samples where biases are minimal. The most common biases affecting the parameters of nearby young clusters are observational and thus related to the survey characteristics, like its completeness, limiting magnitude, and spatial extent. Methodological biases can also appear, for example, due to cuts in the observational space \citep{2018A&A...616A...9L} or due to the membership selection. In young clusters, for example, activity related to youth can result in variable luminosity and colour indices, which can impact the membership analysis and derived properties. Moreover, in the analysis of young star clusters and star-forming regions, the remnants of dust and gas from the parent molecular cloud can also extinct the light of the newborn stars and introduce further biases in the inferred parameters of these young populations.

In this article, we analyse the properties of the stellar groups in the nearby Perseus star-forming region using the \textit{Gaia} Data Release 3 \citep[DR3,][]{2022arXiv220800211G} complemented with publicly available catalogues of radial velocity from the Apache Point Observatory Galactic Evolution Experiment \citep[APOGEE,][]{2022ApJS..259...35A} and the Set of Identifications, Measurements and Bibliography for Astronomical Data \citep[SIMBAD,][]{2000A&AS..143....9W}, and photometry from the Two Micron All Sky Survey \citep[2MASS,][]{2006AJ....131.1163S} and the Panoramic Survey Telescope and Rapid Response System \citep[PanSTARRS,][]{2016arXiv161205560C}. In particular, we focus on the phase-space (the joint space of 3D positions and 3D velocities), mass, and energy distributions of \object{IC348} and \object{NGC1333}, which are two of the youngest nearest and richest star-forming clusters in the solar vicinity \citep{2016ApJ...827...52L}. The phase-space structure, mass, and energy distributions of the young Perseus groups provide observational constraints to compare competing star-formation scenarios, theories about the origin and variability of the initial mass function, and models for the star formation history of this region.

The rest of this article is organized as follows. In Sect. \ref{review}, we review the recent publications about the membership, the spatial, kinematic, and mass distributions, as well as the star formation history of young groups in the Perseus region. In Sects. \ref{dataset} and \ref{methods}, we introduce the data set and methodologies we use. Then, in Sect. \ref{results}, we describe our results about the membership of the groups, their phase-space structure, mass, extinction, and energy distributions, and their dynamical state. Afterwards, in Sect. \ref{discussion}, we compare our results with those from the literature and discuss the differences and implications. In Sect. \ref{conclusions}, we present our conclusions and future perspectives. Finally, Appendix \ref{appendix:assumptions} lists the assumptions that we take throughout this work.

\section{Literature review}
\label{review}
\subsection{Membership}
\label{intro:membership}
IC348 and NGC1333 have been the subject of several literature studies. Concerning the membership status, we refer the reader to the excellent review by \citet{2016ApJ...827...52L}. These authors not only compiled and assessed the membership status of previously known candidate members but also identified new ones based on several indicators, in particular near-IR spectroscopy. With the updated lists of candidate members, these authors analysed the clusters' ages, mass functions, disk fractions and spatial distributions. As a result, they identified 478 and 203 candidate members of IC348 and NGC1333, respectively (see Tables 1 and 2 of the aforementioned authors). Their survey is complete down to Ks<16.8 mag and Ks<16.2 mag, and in sky regions of 14' and 9', for IC348 and NGC1333 respectively. Shortly after, \citet{2017AJ....154..134E} obtained spectra of 11 members from \citet{2016ApJ...827...52L} and confirmed that two and six are candidate members of IC348 and NGC1333, respectively. However, only 364 and 93 of \citet{2016ApJ...827...52L} members of IC348 and NGC133, respectively, have \textit{Gaia} DR3 parallax, proper motions, and photometry. 

\citet{2018A&A...618A..93C} used \textit{Gaia} Data Release \citep[DR2,][]{2018A&A...616A...1G} astrometry and a modified version of the Unsupervised Photometric Membership Assignment in Stellar Clusters algorithm \citep[UPMASK,][]{2014A&A...561A..57K} to identify candidate members of hundreds of open clusters in the Milky Way. These authors found 144 and 50 candidate members of IC348 and NGC1333, respectively.

\citet{2018A&A...618A..59C} utilised \textit{Gaia} DR2 astrometry and the Density Based Spatial Clustering of Applications with Noise algorithm \citep[DBSCAN,][]{10.5555/3001460.3001507} in combination with an artificial neural network to discover 31 new candidate open clusters. In the Perseus region, they found three cluster candidates: UBC4, UBC19 and UBC31, with 44, 34, and 84 candidate members, respectively. They proposed that these clusters are substructures of the Per OB2 complex, although they noticed that UBC4 is located farther away at 570 pc (see their Sect. 5.4).

\citet{2018ApJ...865...73O} combined observations of the Very Long Baseline Array (VLBA) with \textit{Gaia} DR2 and measured the distance and kinematics of IC348 and NGC1333. Based on $3\sigma$ clipping in the independent spaces of parallax and proper motions these authors identified 133 and 31 members of IC348 and NGC1333, respectively (see their Table 7), of which 162 have \textit{Gaia} DR3 parallax, proper motions and photometry.  

\citet{2020AJ....160...57L} identified 12 new candidates for planetary-mass brown dwarfs in IC348 based on infrared images obtained with the Wide Field Camera 3 of the \textit{Hubble Space Telescope}. Their candidates have spectral types later than M8, while their faintest candidate reaches down to 4-5 $M_{Jup}$. Unfortunately, none of these sources has proper motions nor parallax in the \textit{Gaia} DR3 catalogue as they are too faint.

\citet{2020PASP..132j4401A} designed and implemented a medium band near-IR filter to detect low-mass stars and brown dwarfs. Using this filter, these authors survey $1.3$ square degrees of IC348 and NGC1333 clusters, for which they identify 19 and 9 candidate members; however, only 13 and 3 of these sources, respectively, have \textit{Gaia} DR3 parallax, proper motions and photometry.

\citet{2021MNRAS.503.3232P} used the \textit{Gaia} DR2 data to study the entire Perseus star-forming region.
Through successive cuts and clusterings in the astrometric and photometric features, these authors report the discovery of several hundred members of five stellar groups with ages of 1-5 Myr. Their list of members recovers 50\% and 78\% of \citet{2016ApJ...827...52L} members of NGC1333 and IC348, respectively. In addition, they identify 170, 27, 329, 85, and 302 candidate members of the groups Alcaeus, Autochthe, Electryon, Heleus, and Mestor, respectively. Although these authors claimed the discovery of the previous five groups, 29 members of Electryon and another 29 members of Heleus belong to the UBC31 and UBC19 clusters found by \citet{2018A&A...618A..59C}.

\citet{2021ApJ...917...23K} used the \textit{Gaia} DR2 data to identify $\sim\!3\times10^{4}$ young stars within a distance of 333 pc. Applying the Hierarchical Density Based algorithm \citep[HDBSCAN,][]{2017JOSS....2..205M} to this sample, the authors recover young associations like Orion, Perseus, Taurus, and Sco-Cen. They analyse the star-formation history of each group and find evidence of sequential star-formation propagating at a speed of $\sim\!4 \ \ \rm{km\, s^{-1}}$. In Perseus, they identified 264 candidate members that were broadly classified into groups 1A, 1B, 2A and 2B, based on cuts in the plane of the sky to separate the eastern populations (Per 1A and Per 1B) from the western ones (Per 2A and Per 2B), and in age to separate the youngest (Per 1A and Per 2A) from the oldest (Per 1B and Per 2B).  

\citet{2022ApJ...931..156P} identified members of 65 open clusters using \textit{Gaia} Early Data Release 3 \citep[EDR3,][]{2021A&A...649A...1G} and an unsupervised algorithm based on the technique of self-organising maps after proper motions cuts. In the sky region of Perseus, these authors found 211 candidate members in IC348, 353 in UBC31, and 80 and 230 in two substructures related to UBC31, which they claimed as new and called UBC31 group 1 and UBC 31 group 2, respectively. However, comparing their candidate members with those of \citet{2021MNRAS.503.3232P}, we find that 177 (50\%) of UBC31 belong to Electryon, and 176 (82\%) of UBC31 group 2 belong to Mestor. Except for IC348, all these groups fall outside the sky region analysed here.

\citet{2022AJ....164...57K} identified 810 members in the Perseus region using \textit{Gaia} EDR3 and HDBSCAN. In a previous application of HDBSCAN to other star-forming regions \citep{2019AJ....158..122K}, the authors normalised the data and used the same value for the HDBSCAN parameters of minimum sample size and minimum cluster size. However, in the Perseus regions, they did not normalise the data and used a minimum sample size of 10 stars and a minimum cluster size of 25 stars. As a result, they found 43 Perseus groups, out of which they selected nine based on their absolute colour-magnitude diagram; the remaining groups were deemed unrelated to the Perseus region. Out of their nine groups, seven correspond to the groups identified by \citet{2021MNRAS.503.3232P}, while the other two correspond to the California cloud and a new one called Cynurus and located outside the region analysed by \citet{2021MNRAS.503.3232P}. 

\citet{2022ApJ...936...23W} identified 211 members in the Perseus region by applying astrometric, photometric, radial velocity, and quality cuts to the \textit{Gaia} EDR3 data and confirming membership with spectroscopy from the Large Sky Area Multi-Object Fiber Spectroscopic Telescope (LAMOST). In addition to IC348 and NGC1333, these authors also identified two subgroups corresponding to the Autochthe one from  \citet{2021MNRAS.503.3232P} and a new one associated with the Barnard 1 cloud.

\citet{2022AJ....164..125L} searched for substellar objects in the Perseus cloud regions of IC348 and Barnard 5 using narrow-band imaging centred on the water absorption feature at 1.45 $\mu$m. They confirm three brown dwarfs in IC348 and discover the first one in Barnard 5. In addition, these authors used \textit{Gaia} EDR3 to analyse the distance and proper motions of the Perseus regions of Barnard 1, L1448, and NGC1333. They confirm that the western part of the region is closer than the eastern part, with Barnard 5 standing alone 100 pc away from the rest of the groups.

\subsection{Spatial distribution}
\label{intro:spatial_distribution}

The cluster radius of IC348 was measured by \citet{1999A&AS..137..305S}, \citet{2000AJ....120.3139C}, and \citet{2003AJ....125.2029M}, with all finding consistent values of about 10 to 15 arcmin. However, \citet{2003AJ....125.2029M}, studying the radial profile of the cluster, noticed that it showed two subregions (see their Fig. 4): a core within 5\arcmin, and a halo between 5\arcmin and 10.33\arcmin, with this latter value corresponding to the limit of their survey. Nevertheless, the authors mentioned that they could not exclude the possibility that the cluster extends beyond their survey area coverage to larger radii. Transforming the previous measurements with the cluster distance \citep[320 pc,][]{2018ApJ...865...73O}, we obtain cluster radii between 0.9-1.4 pc, while the core radius of \citet{2003AJ....125.2029M} corresponds to 0.5 pc. \citet{2003AJ....125.2029M} also warned about the possibility of halo sources within the core radius due to projection effects.

The core and halo populations of IC348 were identified long ago by the pioneering work of \citet{1998ApJ...497..736H}. He identified these populations based on their $H_{\alpha}$ emissions, age, and spatial location and concluded that the core of IC348 "is projected upon a population of somewhat older pre-main-sequence stars whose $H_{\alpha}$ emission has largely decayed away". He also observed an age gradient in which older ages are located at larger distances. Moreover, he proposed that this extended and somewhat older population formed in the same molecular cloud that gave birth to IC348 and NGC1333. 

\citet{2002A&A...384..145B} created a Compiled Catalogue in the Perseus region with astrometry and photometry of about 30000 stars distributed in an area of 10\degr~ radius. The identified $\sim$1000 members of the Perseus OB2 complex are distributed in two populations occupying the same sky region but differing in proper motions and distance (see Table 8 of those authors).

\citet{2007A&A...471L..33K} used deep infrared K-band data to analyse the spatial distribution of IC348 with the minimum spanning tree (MST) method. They conclude that the stellar population displays a clustered distribution while the substellar one is homogeneously distributed in space within two times the cluster core radius. Although the substellar population is unbounded, it remains within the cluster limits. We notice that these authors establish the cluster limiting radius as the point where the density of the radial profile merges with that of the background, which is not an intrinsic property of the cluster but rather only of its contrast with the background.

\citet{2008MNRAS.389.1209S} analyse the spatial distribution of IC348 and NGC1333 by applying the nearest-neighbour and MST methods to infrared data. They found that both clusters are centrally concentrated and assembled from a hierarchical filamentary configuration that eventually built up to a centrally concentrated distribution. They also found that the stellar population of both clusters is mass segregated.

\citet{2011ApJ...727...64K} identified substructures in nearby star-forming regions (including IC348) by applying the MST method to catalogues of young stellar objects (YSO) from the literature. In all the groups, particularly in the two at IC348, they found that the maximum mass member is typically more than five times more massive than the median mass member. Furthermore, these massive members are clustered in the central region and are associated with a relatively large density of low-mass stars. Given that all these groups are young, \cite{2011ApJ...727...64K} conclude that the observed configurations should be similar to the primordial ones.

\citet{2017MNRAS.468.4340P} analysed IC348 and NGC133 using the membership lists of \citet{2016ApJ...827...52L}. They estimated the spatial structure, mass segregation and relative local surface density of these two systems using methods like the MST and the $\mathcal{Q}$-parameter. They found that both clusters are centrally concentrated with no evidence of mass segregation. They argue that the results of \citet{2011ApJ...727...64K} can be biased by the smaller list of members and their binning method. Afterwards, they compare their results with numerical simulations to estimate possible initial values for the density and velocities.
   
\citet{2018ApJ...865...73O} studied the distance and structure of the Perseus region by combining data from the VLBA and \textit{Gaia} DR2. They estimated distances of $320\pm26$ pc and $293\pm22$pc to IC348 and NGC1333, respectively, and concluded that given the large median value of the parallax uncertainties, which is larger than the parallax dispersion, then the depth of the groups cannot be extracted from these measurements. 
 
\citet{2018ApJ...869...83Z} determined distances to major star-forming regions in the Perseus molecular cloud using a new method that combines PanSTARRS and 2MASS stellar photometry, astrometric \textit{Gaia} DR2 data, and $\rm{^{12}CO}$  spectral-line maps. For IC348 and NGC1333, they estimated distances of $295\pm18$ pc and $299\pm17$ pc, respectively. These uncertainties result from simple addition of the statistical and systematic ones reported in their Table 3.

\citet{2021A&A...647A..14G} analysed the spatial and kinematic structure of four star-forming regions, including IC348. The authors applied their newly developed Small-scale Significant substructure DBSCAN Detection (S2D2) algorithm to  \citet{2016ApJ...827...52L} list of members (after removing binaries). They found that: i) the densities of the six identified substructures indicate that these are Poisson fluctuations rather than imprints of star formation sites, and ii) the members are centrally concentrated in a radial distribution with an exponent between one and two. The results of these authors are thus consistent with a single  and centrally concentrated star-formation event in IC348. 

\subsection{Velocity distributions}
\label{intro:velocity}

\citet{2015ApJ...807...27C} analysed the velocity distribution of IC348 utilising a subpopulation of 152 members with APOGEE radial velocity measurements. They fitted these measurements using Gaussian Mixture Models \citep[GMMs, see][for examples of applications of mixture models in astronomy]{2017arXiv171111101K} with one and two components and found that the second component improved the fit. Their method allows them to marginalise the contribution of binaries thanks to a binary model that incorporates a wide range of binary configurations into their expected radial velocities. The authors hypothesised that this second component could arise from: i) contaminants from nearby groups, ii) a halo of dispersed/evaporated stars, or iii) the cluster has not yet relaxed to a single Gaussian distribution. The authors argue that the measured velocity dispersion ($0.72\pm0.07\,\rm{km\, s^{-1}}$, or $0.64\pm0.08\,\rm{km\,s^{-1}}$ with two components) implies a super-virial state unless the gravitational potential has been underestimated by, for example, unaccounted gas. They accounted for the gas and dust mass by adding 40$\rm{M_\odot}$ and 210$\rm{M_\odot}$ as lower and upper limits. They found no evidence of a gradient in the velocity dispersion regarding distance to the cluster centre or stellar mass. However, they found evidence of convergence along the line of sight, which the small cluster rotation ($0.024\pm0.013\,\rm{km\,s^{-1}\,arcmin^{-1}}$) cannot explain.

Using APOGEE radial velocity measurements and a similar methodological analysis as that of \citet{2015ApJ...807...27C}, \citet{2015ApJ...799..136F} determined the velocity distribution of NGC1333 based on a sample of 70 members. They found a radial velocity dispersion of $0.92\pm0.12\,\rm{km\, s^{-1}}$, which is consistent with the virial velocity of the group. 
 
\citet{2018ApJ...865...73O} analysed the velocity distribution of IC348 and NGC1333 using VLBA and \textit{Gaia} DR2 data. They conclude that there is no evidence of expansion or rotations given that the velocities they measure ($V_{\rm{Exp;IC348}}\!=\!-0.06\,\rm{km\, s^{-1}}$, $V_{\rm{Exp;NGC1333}}\!=\!0.19\,\rm{km\, s^{-1}}$, $V_{\rm{Rot;IC348}}\!=\![-0.16, 0.0, -0.10]\, \rm{km\, s^{-1}}$ and $V_{\rm{Rot;NGC1333}}\!=\![-0.10, 0.10, 0.19]\, \rm{km\, s^{-1}}$) are smaller than the observed velocity dispersions: $\sigma_{\rm{IC348}}=2.28\,\rm{km\, s^{-1}}$ and $\sigma_{\rm{NGC1333}}=2.0\,\rm{km\, s^{-1}}$.

\citet{2019ApJ...870...32K} studied the internal kinematics of 28 young ($\lesssim$ 5 Myr) clusters and associations with \textit{Gaia} DR2 data. Using proper motions and distances these authors computed transverse velocities in the plane of the sky. After correcting for perspective effects, they derived outward and azimuthal velocities, which are 2D proxies for the internal motions of rotation and expansion. In IC348 and NGC1333, these authors found no evidence of contraction or expansion. With respect to rotation, although they found non-zero azimuthal velocities of $-0.45\pm0.21\,\rm{km\, s^{-1}}$ and $-0.27\pm0.23\,\rm{km\, s^{-1}}$ in NGC133 and IC348, respectively, they deemed these values not significant under the large values of the observational uncertainties, and identified only one system, Tr 15, as having significant rotation.

\subsection{Age distribution}
\label{intro:ages}
\citet{2003ApJ...593.1093L} determined the age of IC348 by comparing \citet{1998A&A...337..403B} and \citet{2000ApJ...542..464C} theoretical isochrones of 1, 3, 10, and 30 Myr to their observational Hertzprung-Russell (HR) diagram, which they obtained with spectral measurements of effective temperatures and bolometric luminosities. These authors found that the cluster members have ages compatible with 1 to 10 Myr, with a median value of 1-2 Myr (see their Fig. 9). 

\citet{2015MNRAS.454..593B} obtained an age of 6 Myr for IC348 based on photometric data from a list of confirmed members from the literature. These authors derived ages comparing the extinction-corrected photometry to main sequence evolutionary models as well as their own semi-empirical pre-main sequence isochrones. They conclude that for star-forming regions younger than 10 Myr, the age estimates from the literature are younger by a factor of two.

\citet{1996AJ....111.1964L} estimated the age of NGC1333 at 1-2 Myr based on its large fraction of members bearing a disk (61\%). Later on, \citet{2008ApJ...674..336G} measured a fraction of $83\pm11$\% of members with a disk, confirming the young age of this group.

\citet{2021MNRAS.503.3232P} derived the ages of the newly identified groups by comparing the photometry of their candidate members to those predicted by theoretical isochrones. According to their results, Authochte is coeval with NGC1333, while Heleus and Mestor are slightly older, with some of their members close to or below the 1 Myr isochrone. Finally, Electryon and Alcaeus appear older with lower fractions of discs and their members being compatible with the 5 Myr isochrone. We notice that their age estimates were obtained without applying any extinction correction. These authors assumed that extinction was negligible based on the observation that the sequences of their candidate young stars were close to the theoretical isochrones of young stars.

\cite{2021ApJ...917...23K} used PAdova TRieste Stellar Evolution Code \citep[PARSEC,][and references therein]{2012MNRAS.427..127B,2015MNRAS.452.1068C}, models and \textit{Gaia} photometry to determine isochrones ages of $6.0\pm3.2$ Myr and $4.7\pm0.5$ for NGC1333 (Per1A) and IC348 (Per2A), respectively. These authors also derived ages $\sim$17 Myr for the older and eastern populations of Perseus (Per1B and Per2B).

Recently, several groups have estimated isochrone ages for the Perseus groups. \citet{2022AJ....164...57K} estimated group and individual star ages and found that the age of the new Cynurus group is 7 Myr, whereas the rest of the Perseus groups have ages similar to those of \citet{2021MNRAS.503.3232P}. Also, \citet{2022ApJ...936...23W} used ancillary data from several photometric surveys to estimate ages of 5.4 Myr, 2.9 Myr, and 5.7 Myr for IC348, NGC1333, and the remaining cloud regions (i.e., Autochthe and Barnard 1 group), respectively. Moreover, \citet{2022AJ....164..125L} estimated ages of 5 Myr for IC348 and the Barnard 5 group.

\subsection{Mass distribution}
\label{intro:mass}

\citet{2003AJ....125.2029M} derived the mass distribution of IC348 down to 10 $M_{Jup}$ based on infrared (JHK) photometry. They found a mass distribution similar to that of the Trapezium, a brown-dwarf-to-stars ratio of 25\%, and radial variations of the mass distribution on the parsec scale. They identified two distinct peaks in the mass distribution attributed to the core and halo populations. They mentioned that to reconcile these two peaks, the age of the halo needed to be 5-10 Myr and that although the age gradient reported by \cite{1998ApJ...497..736H} was in the correct direction, it was not big enough to account for this difference.

\citet{2007ApJ...671..767T} analysed the mass distribution of the Trapezium, Taurus-Auriga, IC348 and the Pleiades. They found evidence for correlated but disjoint populations of stars on the one hand and very low-mass stars and brown dwarfs on the other hand, which suggests different dynamical histories for both populations. They obtain a ratio of one brown dwarf for every five stars, although, in IC348, this ratio reaches up to 30\%. 

\citet{2012ApJ...745..131K} analysed the mass distributions of Taurus, Lupus3, ChaI, and IC348 based on their previous results on the spatial distributions of these star-forming regions \citep{2011ApJ...727...64K}. They found that massive stars are more easily located in regions with higher stellar surface density. Their results suggest strong evidence of this effect in Taurus and IC348, where stars typically have 10-20\% higher mean mass in the clustered environments.
 
\citet{2013A&A...549A.123A} performed a large survey of IC348 to uncover its brown dwarf population, for this, they used deep optical and near-infrared images of MegaCam and Wide-Field Infrared Camera (WIRCam) to photometrically select candidate members. They also conducted a spectroscopic follow-up of their candidate members, of which 16 new members were confirmed, including 13 brown dwarfs. Five of these new members have L0 spectral types corresponding to masses of about 13 $M_{Jup}$. Combining their new members with those from the literature, they constructed the cluster mass distribution and found no significant differences with the mass distributions of other young clusters. Based on a Kolmogorov-Smirnov (KS) test, they conclude that the IC348 mass distribution is well-fitted by a \citet{2003PASP..115..763C} log-normal distribution. However, we notice that their mass bin at log Mass [$M_{\odot}$] $\sim$ -1.2 shows a deficit with respect to the density predicted by \citet{2003PASP..115..763C} mass distribution. Interestingly, this feature is predicted by the mass distribution of \citet{2007ApJ...671..767T}, as can be observed when comparing Fig. 7 of \citet{2007ApJ...671..767T} with Fig. 11 of \citet{2013A&A...549A.123A}. 

\citet{2013ApJ...775..138S} determined the mass distributions of IC348 and NGC1333 for a wide interval of isochrone ages, models, extinction and distances. They warn about the strong dependence of the results on these parameters and point out the importance of comparing under similar assumptions. They found brown-dwarfs-to-stars ratios of 40\% to 50\% in NGC1333 and 25\% to 35\% in IC348. This latter value is in agreement with that of \citet{2007ApJ...671..767T}. Comparing these two clusters, they found differences in their cumulative distributions that resulted from a relative excess of low-mass stars in NGC1333. They conclude that the environment plays an important role, with higher-density regions producing larger fractions of low-mass objects, as predicted by gravitational fragmentation models in which filaments fall into the cluster potential.

\subsection{Extinction}
\label{intro:extinction_distribution}

\citet{2013MNRAS.428.1606F} investigated the shape of the extinction law in two regions of the Perseus molecular cloud. They combined red-optical and near-infrared images of Megacam and the UKIRT Infrared Deep Sky Survey (UKIDSS) to measure the colours of background stars. They developed a Bayesian hierarchical model to simultaneously infer the parameters of individual stars as well as those of the population and found a strong correlation between the extinction ($A_v$) and the slope of the extinction law ($R_v$), which they interpreted as evidence of grain growth. Later on, based on the correlation found by the previous authors, \citet{2018ApJ...869...83Z} adopted an $R_v$=3.3 for moderate extinction values of $A_v$ up to 4 mag. These latter authors determined distances to the Perseus molecular clouds using CO spectral-line maps, photometry and \textit{Gaia} DR2 parallaxes.

\citet{2016ApJ...826...95C} obtained dust emissivity spectral index, dust temperature and optical depth maps of the Perseus molecular cloud from fitting spectral energy distributions to combined \textit{Herschel} and James Clerk Maxwell
Telescope (JCMT) data. They found that the distribution of the dust emissivity spectral index varies from cloud to cloud, which indicates grain growth. This effect was already reported by \citet{1974PASP...86..798S} based on multiband photometry of 20 IC348 bright members. \citet{2016ApJ...826...95C} also found evidence of heating from B stars and embedded protostars, as well as outflows.

\citet{2016A&A...587A.106Z} derived optical depth and temperature maps of the clouds in the Perseus region based on \textit{Planck}, \textit{Herschel} and 2MASS data. Their maps have resolutions from 36 arcsec to 5 arcmin and a dynamic range indicating that the extinction in this region reaches up to 20 mag in $A_K$.

\citet{2019ApJ...887...93G} determined 3D maps of dust reddening based on \textit{Gaia} parallaxes and stellar photometry from PanSTARRS and 2MASS. Thanks to a spatial prior, they obtain smooth maps with isotropic clouds and small distance uncertainties. They made their map available online through the \textit{dustmaps} package. Later on, \citet{2020A&A...639A.138L} used variational inference and a Gaussian process to infer highly resolved 3D dust maps up to 400 pc. A detailed comparison between the 3D dust maps of \citet{2020A&A...639A.138L} and \citet{2019ApJ...887...93G} is shown in Fig. 9 to 11 of the former authors. We notice that although the maps of \citet{2020A&A...639A.138L} show better spatial resolution than those of \citet{2019ApJ...887...93G}, the latter have better 2D sky resolution \citep[see Fig. 11 of][]{2020A&A...639A.138L}.

\cite{2021ApJ...914..122D} used optical and near-infrared stellar polarimetry in combination with \textit{Gaia} DR2 parallaxes to study the magnetic field polarisation in the Perseus molecular cloud. They found a bimodal distribution in the polarisation angles that identified with foreground and background molecular clouds. The foreground cloud is located at $\sim\!150$ pc and has a contribution to the extinction of $A_G\!\sim\!0.3$ mag. On the other hand, the background cloud is at $\sim\!300$ pc and has a larger contribution to the extinction: $A_G\!\sim\!1.6$. Thus, these authors interpret these two clouds as the edges of an ellipsoidal HI shell of about 100-160 pc in size created by the PerOB2 association, with its foreground edge coinciding with the Taurus molecular cloud.

\subsection{Star-formation history}
\label{intro:history}

\citet{1996AJ....111.1964L} suggested that the ongoing star formation in IC348 and NGC1333, located at the opposite ends of the Perseus cloud complex, is produced by a similar physical mechanism, with the main difference between both clusters being that IC348 has produced stars for a longer time than NGC1333.

\citet{1998ApJ...497..736H} proposed the following four-point scenario for the star-formation history of the Perseus region. First, the formation of stars in this region has been taking place for at least 10-20 Myr, with the latest episodes corresponding to IC348 and NGC1333. This first star-formation episode created the OB stars $o$ Per and $\zeta$ Per together with low-mass members that are expected to be spread over a large region. Most of this region has been emptied of molecular gas and dust except for the Perseus ridge. Second, there is a population of young stars with ages 1-12 Myr that is in and around IC348. It is composed of the bright A, B, and F stars as well as "field" ones entangled with IC348 and at its boundaries. The $\rm{H_{\alpha}}$ emission lines of these stars have decayed below the limit of detection. Third, within IC348, there is a population of stars with $\rm{H_{\alpha}}$ emission that are probably younger than the population of the previous point but entangled with it. Fourth, in the densest parts of the Perseus ridge, there is ongoing star formation, as suggested by the highly embedded sources. 

\citet{1999AJ....117..354D} compiled a comprehensive census of OB associations within 1 kpc of the Sun based on \textit{Hipparcos} data. The results of their census are in qualitative agreement with a large-scale star-formation scenario in which the Scorpio-Centaurs-Lupus-Crux, Orion, Perseus, and Lacerta associations formed $\sim$20 Myr ago out of the bubble blown by the high-mass stars of the Cas-Tau association.

\cite{2002A&A...387..117B} considered that the two populations they identified in \cite{2002A&A...384..145B} constitute an example of propagated star formation. It started in the Per OB2b region approximately 30 Myr, continued in Per OB2a 10 Myr ago, and is now in progress in the southern border of Per OB2, where IC348 is located.

\cite{2021MNRAS.503.3232P} suggested that the older groups (Alcaeus, Electryon, Heleus, and Mestor) are closer to the Galactic plane with low latitudes (>-19), while the younger ones (NGC1333 and Autochthe) are at higher latitudes (<-19). They also point out that NGC1333 and Autochthe are part of the same star formation event due to the similarity in their properties and their close location particularly. These two groups, together with IC348, are the only ones with ongoing star formation, while the other older groups in the region have stopped forming stars.

\cite{2021ApJ...917...23K}, using \textit{Gaia} and HDBSCAN, identified two populations in PerOB2: Per A and Per B, corresponding to the western and eastern regions in the sky. They subdivided each of these populations into two distance subgroups, Per-1 at 283 pc and Per-2 at 314 pc, with Per-1A and Per-2A corresponding to NGC1333 and IC348, respectively, and Per-1B and Per-2B their corresponding eastern extensions. While Per A is young  (NGC1333: $6.0\pm3.2$ Myr and IC348: $4.7\pm0.5$) and concentrated, Per B is older (Per1B: $17.5\pm0.9$ Myr and Per2B: $17.1\pm1.1$ Myr) and sparser (35 pc). These authors suggest that due to their kinematic similarities, Per 1 and Per 2 most likely formed in the same star-forming process. However, they notice that a continuous star-forming process between the two populations seems unlikely due to the considerable time lag and lack of age gradient. Instead, they hypothesise that in a parallel fashion, the feedback from the first generation (Per B) dispersed the gas of the parent cloud but did not prevent the continuous in-falling flow of external material, which resulted in a new star-formation burst (Per A). In the end, this process produced two distinct epochs of star formation in both Per 1 and Per 2.

\cite{2021ApJ...919L...5B} used different indicators (e.g., X-rays, HI and $^{26}$Al) to identify a dust-free cavity between the Perseus and Taurus star-forming regions which they call the Per-Tau shell. This most likely formed through one or multiple supernovae episodes that created a super-bubble which sweep up the interstellar medium and created today's extended shell, with the age of this shell being $\simeq$6-22 Myr. These authors hypothesise that the supernova that created the shell may have had its origin in: a) a young (<20 Myr) star cluster with a total mass between 800 and 3300 $M_\odot$, b) a single supernova from a dynamically ejected O or B star, or c) an ultra-luminous X-ray source. They mention that the most likely scenario is the first one, given that evidence supporting the existence of a young (<20 Myr) population has been found in Taurus and Perseus. In the latter case, this young population corresponds to the Perseus groups identified by \citet{2021MNRAS.503.3232P}.  

\citet{2022Natur.601..334Z} found that almost all the star-forming complexes in the solar neighbour lie on the surface of the Local Bubble, with the young stars showing expansion with their motions being perpendicular to the Bubble surface. Their trace-back analysis supports a scenario in which the Local Bubble was formed by supernovae at approximately 14 Myr age. The only nearby star-forming complex that does not lie at the Local Bubble surface is the Perseus complex, which is related to the Local Bubble through the Taurus star-forming complex and the Per-Tau shell.

\citet{2022AJ....164...57K} analysed three star-formation scenarios to explain the observed kinematics of the Perseus region: a supernova explosion, a cloud-cloud interaction, and the first generation of stars from the Per-Tau shell, with the most likely scenario being the collision of two clouds.  According to these authors, the evidence supporting the other two scenarios is not conclusive.

\citet{2022ApJ...936...23W} found that the Perseus region also follows the star-formation scenario of the expanding Local Bubble \citep{2022Natur.601..334Z,1987ARA&A..25..303C}, except that in this case, the Per-Tau shell interacts with it and with the Perseus molecular cloud and results in the elongated shape of the latter. The ages and positions of the Perseus groups suggest that Electyryon, Heleus, and Mestor are far away from the Per-Tau shell and unrelated to it. On the contrary, Alcaeus, Autochthe, IC348 and NGC1333 are near the edge of the shell and may have formed during the same star-formation event. 

\cite{2022AJ....164..125L} propose a star-formation scenario in which a supernova explosion in the Perseus region triggered the star-formation of the region through the driving of a HI super-shell. The latter has at its centre the eastern part of the Perseus region.

\section{Data}
\label{dataset}

\subsection{\textit{Gaia} Data Release 3}
\label{dataset:GDR3}
We downloaded\footnote{\url{https://gea.esac.esa.int/archive/}} the astrometry and photometry of 164 502 \textit{Gaia} DR3 sources within the sky region: 51\degr<\texttt{ra}<59\degr, and 30\degr<\texttt{dec}<33\degr, and proper motions within -100 $\rm{mas\, yr^{-1}}$ < \texttt{pmra} < 200 $\rm{mas\, yr^{-1}}$ and -200 $\rm{mas\, yr^{-1}}$ < \texttt{pmdec} < 100 $\rm{mas\, yr^{-1}}$. From these, 163 178 sources (99.2\%) have observed proper motions and photometry, which are necessary to apply our membership methodology. Our initial list of members comprises the 194 candidate members, with a probability > 0.5, found by \citet{2018A&A...618A..93C} on the IC348 and NGC1333 open clusters. We use this sample of members due to its purely \textit{Gaia} origin and the simplicity of its membership algorithm. Although the pre-\textit{Gaia} sample of \cite{2016ApJ...827...52L} is the most extensive one from the literature, we do not use it because it contains several contaminants in the form of astrometric outliers (see Sect. \ref{discussion:members}).

We processed the astrometric data by only applying a parallax zero point correction of $-17\,\rm{\mu as}$ \citep[see Sect. 2.2 of][]{2021A&A...649A...1G}. As stated in Sect. 7 of \citet{2021A&A...649A...1G} the current parallax bias correction as a function of magnitude, colour, and ecliptic latitude is only a tentative recipe.

\subsection{Complementary data}
\label{dataset:complementary_data}

We complement the \textit{Gaia} DR3 data of our candidate members with APOGEE and SIMBAD radial velocities, as well as 2MASS and PanSTARRS photometry. Out of our 1052 candidate members (see Sect. \ref{results:membership}), 428, 407, and 149 have radial velocity entries in APOGEE, SIMBAD, and Gaia DR3, respectively. Whenever a source has multiple radial velocity entries, we select based on the following ordered preference: APOGEE, \textit{Gaia} DR3, and SIMBAD. In the case of APOGEE, we use as radial velocity uncertainty the dispersion of several measurements (i.e., VSCATTER column), except when it was zero, in which case we use the individual uncertainty (i.e., VERR column). In some cases, the radial velocity catalogues report missing or zero value uncertainties for sources with non-missing radial velocity. Thus we process the radial velocities as follows. If the star has either a missing uncertainty but a non-missing value or an uncertainty larger than 50 $\rm{km\, s^{-1}}$, then we set the uncertainty to 50 $\rm{km\, s^{-1}}$. This large value diminishes the contribution of the source and avoids discarding it. Similarly, if the uncertainties are smaller than 0.01 $\rm{km\, s^{-1}}$, we replace them with the latter value. This uncertainty soil avoids convergence issues in our kinematic inference methodology (see Sect. \ref{methods:6D_structure}). After the previous processing, a total of 626 (60\%) of our candidate members have radial velocity measurements, with a median uncertainty of 0.75 $\rm{km\, s^{-1}}$.

In addition, we query the \textit{Hipparcos} \citep{1997A&A...323L..49P} data in the same sky region as defined above, and we find 47 sources. Out of these, 46 have a cross-match in \textit{Gaia} DR3, and the remaining one has parallax and proper motions ($\mu_{\alpha}=-139.6\pm0.89\, \rm{mas\, s^{-1}}$, $\mu_{\delta}=20.19\pm1.05\, \rm{mas\, s^{-1}}$, $\varpi=18.32\pm0.91$ mas) clearly incompatible with the Perseus groups. Due to the previous reason, we assume that the \textit{Gaia} DR3 of the Perseus region is complete on the bright side.

\section{Methods}
\label{methods}
The following sections describe the methodology that we use to determine the properties of the stellar content of the Perseus region. We start by describing the membership methodology, and afterwards, we describe the steps to identify the distinct physical populations, as well as their properties: empirical isochrones, magnitude and mass distributions. We base the inference of the properties of a physical group on the obtained list of members (see Assumption \ref{assumption:groups_independency} in Appendix \ref{appendix:assumptions}).

\subsection{Membership selection}
\label{methods:membership}
We determine members of the Perseus star-forming region using the \textit{Miec} code \citep{2021A&A...649A.159O}, which is an improvement over the Bayesian hierarchical methodology developed by \citet{2018A&A...617A..15O} and focuses on the analysis of extincted nearby stellar clusters. Briefly, \textit{Miec} is a statistical model that describes the observed astrometry and photometry of possible hundreds of thousands of sources in a sky region that encompasses an open cluster. It delivers candidate members of the open cluster as well as its astrometric (proper motions and parallax, if available) and photometric distributions (colour-index, and photometric bands). The likelihood of the data is a mixture of the field and cluster models, where the former consists of independent and multivariate GMMs in the astrometric and photometric spaces, and the latter is also made of a GMM in the astrometric space and a multivariate Gaussian in the photometric one. The median value of the latter corresponds to the cluster photometric sequence, in which each photometric band is modelled by a spline function of the colour index. The model is Bayesian because it infers the posterior distribution of the cluster parameters given the likelihood of the data and the prior distribution. This latter is constructed from the initial list of members. Once the posterior distributions of model parameters have been inferred (through Markov Chain Monte Carlo methods) the cluster membership probability of each source in the data set is computed using Bayes' theorem and the cluster and field likelihoods as follows:
\begin{equation}
\label{equation:probability}
\begin{split}
& Probability(cluster|data,\mathcal{M}_{cluster},\mathcal{M}_{field})= \\
& \frac{\mathcal{L}(data|\mathcal{M}_{cluster})\cdot \mathcal{P}(cluster)}{\mathcal{L}(data|\mathcal{M}_{cluster})\cdot \mathcal{P}(cluster)+\mathcal{L}(data|\mathcal{M}_{field})\cdot \mathcal{P}(field)},
\end{split}
\end{equation}
where $\mathcal{L}$, $\mathcal{M}$, and $\mathcal{P}$ stand for likelihood, model, and prior, respectively. We use as prior probabilities for the field and cluster their fractions of sources in the entire dataset.

The \textit{Miec} code has been designed for open clusters, and thus, it models the astrometric features of the representation space using GMMs in which all Gaussian components share the same mean value, this is, they are concentric \citep[for more details see][]{2018A&A...617A..15O,2021A&A...649A.159O}. However, the members of star-forming regions have proper motions and parallax distributions that are not necessarily concentric (see, for example, the parallax and proper motions of the Taurus candidate members depicted in Fig. 5 of \citealt{2019A&A...630A.137G}). For this reason, we modify the \textit{Miec} code to deal with the dispersed populations present in star-forming regions by allowing non-concentric GMMs in the proper motions and parallax features. 

In the \textit{Miec} methodology, the representation space (i.e., set of observable features) is of paramount importance since it allows the disentanglement of the field and target populations. Given the known issue of the overestimated \textit{Gaia} \texttt{BP} flux for faint red sources \citep[see Sect. 8.1 of][]{2021A&A...649A...3R}, we use as colour index \texttt{G-RP} instead of \texttt{BP-RP}. Thus, our choice for the representation space comprises the following \textit{Gaia} features: \texttt{pmra, pmdec, parallax, G-RP, BP,} and \texttt{G}. The \textit{Miec} code requires that the spline functions describing the cluster photometric sequence be injective functions of the colour index \citep{2018A&A...612A..70O,2021A&A...649A.159O}. However, the previous representation space only allows for this condition to be fulfilled for \texttt{G} values fainter than 5-6 mag. For this reason, we search for Perseus candidate members brighter than this magnitude limit using only their astrometric membership probabilities, which are also delivered by \textit{Miec}. We classify sources brighter than \texttt{G}$\sim$5 mag as candidate members if their astrometric membership probability is larger than $3\sigma$ (0.997). We choose this highly conservative probability threshold given that, in these cases, the discriminant photometric information is not taken into account.

In addition to the membership probabilities, the code delivers the astrometric and photometric distributions of the target population. While the astrometric distributions are multivariate mixtures in the joint space of proper motions and parallax, the photometric ones are multivariate mixtures in the joint space of colour index and photometric bands. More details about methods to obtain the astrometric, colour index and magnitude distributions can be found in \citet{2018A&A...617A..15O}.

Since the Perseus region contains several dust and gas clouds, we use the extinction module of \textit{Miec} \citep[see Sect. 2.2 of][]{2021A&A...649A.159O}. Briefly, this module permits the extraction of the extinction-free population parameters (i.e., those that describe the cluster's or group's colour-index and magnitude distributions) by marginalising the possible extinction values, $A_v\in[0,A_{v,max}]$ of each source. For the maximum extinction value, $A_{v,max}$, we use the 3D extinction map of \citet{2019ApJ...887...93G}\footnote{We query the extinction map at each source position using the \textit{dustmaps} python package \citep{2018JOSS....3..695M}.} at the group distance. We prefer the maps of \citet{2019ApJ...887...93G} over those of \citet{2020A&A...639A.138L}, given their better 2D sky resolution and because we are interested in individual stars rather than in the 3D structure of the dust clouds.

As explained in \citet{2021A&A...649A.159O}, the extinction module of the \textit{Miec} code faces two main caveats: an increased contamination rate due to sources with missing values and a reduced recovery rate in sources with high extinction values. Given that our data set comprises sources with complete astrometry and photometry (see Sect. \ref{dataset}), that the mean value of extinction to the Perseus region provided by the 3D extinction map is $A_{v,max}\sim 3\pm2$ mag, and that we classify members based on an optimum probability threshold that is optimized as a function of the \texttt{G} magnitude, then we expect that the performance of the \textit{Miec} code in our specific conditions will be better than the extreme conditions reported in  Table C.3 of \citet{2021A&A...649A.159O}. In other words, we expect a recovery rate better than 87\% and a contamination rate less than 7\%. 

Although the physical members of distant open clusters can be identified using clustering methods working in the proper-motions-parallax space, nearby clusters and dispersed stellar populations extending several degrees on the sky are affected by projection effects that distort the proper-motions-parallax space and make difficult the identification of their members. Thanks to the improved non-concentric GMM, the \textit{Miec} code is now flexible enough to accommodate possible distortions in the proper-motions-parallax space created by these projection effects. Nonetheless, these distortions increase the mixing of the physical groups in the proper-motions-parallax space and difficult their disentanglement. This effect can be seen in Fig. 10 of \citet{2021MNRAS.503.3232P}, where the proper motions of the Perseus groups are heavily mixed. To disentangle these populations, we iteratively run the  \textit{Miec} and \textit{Kalkayotl} \citep{2020A&A...644A...7O} codes. The first one identifies the candidate members in the astro-photometric space, while the second one separates the group in the phase-space (more below). 

\subsection{Phase-space structure}
\label{methods:6D_structure}
We analyse the phase-space distribution of the Perseus star-forming region using the \textit{Kalkayotl} code \citep[][a new version of the code is in preparation]{2020A&A...644A...7O}. This code implements a Bayesian hierarchical model that allows the joint inference of stellar positions, velocities, and population parameters without imposing a fixed prior. On the contrary, the code allows us to test different 1D (distance), 3D (positions) or 6D (positions+velocities) prior families and infer their parameters based on the \textit{Gaia} data. Moreover, it corrects for the parallax and proper motions angular spatial correlations and zero points using the values provided by \citet{2021A&A...649A...2L}. We notice that the output phase-space Cartesian coordinates returned by \textit{Kalkayotl} are in the Equatorial ICRS reference system rather than a Cartesian Galactic one. Throughout the rest of this work, unless stated otherwise, we will use this Equatorial reference system and the names X, Y, and Z for the 3D positions and U, V, and W for the 3D velocities.

As mentioned in the previous section, the identification of the physical groups is a mandatory step for the subsequent astrophysical analyses. Given that the region is known to host several populations (see Sect. \ref{intro}), we use the \textit{Kalkayotl} code to probabilistically disentangle the possible physical groups.

We model the stellar positions and velocities of the Perseus star-forming region using 6D GMMs (see Assumption \ref{assumption:gaussian} in Appendix \ref{appendix:assumptions}). We classify the candidate members into the physical groups by maximising membership probability to each Gaussian component. It is important to notice that the modelling of the phase-space structure  is computationally expensive because the number of inferred parameters, $N_p$, grows linearly with the number of sources: $N_p=(6\times N_s) + (28\times N_c - 1)$, with $N_s$ and $N_c$ the number of sources and Gaussian components, respectively. In the latter equation, the first term corresponds to the parameters of the individual sources, with six phase-space coordinates for each one of them, and the second term corresponds to the global or population parameters. In these latter, each Gaussian component needs 28 parameters: 21 for the covariance matrix, six for the median, and one for the weight. Given that the components' weights are restricted to add to one, there is one non-free weight.

Our iterative approach to identifying candidate members and physical groups proceeds as follows. Once the candidate members of the entire region have been identified by the \textit{Miec} code, we use the \textit{Kalkayotl} code to fit their observables using 6D GMMs with one to six components. We consider that several Gaussian components pertain to the same physical group if their medians are mutually contained within one Mahalanobis distance (see Assumption \ref{assumption:gaussian} in Appendix \ref{appendix:assumptions}). Otherwise, each Gaussian component corresponds to a physical group. We reject as physical groups those Gaussian components for which: i) the Hamiltonian Monte Carlo sampler does not converge \citep[see][]{2020A&A...644A...7O}, or ii) the contribution to the mixture is $\lesssim 5\%$. We recursively fit 6D GMM to each physical group until Assumption \ref{assumption:gaussian} (see Appendix \ref{appendix:assumptions}) is fulfilled. This recursive fit allows us to reject non-physical members in the 3D space that, due to projection effects, have similar astrometric features as the bulk of the group. Once the physical groups have been disentangled, we join the list of members of each of them to the field population and run again the \textit{Miec} code independently on the resulting data set. The independent run of \textit{Miec} on each identified group ensures that the candidate members' uncertainties (the photometric ones in particular) are propagated into the group's empirical isochrone and mass distribution (see Sect. \ref{methods:isochrones} and \ref{methods:mass}). We iterate the procedure until the number of candidate members of each physical group converges under Poisson uncertainties.

To avoid the radial velocity of unresolved binary stars from biasing the group-level parameters (e.g., internal velocity dispersion) of the Perseus groups, we make an additional run of \textit{Kalkayotl} using as input list the final members of each group, however, this time, we set as missing the radial velocities of sources lying more than 3-$\sigma$ away from any of the group's mean U, V, and W space velocities. Setting as missing the radial velocity rather than removing the source avoids discarding the astrometric information of these sources.

Finally, we also analyse the internal kinematics of the identified physical groups by searching for evidence of their expansion or rotation. To do this, we compute the dot and cross product of the positions and velocity vectors of each candidate member on the reference system of its parent group. The average values of these vector products are proxies for the expansion and rotation rates of stellar groups \citep[see][]{2019A&A...630A.137G,2021A&A...654A.122G}.

\subsection{Empirical isochrones and age estimates}
\label{methods:isochrones}
The empirical isochrones of the physical groups are inferred from the data and delivered by \textit{Miec} as a by-product. These empirical isochrones are cubic spline functions that model the mean value of the \texttt{BP} and \texttt{G} magnitudes as functions of the colour index \texttt{G-RP}. We notice that thanks to the use of the extinction module (see Sect. \ref{methods:membership}), these empirical isochrones are free of extinction.

We estimate the age of each physical group by comparing its extinction-free empirical isochrone with the theoretical ones from the PARSEC \citep{2020MNRAS.498.3283P,2013MNRAS.434..488M}, MESA Isochrones \& Stellar Tracks \citep[MIST,][]{2016ApJS..222....8D,2016ApJ...823..102C}, and BT-Settl \citep{2014IAUS..299..271A} models. We are aware that this dating method provides only a rough estimate of  the group age due to well -known issues of theoretical isochrones to reproduce the observed colour-magnitude diagram of young clusters \citep[e.g.,][]{2015MNRAS.454..593B,2015A&A...577A.148B,2022MNRAS.513.5727B}.

\subsection{Mass distributions}
\label{methods:mass}

We infer the mass distribution of each physical group using two different methods, but both based on the PARSEC, MIST and BT-Settl theoretical isochrones at each group's estimated age (see Sect. \ref{methods:isochrones}). The first method uses the \textit{Sakam} code \citep{2019A&A...625A.115O} to independently infer the mass of each candidate while the second method transforms the group's magnitude distributions delivered by \textit{Miec} into mass distributions.

None of the theoretical isochrone models that we use fully covers the magnitude interval of our candidate members. Thus, when used independently, these models introduce border artefacts in the resulting mass distributions. We overcome this problem by computing a unified theoretical model that we call PMB (standing for PARSEC-MIST-BT-Settl). In this, we fitted cubic splines to the grid values of mass and magnitudes provided by the three theoretical models. We use cubic splines because they provide continuous derivatives of the magnitude-mass relations and thus avoid the typical problems of simple polynomials (i.e., Runge's phenomenon).     

\textit{Sakam} is a Bayesian inference code that samples the joint posterior distribution of mass, $A_v$ and $R_v$ of individual stars based on theoretical isochrones and the star's distance (see Sect. \ref{methods:6D_structure}) and available photometry (see Sect. \ref{dataset:complementary_data}). As prior distributions of the mass, $A_v$ and $R_v$ we use the \citet{2005ASSL..327...41C} distribution, a uniform distribution ($A_v\in[0,10]$ mag), and a Gaussian distribution ($R_v\sim\mathcal{N}(3.1,0.5)$), respectively. Once the posterior distributions of all the candidate members are inferred, we compute the group's mass distribution as a kernel density estimate on the aggregated mass samples of all the group members. In the second method, we use the mass-magnitude relations provided by the unified theoretical isochrones (at the group's estimated age, see Sect. \ref{methods:isochrones}) together with the group distance to transform \textit{Miec}'s magnitude distributions of each group into mass distributions. We notice that the theoretical mass-magnitude relations are not one-to-one and have abrupt changes of slope in the 1.5--2.5 $M_{\odot}$ mass interval, resulting in mass distributions with a large scatter in this region (see discussion in Sect. \ref{discussion:mass}).

Working with the previous two methods offers the following advantages. First, the \textit{Sakam} method allows us to study possible variations in the $R_v$ value across the Perseus groups. These variations have already been suggested in the literature, see, for example, \citet{2013MNRAS.428.1606F,2018ApJ...869...83Z}. On the other hand, the \textit{Miec} method has two advantages over the \textit{Sakam} one. First, it obtains the group's magnitude distributions using the entire dataset, weighting each source by its membership probability to the group. This approach removes the sample bias introduced when working on the subsample of the most probable group members.  Second, it does a full propagation of the model uncertainties and observational uncertainties to the magnitude and mass distributions, whereas the \textit{Sakam} method only propagates the photometric and distance uncertainties. Although the \textit{Miec} method offers a statistically more robust approach than that of the \textit{Sakam} method, none of them is a perfect solution to the mass distribution inference problem. Nonetheless, until the arrival of a complete and spectroscopically confirmed and characterised list of the group's members (see Assumption \ref{assumption:groups_independency} in Appendix \ref{appendix:assumptions}), the comparison of these two methods offers what we consider the best strategy to derive the mass distribution of the groups.

\subsection{Dynamical analysis}
\label{method:dynamical_analysis}

We perform a dynamical analysis of the Perseus groups based on the source and group level parameters inferred with the methods presented in the previous subsections. We determine the dynamic state of each physical group with two methods. 

The first method takes the mass, position, and velocity posterior distributions of each candidate member and computes its energy distribution with respect to its parent group, under the assumption that the latter are self-gravitating (see Assumption \ref{assumption:self_gravitating} in Appendix \ref{appendix:assumptions}). We propagate uncertainties by taking samples from the posterior distributions of each candidate member and computing the energy of each sample as follows:

\begin{equation}
\label{equation:energy}
E=\frac{1}{2}m \cdot v^2 - \frac{G\cdot M\cdot m}{r},
\end{equation}
where $r$ and $v$ are the distance and speed in the {reference system} of the stellar group, $M$ is the total group's mass enclosed within the distance $r$ from its centre, $m$ is the sample's mass, and $G$ is the gravitational constant. To obtain $r$ and $v$ in the reference system of the stellar group, we use the population parameters delivered by \textit{Kalkayotl} (see Sect. \ref{methods:6D_structure}).

The second method compares the observed velocity dispersion of each group with the theoretical one expected if the stellar system were at virial equilibrium. To compute the latter, we follow the approach that \citet{2015ApJ...807...27C} took in the analysis of IC348 (see their Sect. 4.3.3). Briefly, these authors assumed that the velocity dispersion at virial equilibrium, $\sigma_{vir}$, can be estimated using the total mass of the cluster $M$, its half-mass radius, $r_{hm}$, and a structural parameter, $\eta$, (see Assumption \ref{assumption:virial_equilibrium} in Appendix \ref{appendix:assumptions}) following Eq. 4 of \citet{2010ARA&A..48..431P}. Moreover, we follow \citet{2015ApJ...807...27C} additional assumption that these parameters can be obtained by fitting an \citet{1987ApJ...323...54E} profile (hereafter EFF) to the 2D stellar number density of the system (see Assumption \ref{assumption:EFF_profile} in Appendix \ref{appendix:assumptions}). We do this fitting for the Perseus physical groups with the free and open-source code \textit{PyAspidistra} \citep{2018A&A...612A..70O}. In addition, given that our methods also deliver mass and 3D positions for each member in the Perseus groups, we also estimate the half-mass radius by finding the radial distance at which the group's mass reaches 50\%.

Furthermore, we notice the following two aspects of the Perseus star-forming region. First, it is known that this region is still embedded in the dust and gas of its parent molecular cloud and that the contribution of this non-stellar mass to the total mass of the groups is non-negligible. \citet{2015ApJ...807...27C} accounted for this non-stellar mass (see Sect. \ref{intro:velocity}), assuming that the dust and gas still follow the observed distribution of stars and that its total mass contribution has lower and upper limits equal to 65\% (80 $\rm{M_\odot}$) and 169\% (210 $\rm{M_\odot}$), respectively, of the total stellar mass of IC348. Here, we also take the previous two assumptions and extend them to the rest of the Perseus groups (see Assumption \ref{assumption:dust_mass} in Appendix \ref{appendix:assumptions}). Second, it is known that the fraction of binary systems in open clusters varies between 11\% to 70\% \citep{2010MNRAS.401..577S}, with the fraction of unresolved binaries between 12\% to 20\% \citep[e.g.,][]{2021AJ....162..264J}. However, our methodologies are unable to identify and infer the mass of these possibly unresolved binaries. Therefore, it follows that our gravitational potential will be underestimated due to the unaccounted mass of the unresolved binaries. Thus, we correct this bias by increasing by 20\% the mass contribution of the individual stars when computing the gravitational potential of the groups (see Assumption \ref{assumption:binaries_mass}  in Appendix \ref{appendix:assumptions}). 

\section{Results}
\label{results}

\subsection{Membership}
\label{results:membership}

We iteratively apply the \textit{Miec} and \textit{Kalkayotl} codes (as described in Sects. \ref{methods:membership} and \ref{methods:6D_structure}) to the Perseus data set (see Sect. \ref{dataset}). In the first iteration, the code recovered 920 candidate members, and after successive iterations utilising the extinction module, we recover 130 more candidate members. Our search for bright (\texttt{G}>5 mag) members delivered only two astrometric candidate members: $\zeta$ Per and $o$ Per, with astrometric membership probabilities of 0.99989 and 0.99988, respectively, which implies a $\sim\!4\sigma$ discovery.

\begin{figure*}[ht!]
    \resizebox{\hsize}{!}{\includegraphics{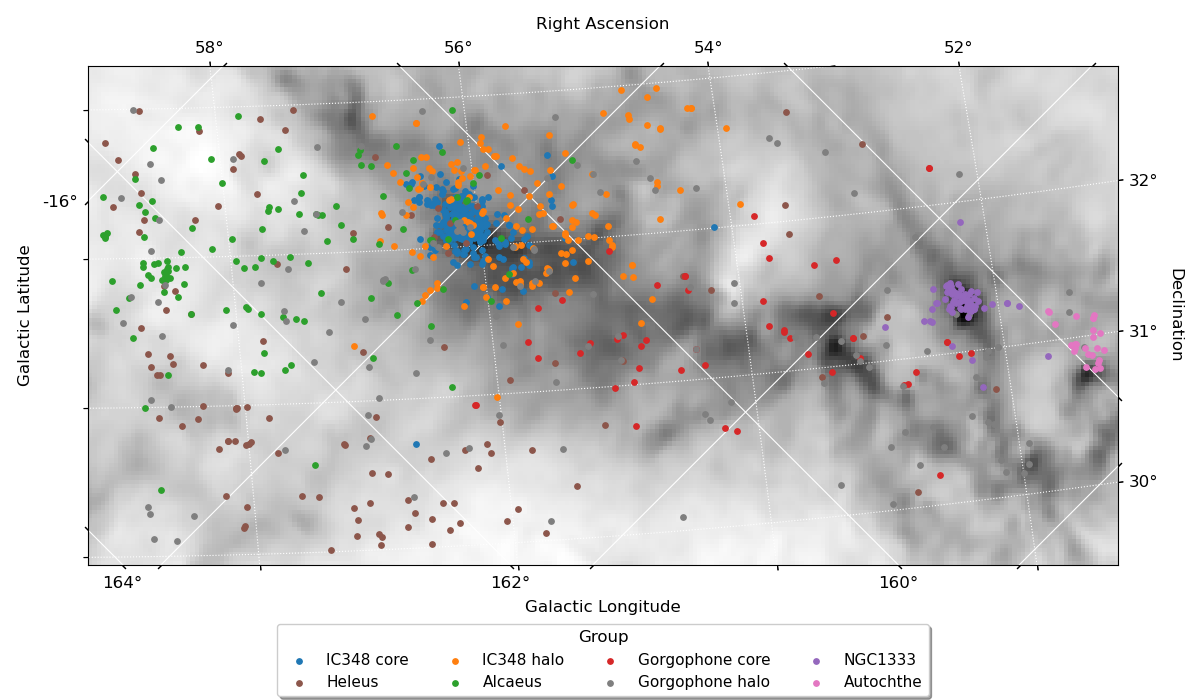}}
     \caption{Sky coordinates of the Perseus candidate members. The colour code shows the probabilistic classification, and the background image shows the thermal dust emission (545 GHz) from \citet{2020A&A...643A..42P}.}
\label{fig:sky}
\end{figure*}

According to our iterative methodology, the final 1052 candidate members (see Table \ref{table:list_of_members}) are distributed into eight statistical groups (see Sect. \ref{results:6D_structure}). Table \ref{table:groups_members} shows the names, number of members, mean distance, age, and mass estimates of these groups. In Sect. \ref{results:core_and_halo}, we will show that two of these statistical groups pertain to the same physical one, thus effectively reducing the number of physical groups to seven. We identify the well-known IC348 (with its core and halo) and NGC133 young clusters (see Fig. \ref{fig:sky}) and three of the recently discovered populations of \cite{2021MNRAS.503.3232P}: Heleus, Alcaeus and Autochthe. In addition, we discover a putatively new young physical group of $\sim$7 Myr and 191 candidate members that is composed of a core and halo populations. Following the nomenclature style of \cite{2021MNRAS.503.3232P}, we call this group Gorgophone. In Sect. \ref{discussion:members}, we present a detailed comparison between the candidate members that we find in this work and those from the literature. 

\begin{table}[ht!]
\caption{The name, number of members, mean distance, age estimate, and mass lower limit of the Perseus groups.}
\label{table:groups_members}
\centering
\begin{tabular}{c|c|c|c|c}
\toprule
Name & Number & Distance & Age & Mass \\
{}   &      {}    &  [pc]    & [Myr] & [$\rm{M_{\odot}}$] \\
\midrule
IC348 core      & 329 & $315\pm  1$ & 3  & $146\pm  5$\\
IC348 halo      & 172 & $312\pm  6$ & 5  & $125\pm  5$\\
Heleus          & 124 & $365\pm 30$ & 5  & $ 47\pm  2$\\
Alcaeus         & 127 & $286\pm  5$ & 10 & $ 93\pm  4$\\
Gorgophone core & 46  & $291\pm  5$ & 7  & $ 36\pm  2$\\
Gorgophone halo & 145 & $290\pm 19$ & 7  & $109\pm  4$\\
NGC1333         & 84  & $292\pm  1$ & 3  & $ 39\pm  2$\\
Autochthe       & 25  & $295\pm  2$ & 3  & $ 19\pm  2$\\
\bottomrule
\end{tabular}
\end{table}

\subsection{Phase-space structure}
\label{results:6D_structure}

\begin{figure}[ht!]
    \centering
     \includegraphics[width=\columnwidth,page=1]{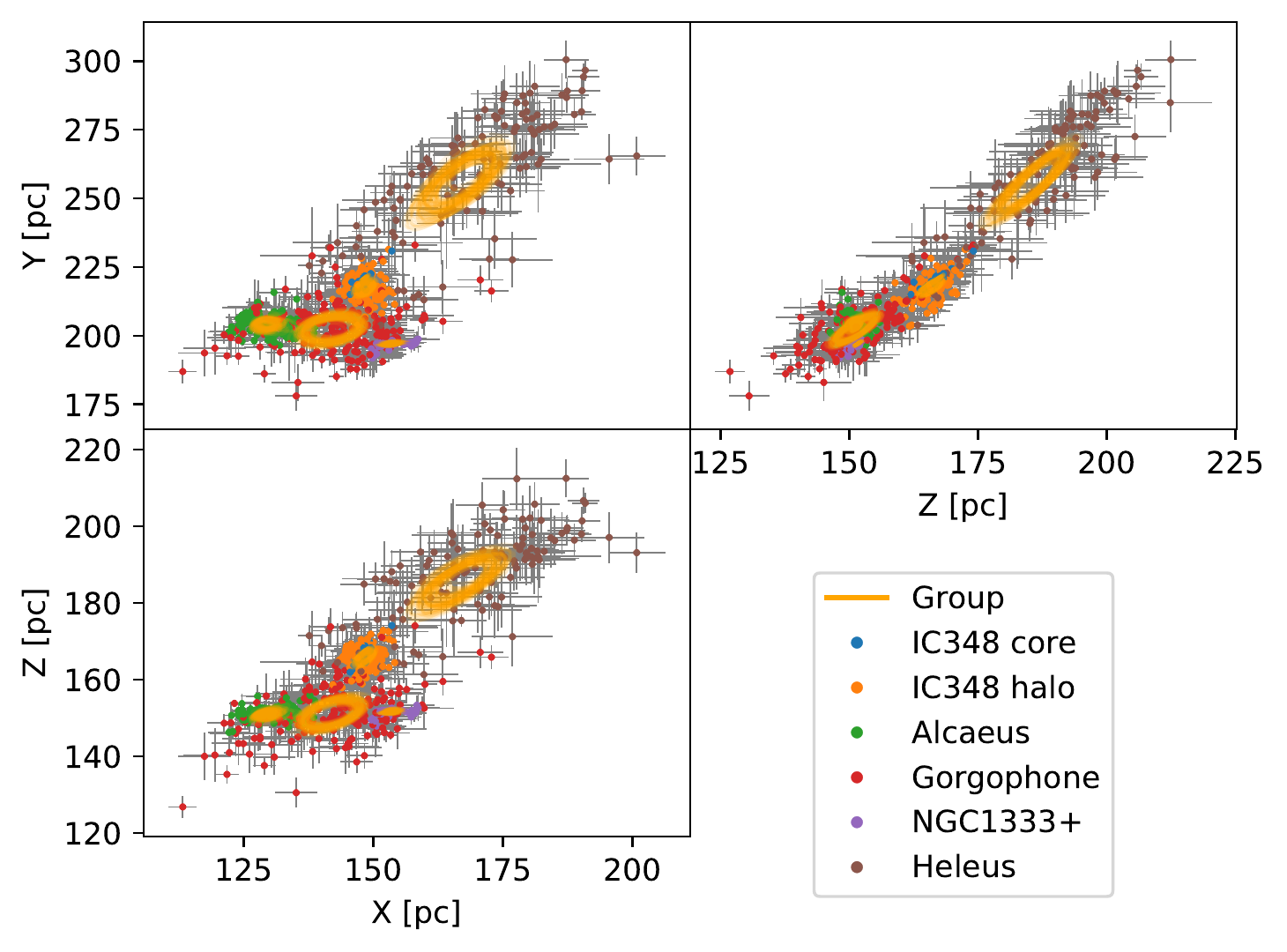}
     \includegraphics[width=\columnwidth,page=2]{Figures/Model.pdf}
     \caption{Cartesian equatorial (ICRS) positions (top panel) and velocities (bottom panel) of the 1052 Perseus candidate members. The colour code shows the probabilistic classification, and the orange ellipses depict samples from the posterior distribution of the group level parameters corresponding to the one-sigma covariance matrix, and are centred at the mean position and velocity of the groups.}
\label{fig:kalkayotl}
\end{figure}

As explained in Sect. \ref{methods:6D_structure}, we infer the phase-space structure of the Perseus groups by fitting 6D GMMs using the \textit{Kalkayotl} code. We jointly inferred the parameters of all candidate members and choose the model with six components as the best one. We based this decision on the convergence properties of the sampler and the weights of the components. Models with more than six components resulted in inefficient sampling and negligible weights for the additional components. Figure \ref{fig:kalkayotl} shows the inferred phase-space coordinates of the candidate members as well as the six-components GMM (the orange lines show samples from the posterior distribution of the one-sigma covariance matrices). The colour code shows the probabilistic classification of each of the components (i.e., IC348 core, IC348 halo, Alcaeus, Heleus, Gorgophone, and NGC1333+). 

The joint inference with the 1052 candidate members, our method was unable to completely disentangle the NGC1333 members from those of Autochthe (see the group called "NGC1333+" in Fig. \ref{fig:kalkayotl}). To overcome this problem and to break any possible entanglement, we iteratively fitted two-component GMMs to the candidate members of each identified group. At the end of this hierarchical-tree exploration, we find out that all the groups except for Gorgophone required only one Gaussian component to describe their phase-space structure. In the rest of the groups, the additional components showed negligible weights $\leq$5\% and convergence issues. In the two-component GMM of Gorgophone, the weights of the additional component were non-negligible (>5\%).
 
\begin{figure}[ht!]
    \centering
     \includegraphics[width=\columnwidth,page=1]{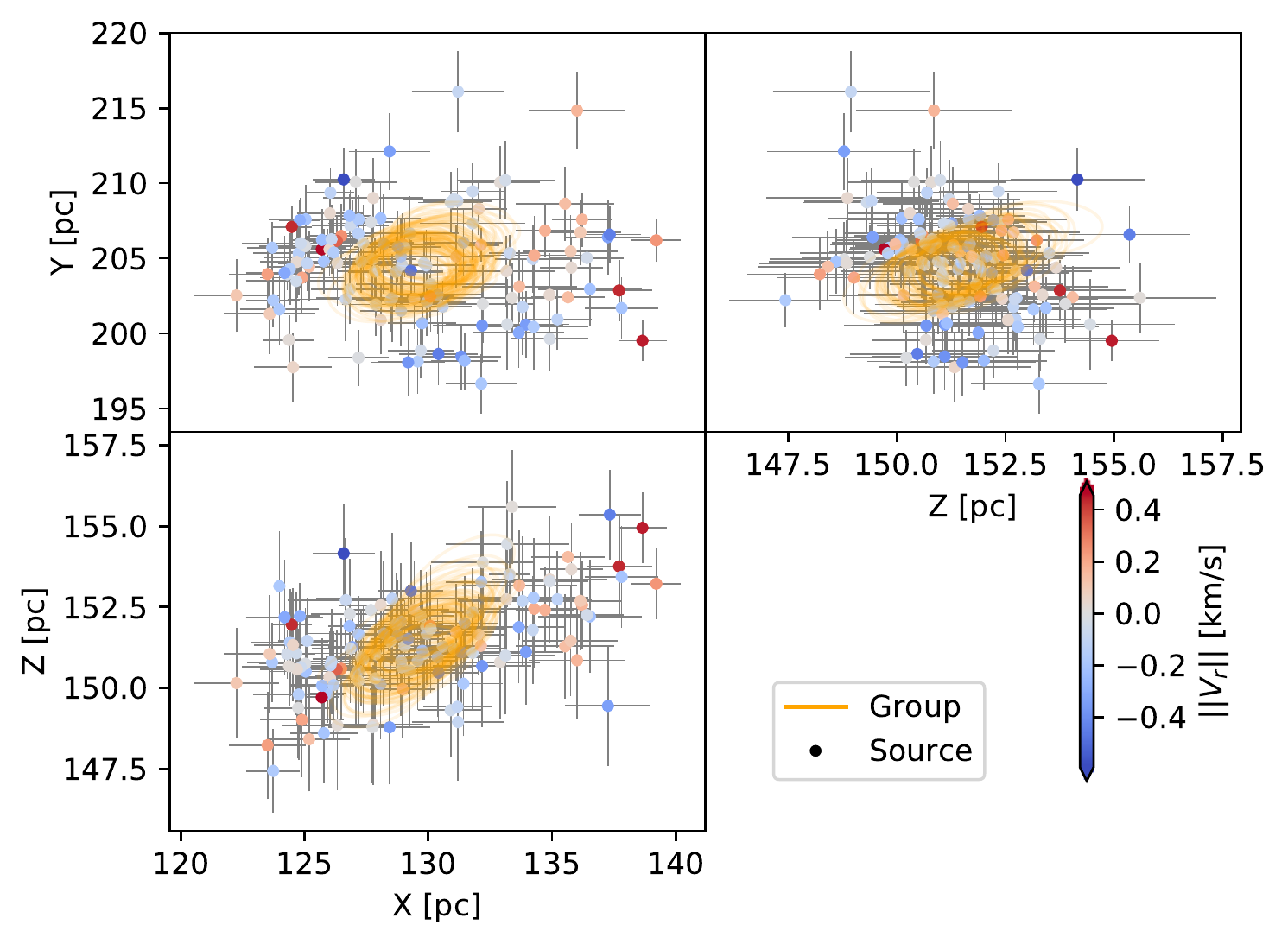}
     \includegraphics[width=\columnwidth,page=2]{Figures/Alcaeus.pdf}
     \caption{Cartesian equatorial (ICRS) positions (top panel) and velocities (bottom panel) of the Alcaeus group. The colour code shows the speed (top panel) and the distance (bottom panel) in the radial direction and both relative to the group centre. The orange ellipses show samples from the posterior distribution of the group-level parameters (see Fig. \ref{fig:kalkayotl}).}
\label{fig:6d_K5}
\end{figure}

Our iterative methodology delivers phase-space parameters of the Perseus groups as well as posterior distributions of the positions and velocities of each candidate member. The group-level parameters (mean and standard deviation) of the identified groups are shown in Tables \ref{table:groups_mean} and \ref{table:groups_sd}. As an example of the inferred group- and source-level parameters, Fig. \ref{fig:6d_K5} shows the cartesian (ICRS) positions and velocities of the Alcaeus group. The dots and error bars show the mean and standard deviation of the posterior distributions of the source-level parameters (i.e., 6D cartesian coordinates of each candidate member), while the orange ellipses show samples from the posterior distribution of the group-level parameters (i.e., the one-sigma covariance matrix centred at the mean position and velocity of the group). The total spatial and velocity  dispersions of the identified groups are shown in the second and third columns of Table \ref{table:kinematic_indicators}. As can be observed from this table, the most distant Heleus group is also the most spread in the XYZ space. On the contrary, the core of IC348 is the most compact one, with only a 0.66 pc radius. 

\begin{table}[ht!]
\caption{The mean values of the group's cartesian coordinates.}
\label{table:groups_mean}
\centering
\resizebox{\columnwidth}{!}{
\begin{tabular}{|c|c|c|c|c|c|c|}
\toprule
{} &              $X$ &              $Y$ &              $Z$ &                      $U$ &                      $V$ &                      $W$ \\
{} &      $\rm{[pc]}$ &      $\rm{[pc]}$ &      $\rm{[pc]}$ & $\rm{[km \cdot s^{-1}]}$ & $\rm{[km \cdot s^{-1}]}$ & $\rm{[km \cdot s^{-1}]}$ \\
Group           &                  &                  &                  &                          &                          &                          \\
\midrule
IC348 core      &  $148.6 \pm 0.5$ &  $221.3 \pm 0.7$ &  $167.6 \pm 0.5$ &            $4.5 \pm 0.1$ &           $18.6 \pm 0.1$ &           $-0.0 \pm 0.1$ \\
IC348 halo      &  $148.9 \pm 0.6$ &  $218.0 \pm 1.0$ &  $166.3 \pm 0.7$ &            $5.0 \pm 0.1$ &           $19.0 \pm 0.1$ &            $0.7 \pm 0.1$ \\
Heleus          &  $170.3 \pm 1.7$ &  $261.8 \pm 2.6$ &  $189.3 \pm 1.6$ &            $7.3 \pm 0.2$ &           $21.5 \pm 0.3$ &            $2.7 \pm 0.2$ \\
Alcaeus         &  $129.6 \pm 0.6$ &  $204.5 \pm 0.9$ &  $151.4 \pm 0.6$ &            $4.0 \pm 0.1$ &           $22.3 \pm 0.2$ &           $-2.3 \pm 0.2$ \\
Gorgophone core &  $146.4 \pm 1.0$ &  $200.9 \pm 1.1$ &  $150.9 \pm 0.7$ &            $3.7 \pm 0.3$ &           $20.9 \pm 0.3$ &           $-1.9 \pm 0.3$ \\
Gorgophone halo &  $140.5 \pm 1.3$ &  $203.1 \pm 1.4$ &  $151.1 \pm 1.1$ &            $3.0 \pm 0.2$ &           $20.4 \pm 0.3$ &           $-2.8 \pm 0.3$ \\
NGC1333         &  $152.5 \pm 0.7$ &  $197.3 \pm 0.9$ &  $151.9 \pm 0.7$ &            $4.3 \pm 0.2$ &           $21.5 \pm 0.2$ &           $-4.3 \pm 0.2$ \\
Autochthe       &  $157.7 \pm 0.7$ &  $197.4 \pm 0.8$ &  $151.7 \pm 0.6$ &            $1.0 \pm 0.3$ &           $19.3 \pm 0.4$ &           $-4.0 \pm 0.3$ \\
\bottomrule
\end{tabular}

}
\end{table}

\begin{table}[ht!]
\caption{The standard deviation of the group's cartesian coordinates.}
\label{table:groups_sd}
\centering
\resizebox{\columnwidth}{!}{
\begin{tabular}{|c|c|c|c|c|c|c|}
\toprule
{} &      $\sigma_X$ &      $\sigma_Y$ &      $\sigma_Z$ &               $\sigma_U$ &               $\sigma_V$ &               $\sigma_W$ \\
{} &     $\rm{[pc]}$ &     $\rm{[pc]}$ &     $\rm{[pc]}$ & $\rm{[km \cdot s^{-1}]}$ & $\rm{[km \cdot s^{-1}]}$ & $\rm{[km \cdot s^{-1}]}$ \\
Group           &                 &                 &                 &                          &                          &                          \\
\midrule
IC348 core      &   $0.3 \pm 0.1$ &   $0.4 \pm 0.2$ &   $0.4 \pm 0.1$ &            $0.9 \pm 0.0$ &            $0.8 \pm 0.0$ &            $0.8 \pm 0.0$ \\
IC348 halo      &   $2.9 \pm 0.4$ &   $5.1 \pm 0.5$ &   $3.7 \pm 0.4$ &            $0.9 \pm 0.1$ &            $1.0 \pm 0.1$ &            $1.0 \pm 0.1$ \\
Heleus          &  $15.0 \pm 1.1$ &  $23.3 \pm 1.7$ &  $13.6 \pm 1.1$ &            $1.4 \pm 0.2$ &            $1.1 \pm 0.3$ &            $1.1 \pm 0.2$ \\
Alcaeus         &   $4.4 \pm 0.4$ &   $4.1 \pm 0.5$ &   $2.1 \pm 0.4$ &            $0.4 \pm 0.1$ &            $0.4 \pm 0.2$ &            $0.5 \pm 0.1$ \\
Gorgophone core &   $5.2 \pm 0.7$ &   $4.5 \pm 0.6$ &   $1.6 \pm 0.4$ &            $1.0 \pm 0.2$ &            $1.2 \pm 0.2$ &            $1.0 \pm 0.2$ \\
Gorgophone halo &  $13.0 \pm 0.9$ &  $13.0 \pm 1.2$ &  $10.4 \pm 0.9$ &            $2.0 \pm 0.2$ &            $2.4 \pm 0.3$ &            $2.2 \pm 0.2$ \\
NGC1333         &   $0.7 \pm 0.2$ &   $0.5 \pm 0.3$ &   $0.3 \pm 0.2$ &            $1.4 \pm 0.1$ &            $1.6 \pm 0.1$ &            $1.2 \pm 0.1$ \\
Autochthe       &   $0.7 \pm 0.3$ &   $0.6 \pm 0.3$ &   $0.8 \pm 0.2$ &            $0.6 \pm 0.2$ &            $0.5 \pm 0.3$ &            $0.6 \pm 0.2$ \\
\bottomrule
\end{tabular}
}
\end{table} 

\subsubsection{Core and halo populations of IC348 and Gorgophone}
\label{results:core_and_halo}
In IC348 and Gorgophone, our methodology finds two Gaussian components that we call core and halo populations, with the core having the smallest dispersion in the 3D positional space and halo the largest (see the value of $\sigma_{XYZ}$ in Table \ref{table:groups_sd}). In IC348, the medians and covariance matrices of these two Gaussians result in Mahalanobis distances between them of 10.16 (halo with respect to the core) and 1.23 (core with respect to halo). Given that these distances are mutually farther away than one Mahalanobis distance, then we conclude that they correspond to independent physical groups (see Assumption \ref{assumption:gaussian} in Appendix \ref{appendix:assumptions}). However, in the case of Gorgophone, the Mahalanobis distances are 1.6 (halo with respect to the core) and 0.96 (core with respect to halo), which prevent us from concluding that they pertain to independent physical groups (see Assumption \ref{assumption:gaussian} in Appendix \ref{appendix:assumptions}). For historical reasons, we continue using \citet{1998ApJ...497..736H} nomenclature of core and halo populations for the two identified physical groups of IC348 (see Sect. \ref{intro:spatial_distribution} and \ref{intro:history}).

\subsubsection{Internal kinematics}

We analyse the internal kinematics of the groups using the inferred positions and velocities of both the sources and the groups. As an example, Fig. \ref{fig:6d_K5} shows the 3D positions (top panel) and 3D velocities (bottom panel) of the Alcaeus group. The colour code of this figure shows the distance (bottom panel) and speed (top panel), in the radial direction, both with respect to the group's centre. As shown in this figure, there are no observable trends of expansion. Appendix \ref{appendix:3D_velocities} shows figures with the Galactic Cartesian positions and velocities of the candidate members in our eight statistical groups. As can also be seen in those figures, there are no observable trends of expansion in any of the Perseus groups.

As explained at the end of Sect. \ref{methods:6D_structure}, to objectively quantify the internal kinematics of the groups, we computed the average magnitude of the dot and cross products of the radial distance and velocity vectors of all the group's members, which are proxies of the group's expansion and rotation, respectively. Columns fourth and fifth of Table \ref{table:kinematic_indicators} show the average values of the dot and cross product vectors, respectively. Although the previous values show some trend of contraction, particularly in NGC1333 and the core of IC348, the current uncertainties do not allow us to draw firm conclusions. Similarly, the uncertainties in the cross product show that the observed trends of rotation are significant only at the one-sigma level, but fail to exceed the two-sigma level.

\begin{table}[ht!]
\caption{Total dispersions and kinematic indicators of the physical groups.}
\label{table:kinematic_indicators}
\centering
\resizebox{\columnwidth}{!}{
\begin{tabular}{c|c|c|c|c}
\toprule
Group  & $\lVert\vec{\sigma}_{XYZ}\rVert$ & $\lVert\vec{\sigma}_{UVW}\rVert$ & $\overline{\hat{\vec{e}}_r \cdot \vec{v}}$ & $\overline{ \lVert\hat{\vec{e}}_r \times \vec{v}\rVert}$ \\
 {} & [pc]  &$\rm{[km \cdot s^{-1}]}$ &$\rm{[km \cdot s^{-1}]}$ & $\rm{[km \cdot s^{-1}]}$\\
\midrule
IC348 core      &$0.65\pm0.20$  & $1.44\pm0.08$ & $-0.2\pm0.8$ & $1.0\pm0.6$ \\
IC348 halo      &$6.89\pm0.75$  & $1.68\pm0.16$ & $+0.1\pm1.0$ & $1.2\pm0.6$ \\
Heleus 	        &$30.84\pm2.32$ & $2.08\pm0.35$ & $+0.5\pm1.5$ & $1.1\pm0.7$ \\
Alcaeus        	&$6.38\pm0.75$  & $0.76\pm0.21$ & $-0.1\pm0.4$ & $0.6\pm0.3$ \\
Gorgophone core &$7.09\pm0.99$  & $1.81\pm0.31$ & $+0.1\pm1.1$ & $1.2\pm0.8$ \\
Gorgophone halo &$21.13\pm1.72$ & $3.79\pm0.41$ & $+0.4\pm2.2$ & $2.6\pm1.4$ \\
NGC1333         &$0.90\pm0.39$  & $2.37\pm0.23$ & $-0.1\pm1.1$ & $1.4\pm0.8$ \\
Autochthe       &$1.24\pm0.44$  & $1.04\pm0.38$ & $+0.2\pm0.8$ & $0.7\pm0.6$ \\
\bottomrule
\end{tabular}
}
\end{table} 

\subsection{Empirical isochrones and age estimates}
\label{results:isochrones}

\begin{figure}[ht!]
    \centering
    \includegraphics[width=\columnwidth]{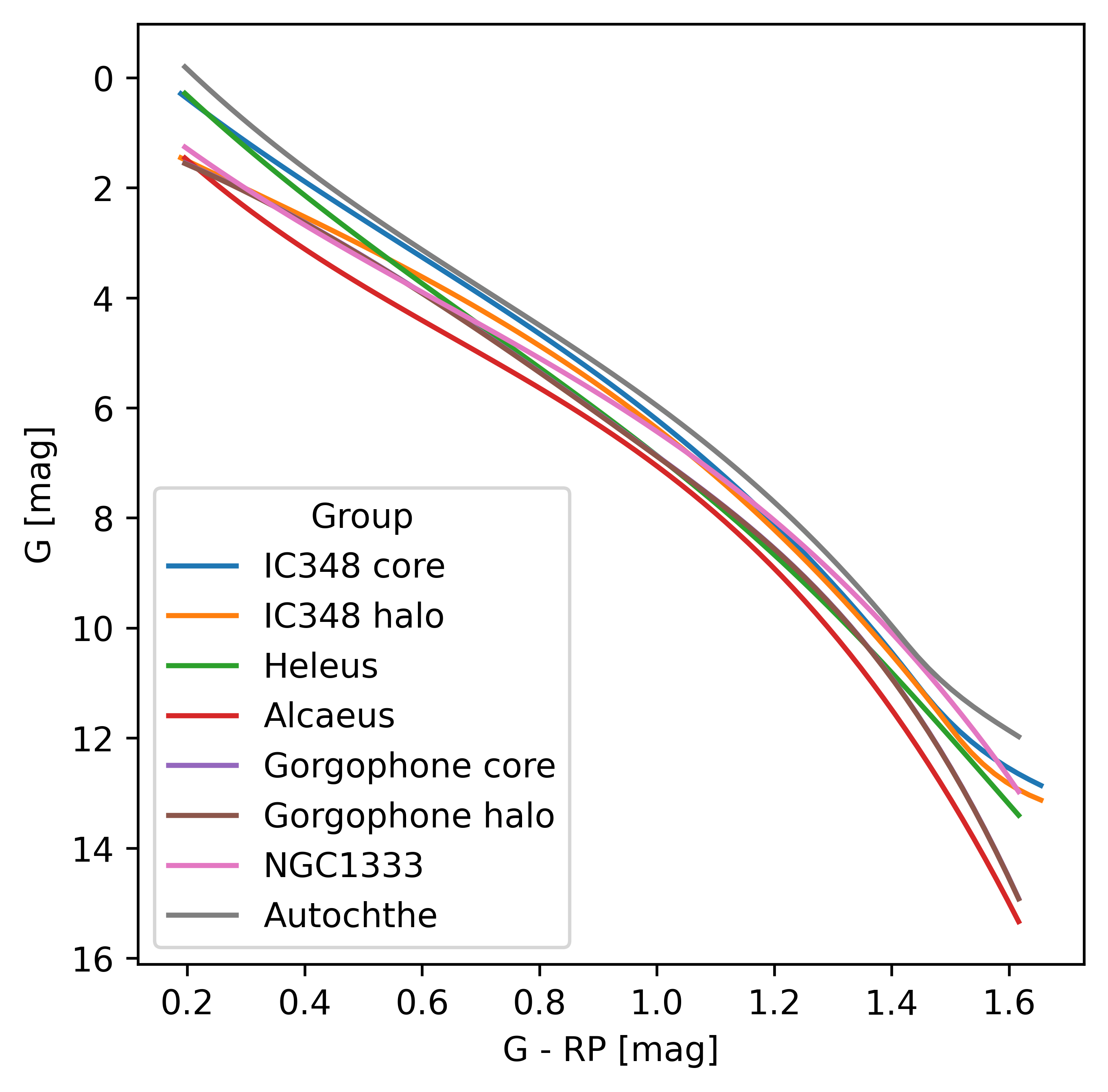}
     \caption{Empirical isochrones of the Perseus groups as obtained by \textit{Miec}.}
\label{fig:relative_ages}
\end{figure}

\begin{figure}[ht!]
    \centering
     \includegraphics[width=\columnwidth]{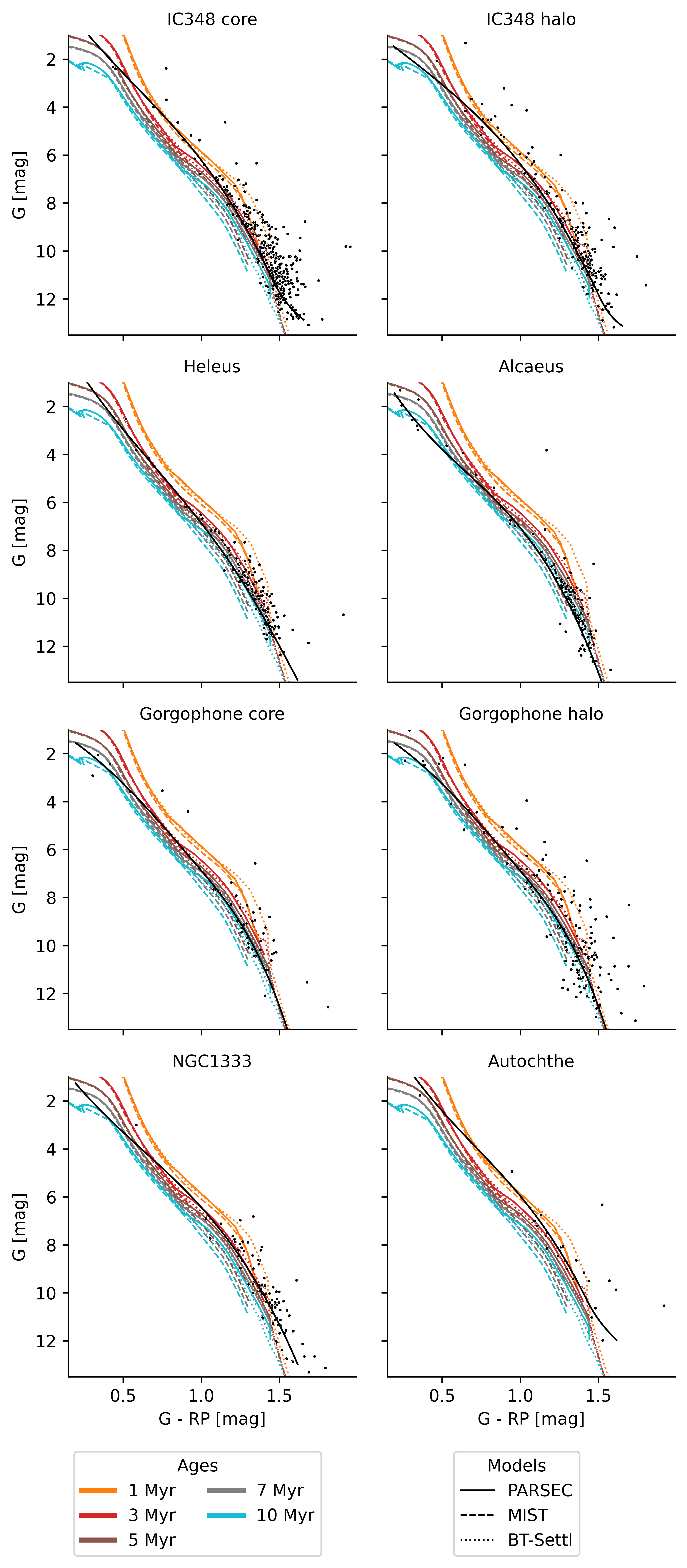}
     \caption{Absolute CMD of the Perseus groups candidate members (black dots). The black and coloured lines show the empirical isochrone and the theoretical ones of different evolutionary models, respectively.}
\label{fig:ages}
\end{figure}

As described in Sect. \ref{methods:isochrones}, the \textit{Miec} code delivers, as a by-product, the extinction-free empirical isochrone of the group under analysis. Figure \ref{fig:relative_ages} shows the empirical isochrones of the identified groups. We notice that in the case of Gorgophone, both the core and halo have the same isochrone. In addition, Fig. \ref{fig:ages} shows, for each of the groups, the absolute CMD of the candidate members (black dots) as well as their extinction-free empirical isochrones (solid black lines). The apparent magnitudes of both the candidate members and the empirical isochrones were transformed into absolute ones using the source- and group-level distances (see Sect. \ref{results:6D_structure}), respectively. The figures also show the theoretical isochrones from the PARSEC, MIST, and BT-Settl models for the ages of 1, 3, 5, 7, and 10 Myr. We notice that due to the scarcity of candidate members in the high-luminosity region (absolute G < 7 mag), the empirical isochrones do not follow the curvature of the theoretical isochrones but that of our prior, which is a simple linear regression \citep[see Fig. B.3 of ][]{2021A&A...649A.159O}.

We estimate the ages of the physical group by comparing their extinction-free empirical isochrone to the theoretical ones (see Sect. \ref{methods:isochrones}) in the faint region (G>8 mag) where the bulk of the candidate members are located. Our age estimates are shown in the fourth column of Table \ref{table:groups_members}. Given that these estimates are based on a simple visual comparison, we do not provide uncertainties. We stress the fact that these ages highly depend on the isochrone models and extinction map and thus will benefit from further refinement. 

\subsection{Mass distributions}
\label{results:mass}
\begin{figure*}[ht!]
    \centering
     \includegraphics[width=\textwidth]{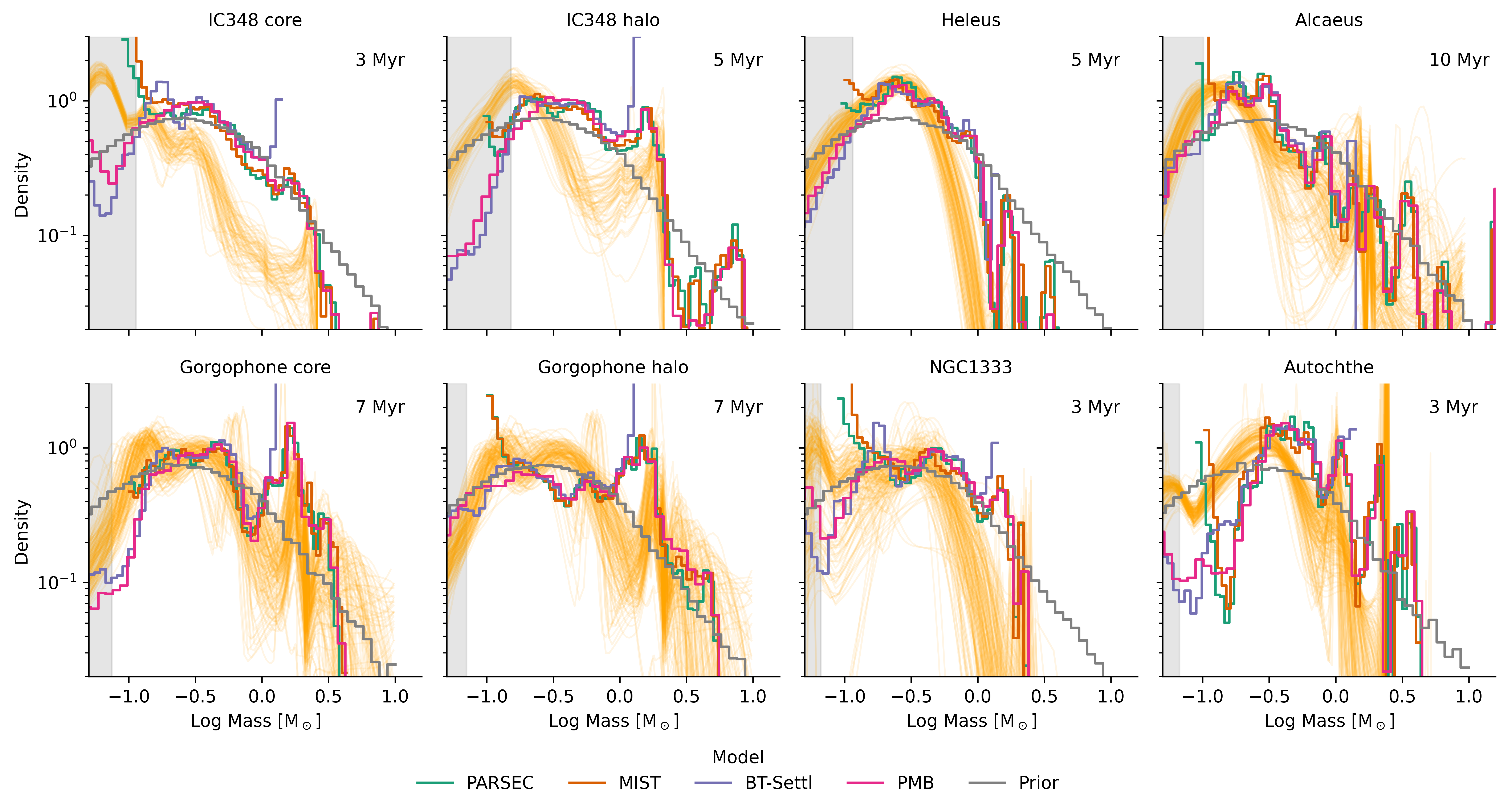}
     \caption{Mass distributions of the Perseus groups. The orange lines show 100 realisations from the mass distribution obtained after transforming \textit{Miec}'s magnitude distributions (see text). The rest of the coloured lines (those of the legend) depict the mass distributions computed with \textit{Sakam}. The grey area shows the \textit{Gaia} incompleteness region.}
\label{fig:mass}
\end{figure*}

We infer the mass distributions of each of the physical groups using the two methods described in Sect. \ref{methods:mass}. Figure \ref{fig:mass}\footnote{The electronic data to reconstruct this figure will be available at \url{www.project-dance.com}.} shows the result of these inferences. The orange lines depict a hundred realisations of the mass distribution that result from the propagation of the same number of samples taken from the posterior distributions of the \textit{Miec} parameters that describe the G band magnitude distribution. The magnitude distributions are transformed into mass distributions using the theoretical mass-magnitude relations of the unified theoretical model PMB (see Sect. \ref{methods:mass}) at the group's age and distance (see Table \ref{table:groups_members}). The mass distributions inferred using the \textit{Sakam} code with the theoretical (i.e., PARSEC, MIST, BT-Settl) and unified (PMB) models are shown in the same figure as coloured lines. The grey line shows the \citet{2005ASSL..327...41C} mass prior, and the grey area depicts the incompleteness region of the \textit{Gaia} data, which corresponds to G=19 mag \citep[see Tables 4 and 5 of][]{2021A&A...649A...2L} and has been extinction corrected.

As can be observed in Fig. \ref{fig:mass}, the mass distributions inferred with the \textit{Sakam} code agree for all the theoretical models except at the borders of their mass domains. Indeed, the peaks observed at Log Mass$[M_{\odot}]\sim-1$ and $\sim0.2$ correspond to border effects introduced by the lower limits of PARSEC and MIST models and the upper one of the BT-Settl model, respectively. As shown in the figure, the unified PMB model does not show these artefacts.

We observe that the uncertainty in the mass distributions obtained with the \textit{Miec} code (dispersion of the orange lines) is proportional to the population size of the group. The uncertainty of IC348 mass distribution is the smallest, while that of NGC1333 and Autochthe are the largest ones.

Comparing the mass distributions inferred with the two methods, we observe that the largest discrepancies are observed in IC348 while the smallest ones are in Gorgophone and Alcaeus. The extent of these discrepancies is proportional to the extinction value of the group. As will be shown in the next Section, the mode of the inferred extinction distributions is the largest in IC348 and the lowest in Gorgophone (see Fig. \ref{fig:av}). The discrepancy in the mass distributions is explained by the difficulties that the \textit{Miec} code has to infer the magnitude distributions of extincted regions under the presence of low-information-content datasets \citep[see the discussion of ][]{2021A&A...649A.159O}, in this particular case, the visual bands of \textit{Gaia}. Thus,  we expect that in the heavily-extinct groups, the mass distributions inferred with the \textit{Sakam} method are more realistic than those of the \textit{Miec} one because \textit{Sakam} uses additional photometric bands (see Sect. \ref{dataset}), in particular the infrared ones that are less affected by extinction. 

\subsection{$A_v$ and $R_v$ distributions}
\label{results:extinction_and_rv}

\begin{figure}[ht!]
    \centering
     \includegraphics[width=\columnwidth]{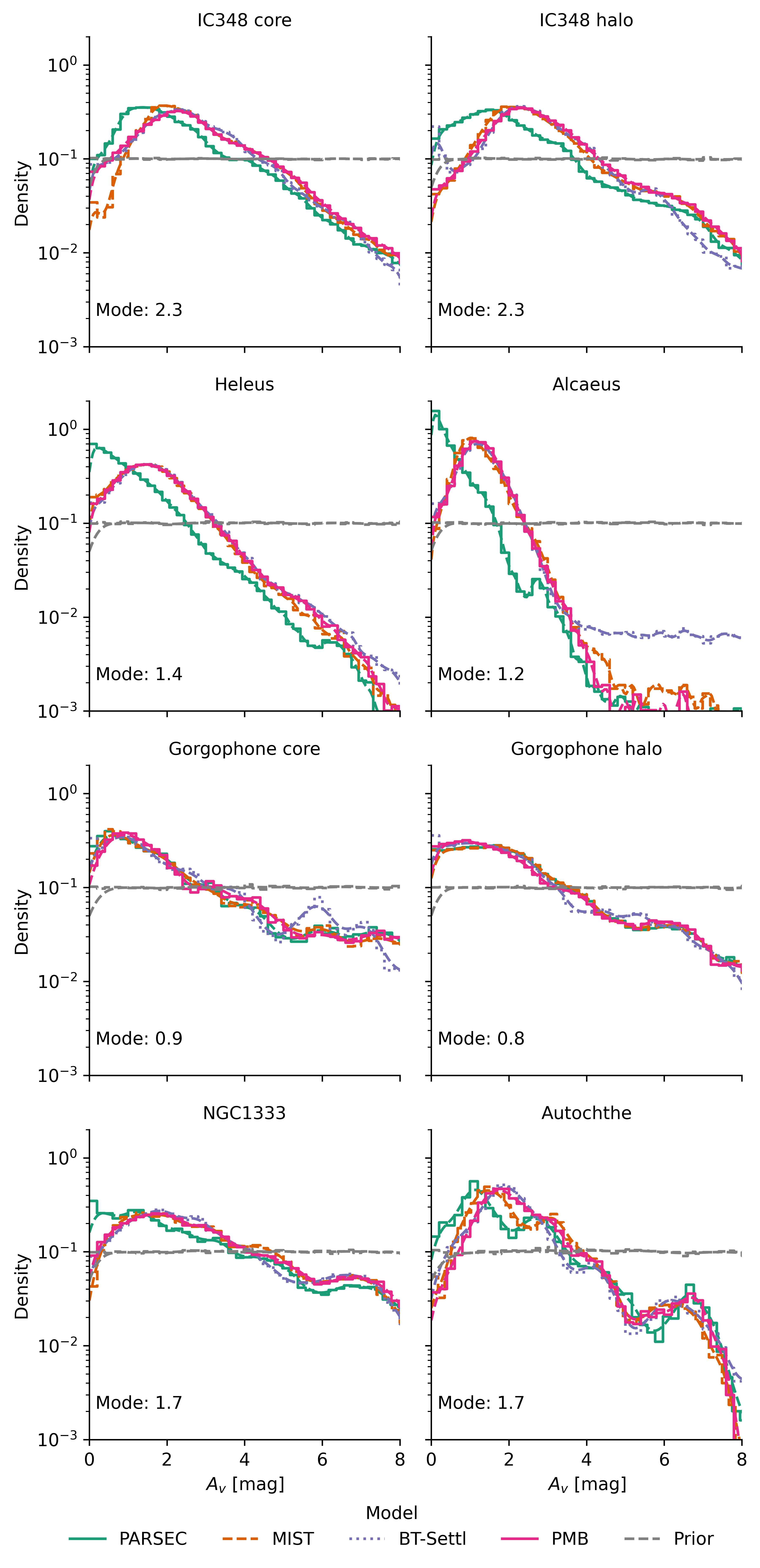}
     \caption{Extinction distributions of the Perseus groups. The color code indicates the results of the theoretical and unified models as well as the uniform prior.}
\label{fig:av}
\end{figure}

\begin{figure}[ht!]
    \centering
     \includegraphics[width=\columnwidth]{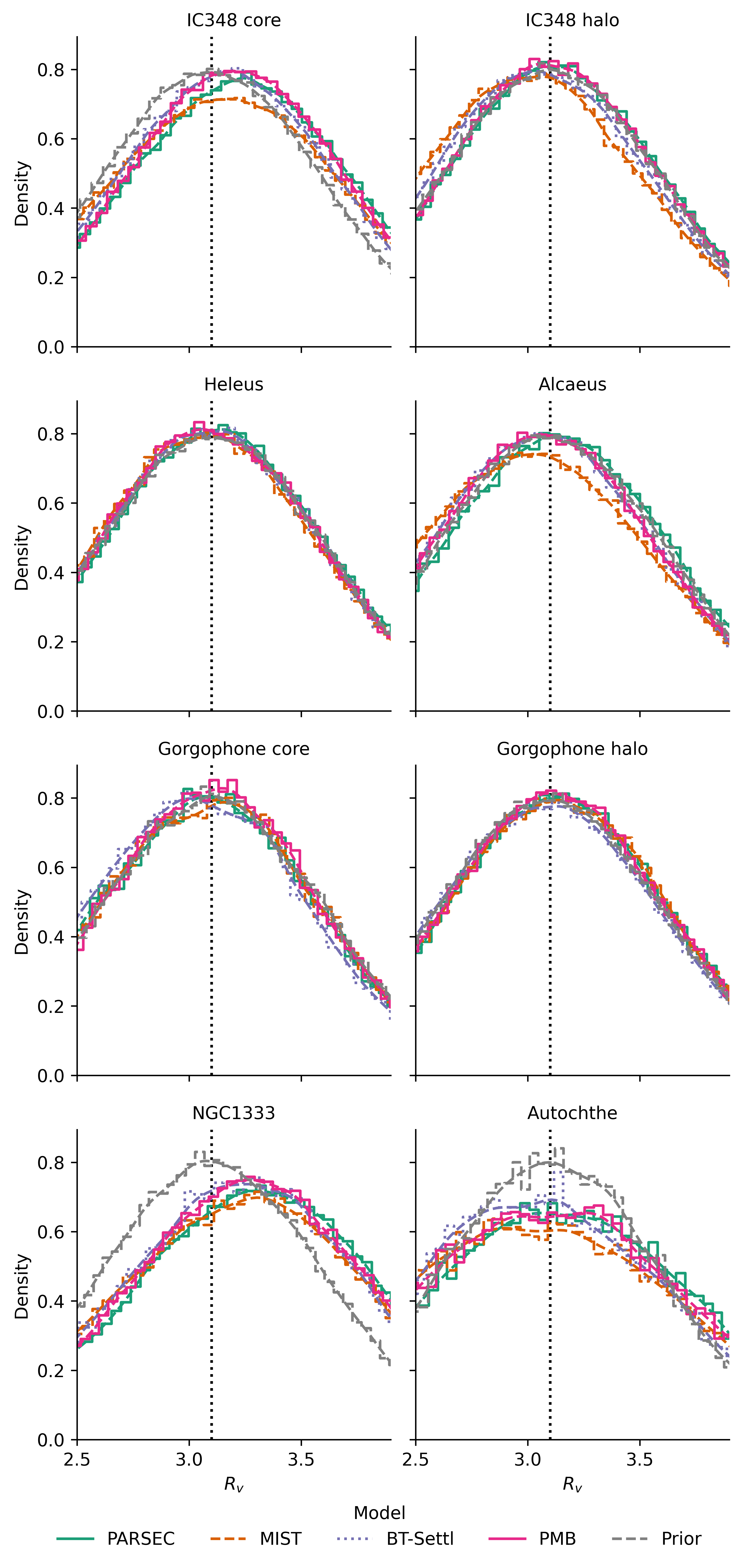}
     \caption{Rv distributions of the Perseus groups. Captions as in Fig. \ref{fig:av}. The vertical dotted line at $Rv=3.1$ depicts the mode of the Gaussian prior.}
\label{fig:rv}
\end{figure}

The mass inference done with the \textit{Sakam} code also delivers samples from the posterior distributions of the $A_v$ and $R_v$ values of each source. Figures \ref{fig:av} and \ref{fig:rv} show histograms and kernel density estimates of the posterior samples of $A_v$ and $R_v$, respectively, of all the candidate members of each physical group. As in Fig. \ref{fig:mass}, the coloured lines indicate the theoretical and unified models, and the assumed prior distribution. 

The figures show that the $A_v$ extinction is highly variable, with the mode ranging from 0.8 mag in Gorgophone to 2.3 mag in IC348. In addition, the within-group differential extinction has a large dispersion, with the exception of the Alcaeus group, in which the internal dispersion is only $\sim0.5$ mag.

We notice that the PARSEC models in all physical groups, except in Gorgophone, for which we use the 7 Myr isochrone, produces lower values of $A_v$ compared to those obtained with the MIST and BT-Settl models. This effect is a consequence of the redder theoretical isochrones of the PARSEC model compared to the other two models, particularly at the faintest magnitudes of the infrared bands (absolute K >4). However, at 7 Myr, the isochrones of the three models completely agree, which explains the negligible difference among the inferred $A_v$ values of the three theoretical models in the Gorgophone group.

Concerning the inferred distributions of $R_v$, we observe that the three theoretical models result in similar distributions in all the groups. Thus, the results are all consistent. We notice, though, that in the case of NGC1333, the mode of the distribution is shifted from that of the prior towards a larger value of 3.4. We will further discuss the consequences of this latter value in Sect. \ref{discussion:extinction}.  

\subsection{Dynamical analysis}
\label{results:dynamical_analysis}

Using the group level parameters inferred in previous sections, we now analyse the dynamical state of each of the Perseus physical groups. As mentioned in Sect. \ref{method:dynamical_analysis}, we make this analysis with two methods: the observed energy distributions and the comparison of the group's internal velocity dispersions with that expected for virial equilibrium.

\subsubsection{Energy distributions}
\label{results:energy}
We compute the energy distributions of the Perseus groups using Eq. \ref{equation:energy} and samples from the posterior mass, position, and velocity distributions of each candidate member. As explained in Sect. \ref{method:dynamical_analysis}, we correct our mass estimates by unresolved binaries (see Assumption \ref{assumption:binaries_mass} in Appendix \ref{appendix:assumptions}) and the non-stellar mass of dust and gas (see Assumption \ref{assumption:dust_mass} in Appendix \ref{appendix:assumptions}). 

\begin{figure}[ht!]
    \centering
     \includegraphics[width=\columnwidth]{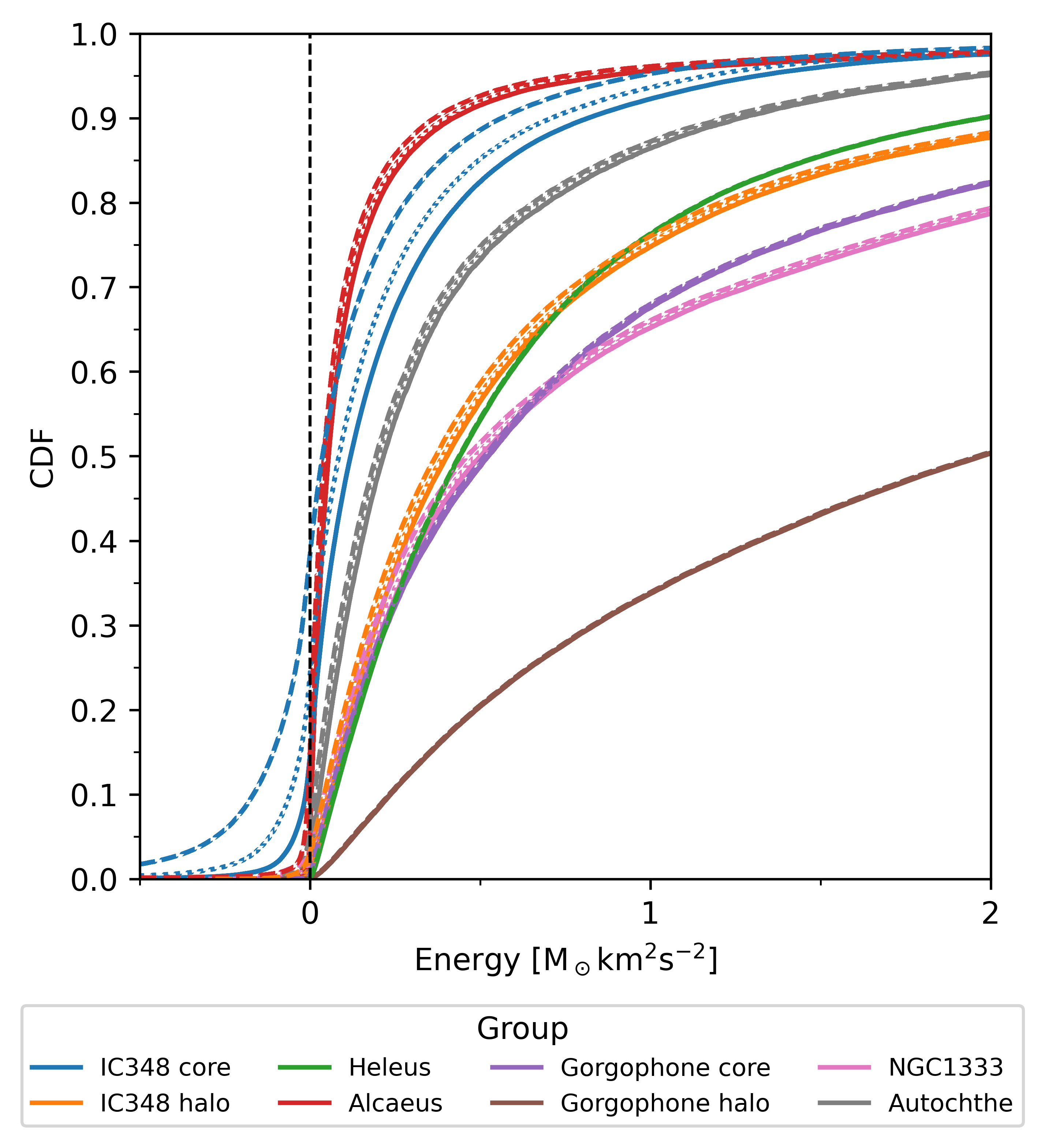}
     \caption{Cumulative energy distributions of the Perseus groups. The dotted and dashed lines depict the results of accounting for the dust and gas mass by increasing the stellar mass in 65\% and 169\%, respectively.}
\label{fig:cedfs}
\end{figure}

Figure \ref{fig:cedfs} shows the cumulative energy distribution functions (CEDFs) of the Perseus physical groups. As can be observed, all the Perseus groups have positive energies except for Alcaeus and the core of IC348. The fraction of gravitationally bound stars varies between 10\% and 25\% in Alcaeus and 15\% to 40\% in the core of IC348. In this latter group, the fraction of bounded stars depends highly on the applied correction for the non-stellar mass fraction (dust and gas), with the larger fraction of bound stars corresponding to the most massive gravitational potential, as expected. We also observe that this non-stellar mass correction has a negligible impact on energy distributions of the other Perseus groups. Given that our non-stellar mass correction follows the distribution of observed stars (see Assumption \ref{assumption:dust_mass} in Appendix \ref{appendix:assumptions}) rather than being massive particles located at the centres of the groups, then its effect is most pronounced in the most compact group, which is the core of IC348 (see the second column of Table \ref{table:kinematic_indicators}).

\subsubsection{Virial state}
\label{results:virial_state}
We now estimate the virial state of the Perseus physical groups. First, we compute the velocity dispersion, $\sigma_{vir}$, that the groups would have if they were in virial equilibrium (see Sect. \ref{method:dynamical_analysis} and Assumptions \ref{assumption:virial_equilibrium} and \ref{assumption:EFF_profile}  in Appendix \ref{appendix:assumptions}). Then, we compare these $\sigma_{vir}$ with the observed velocity dispersions $\sigma_{UVW}$ (see Sect. \ref{results:6D_structure}) and estimate the dynamical state of each group. For completeness reasons, we also compute the total mass that would be needed for the groups to be in virial equilibrium given their observed velocity dispersion.

\begin{table*}[ht!]
\caption{The virial velocity dispersions of the Perseus groups.}
\label{table:virial_state}
\resizebox{\textwidth}{!}{
\begin{tabular}{cccccccccccc}
\toprule
{} &             $r_c$ &           $r_{hm}$ & $\rm{\gamma_{EFF}}$ & $\rm{\sigma_{vir;EFF;0\%}}$ & $\rm{\sigma_{vir;EFF;65\%}}$ & $\rm{\sigma_{vir;EFF;169\%}}$ & $\rm{\sigma_{vir;HM;0\%}]}$ & $\rm{\sigma_{vir;HM;65\%}}$ & $\rm{\sigma_{vir;HM;169\%}}$ & $\rm{\pi_{vir;EFF;0\%}}$ & $\rm{\pi_{vir;HM;0\%}}$ \\
{} &              [pc] &               [pc] &                   - &       $\rm{[km\ \ s^{-1}]}$ &        $\rm{[km\ \ s^{-1}]}$ &         $\rm{[km\ \ s^{-1}]}$ &       $\rm{[km\ \ s^{-1}]}$ &       $\rm{[km\ \ s^{-1}]}$ &        $\rm{[km\ \ s^{-1}]}$ &                        - &                       - \\
\midrule
IC348 core      &   $0.69 \pm 0.18$ &    $1.41 \pm 0.67$ &     $4.38 \pm 0.50$ &             $0.57 \pm 1.10$ &              $0.71 \pm 1.37$ &               $0.89 \pm 1.71$ &             $0.40 \pm 0.43$ &             $0.50 \pm 0.53$ &              $0.62 \pm 0.67$ &                   $   6$ &                  $  13$ \\
IC348 halo      &   $5.78 \pm 2.60$ &    $6.22 \pm 3.36$ &     $8.36 \pm 4.50$ &             $0.18 \pm 0.21$ &              $0.23 \pm 0.26$ &               $0.28 \pm 0.32$ &             $0.18 \pm 0.17$ &             $0.22 \pm 0.21$ &              $0.27 \pm 0.26$ &                   $  84$ &                  $  91$ \\
Heleus          &  $16.91 \pm 6.70$ &  $27.32 \pm 15.69$ &     $9.29 \pm 4.90$ &             $0.07 \pm 0.08$ &              $0.08 \pm 0.10$ &               $0.10 \pm 0.13$ &             $0.05 \pm 0.05$ &             $0.06 \pm 0.06$ &              $0.08 \pm 0.07$ &                   $1005$ &                  $1624$ \\
Alcaeus         &   $5.26 \pm 6.10$ &    $5.90 \pm 2.64$ &     $4.83 \pm 2.40$ &             $0.17 \pm 0.07$ &              $0.21 \pm 0.09$ &               $0.26 \pm 0.12$ &             $0.16 \pm 0.18$ &             $0.19 \pm 0.22$ &              $0.24 \pm 0.28$ &                   $  21$ &                  $  24$ \\
Gorgophone core &   $3.73 \pm 3.70$ &    $6.42 \pm 2.74$ &     $2.60 \pm 3.60$ &             $0.12 \pm 0.07$ &              $0.15 \pm 0.08$ &               $0.19 \pm 0.10$ &             $0.09 \pm 0.11$ &             $0.12 \pm 0.14$ &              $0.14 \pm 0.17$ &                   $ 219$ &                  $ 377$ \\
Gorgophone halo &   $7.16 \pm 5.00$ &  $18.35 \pm 10.74$ &     $2.55 \pm 2.30$ &             $0.15 \pm 0.11$ &              $0.19 \pm 0.14$ &               $0.24 \pm 0.17$ &             $0.10 \pm 0.08$ &             $0.12 \pm 0.10$ &              $0.15 \pm 0.13$ &                   $ 609$ &                  $1562$ \\
NGC1333         &   $0.58 \pm 1.34$ &    $1.37 \pm 0.64$ &     $4.76 \pm 2.88$ &             $0.32 \pm 0.08$ &              $0.40 \pm 0.10$ &               $0.50 \pm 0.12$ &             $0.21 \pm 0.23$ &             $0.26 \pm 0.29$ &              $0.33 \pm 0.36$ &                   $  54$ &                  $ 127$ \\
Autochthe       &   $1.65 \pm 2.69$ &    $1.91 \pm 0.91$ &     $8.25 \pm 6.05$ &             $0.13 \pm 0.05$ &              $0.17 \pm 0.06$ &               $0.21 \pm 0.07$ &             $0.12 \pm 0.14$ &             $0.15 \pm 0.17$ &              $0.19 \pm 0.21$ &                   $  61$ &                  $  70$ \\
\bottomrule
\end{tabular}

}
\end{table*} 

Table \ref{table:virial_state} shows, for each Perseus group, the EFF's core radius ($r_c$) and $\gamma$ parameters, the half-mass radius ($r_{hm}$), the virial ($\sigma_{vir}$) velocity dispersions, and the mass factors ($\pi_{vir}$) that should multiply the observed stellar mass for the group to be in virial equilibrium (assuming a binary mass fraction of 20\%). In the virial velocity dispersions and mass factors, we show the values computed using the core and half-mass radii (sub-indices EFF and HM, respectively) as well as the original, lower, and upper limits of the cluster mass (sub-indices 0\%, 65\%, and 169\%, respectively). As expected, the most massive clusters (i.e., those with the 169\% non-stellar mass correction) have larger virial velocity dispersions. Comparing the observed velocity dispersions (third column of Table \ref{table:kinematic_indicators}) with the virial ones, we observe that all Perseus groups are in a super-virial state. In other words, for the groups to be in virial equilibrium, their stellar mass should be increased in factors ranging from 6 times to a thousand times.  We will further discuss these results in Sect. \ref{discussion:dynamical_analysis}.

\section{Discussion}
\label{discussion}
We now compare our results with those from the literature and discuss their differences. We proceed in the same order as in the previous sections. 
\subsection{Membership}
\label{discussion:members}

\begin{figure}[ht!]
    \centering
     \includegraphics[width=\columnwidth]{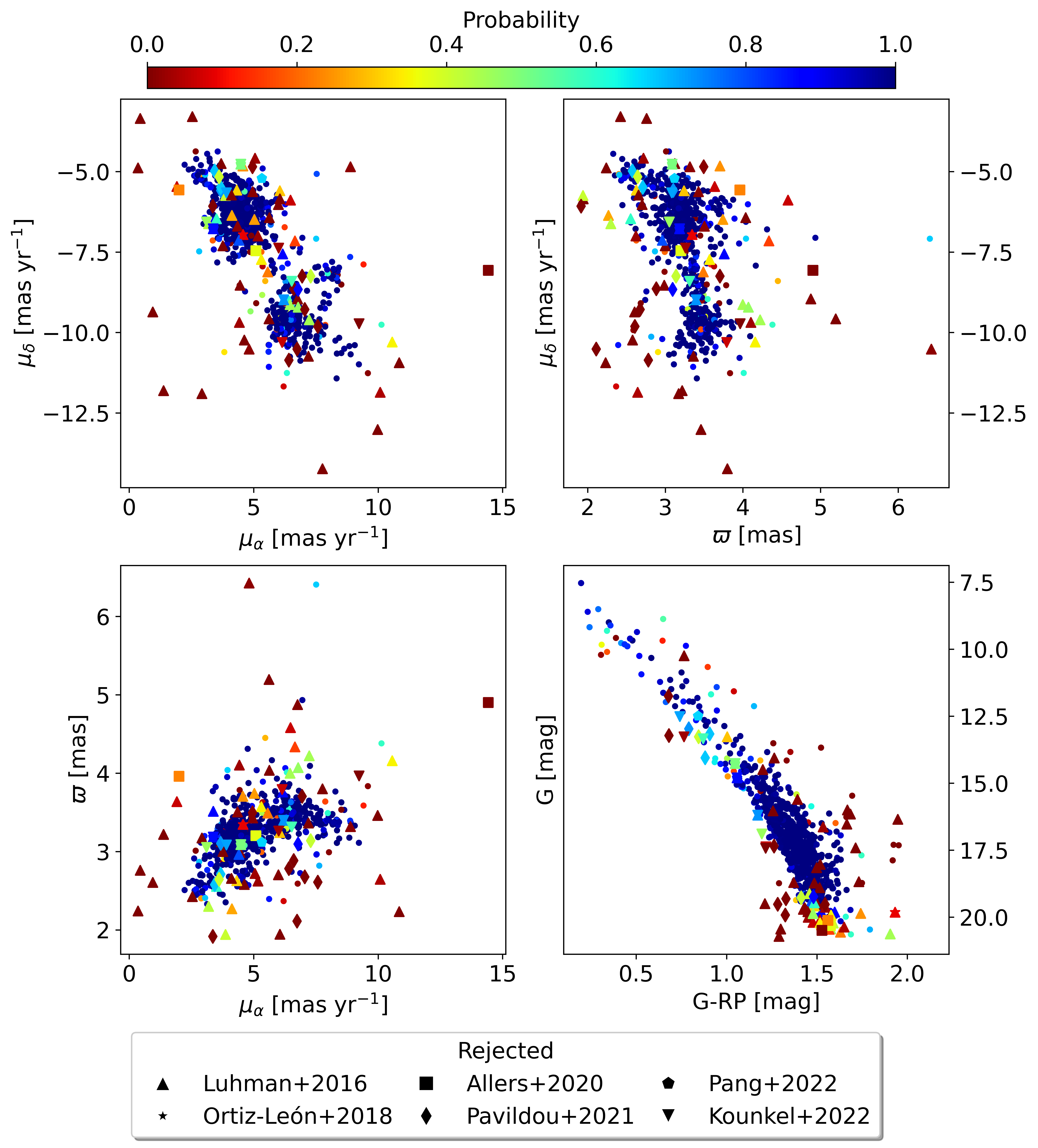}
     \caption{Representation space coordinates of the confirmed (dots) and rejected members from the literature after removal of clear outliers ($\varpi\notin[1.9,6.5]$ mas). The membership probability is shown as a colour code.}
\label{fig:comparison_literature}
\end{figure}

When comparing lists of candidate members is important to highlight that the parallax is the most discriminant feature that \textit{Gaia} DR3 provides to identify candidate members. Thus, we use it to exclude clear outliers from previous results of the literature. We consider that sources with parallaxes lower than 1.9 mas (farther than 526 pc) and higher than 6.5 mas (closer than 178 pc) belong to the field population. We choose this highly conservative parallax interval to ensure the inclusion of possible members within more than -100 pc and +200 pc around the traditional Perseus distance \citep[320 pc,][]{2018ApJ...865...73O}. Figure \ref{fig:comparison_literature} shows the representation space coordinates of our candidate members that are common with (dots) and rejected from the literature works. The colour code shows the membership probability of the sources. The following paragraphs present detailed comparisons of our candidate members with selected works from the literature.

\citet{2016ApJ...827...52L} identified 478 and 203 members in IC348 and NGC133, respectively. Only 450 of these members have \textit{Gaia} DR3 parallaxes, proper motions, and photometry. Of these, 16 are clear outliers according to our parallax limits. Our methodology recovers 380 and rejects 54 of their candidate members. As can be seen in Fig. \ref{fig:comparison_literature}, most of the rejected sources lay on the outskirts of the group's loci.

From the list of members of \citet{2018ApJ...865...73O} 162 have a counterpart in our data set. Out of these, we recover 158 as candidate members and reject one due to its discrepant parallax. We identify the remaining three sources as false negatives of our methodology, given that their coordinates are consistent with the locus of the Perseus groups.

Sixteen of \cite{2020PASP..132j4401A} members are in our data set, and four of them are clear parallax outliers. Of the remaining sources, nine are recovered by our methodology, and three were rejected. Of the latter, only one can be identified as a false negative of our methodology since the other two lay at the outskirts of the groups' loci.

\citet{2021MNRAS.503.3232P} identified 913 candidate members in their five groups, which do not include IC348 or NGC1333. Contrary to the previous authors, our main objective is the analysis of these latter two clusters, and we focus on a smaller region (see Sect. \ref{dataset}). This region contains 183 of their candidate members, and 2 of them are clear parallax outliers. Our methodology recovers 166 of their candidate members and rejects 15. Half of the rejected candidates lay slightly below our probability thresholds (blue and green diamonds) and can be considered false negatives by our methodology.

\citet{2022AJ....164...57K} identified 810 candidate members in the Perseus region, but only 429 are in our sky region and 425 in our data set\footnote{The four missing members in our sky region lack both \texttt{BP} and \texttt{RP} bands, which prevent us from using our membership methodology. Their \textit{Gaia} \texttt{source\_id} are: 123998252751482368, 216690175350832000,
216443575506877824, and 216443884747397504.}. From the latter, we recover 416 and reject nine. Only four of the rejected have membership probabilities above 0.5, which can be considered false negatives of our methodology.

\citet{2022ApJ...931..156P} found 211 candidate members to IC348, 195 of these are within our sky region, and we recover 192 of them. The three rejected sources have probabilities larger than 0.5 and thus can be considered false negatives of our methodology.

\cite{2022ApJ...936...23W} found 207 candidate members, out of which 197 are in our sky region, and we recover 191 of them. Out of the six rejected, two have negligible membership probabilities, while the remaining four have membership probabilities >0.5 but lower than our probability threshold. Unfortunately, these authors do not provide a classification of their candidate members into the Perseus groups. Therefore, we cannot make more detailed comparisons with our candidate members to the Perseus groups.

As can be observed in Fig. \ref{fig:comparison_literature}, there are sources with very low membership probabilities that lay at the outskirts of the cluster locus and that we consider field population. Nonetheless, there are 12 sources from \citet{2016ApJ...827...52L}, three of \citet{2018ApJ...865...73O}, one of \citet{2020PASP..132j4401A}, five of \citet{2021MNRAS.503.3232P}, four of \citet{2022AJ....164...57K} and three of \citet{2022ApJ...931..156P} with coordinates consistent with the cluster locus that lay below our probability thresholds, we consider these members as false negatives of our methodology. Under the conservative assumption that all the rejected members from the literature (excluding clear outliers with discrepant parallaxes, as described above) are indeed true Perseus members, then the true positive rate of our methodology is 93.2\%, which is better than the 88\% estimated by \citet[][see their Table C.3]{2021A&A...649A.159O} for a cluster with $A_v\sim 6$ mag. The better performance of the current application of the \textit{Miec} code results from the use of the highly discriminant parallax feature and the exquisite precision of the \textit{Gaia} DR3 astrometry. Finally, we highlight that our membership methodology recovers 267 candidate members not previously identified as such by the literature. This represents an increase of 31\% with respect to the previous studies of the Perseus region.

\subsubsection{Group's classification}
We now compare the list of members that we recover for IC348 (core and halo) and NGC1333 with those found in other works. Tables \ref{table:ic348_literature} and \ref{table:ngc1333_literature} show, for IC348 and NGC1333, respectively, the number of members from other works that are in our dataset and that we recover and reject. The third column of these tables shows the number of recovered members together with those sources that we still recover as members of the Perseus region but belonging to another group (shown with the first two letters of its name).

\begin{table}[ht!]
\caption{The number of IC348 members in the literature, in our dataset, and in this work.}
\label{table:ic348_literature}
\centering
\resizebox{\columnwidth}{!}{
\begin{tabular}{c|c|c|c|c}
\toprule
Work & Members & In dataset & Recovered & Rejected \\
\midrule
\text{\citet{2016ApJ...827...52L}} & 478 & 357 & 285 + 8Go + 7Al + 6He & 51 \\
\text{\citet{2018ApJ...865...73O}} & 133 & 131 & 126 + 1Go + 1He       &  3 \\
\text{\citet{2018A&A...618A..93C}} & 144 & 144 & 140 + 1He + 1Al       &  2 \\
\text{\citet{2020PASP..132j4401A}} & 19  & 13  & 7                     &  6 \\
\text{\citet{2022ApJ...931..156P}} & 211 & 195 & 190 + 1Go + 1He       &  3 \\
\text{\citet{2022AJ....164...57K}} & 270 & 263 & 250 + 4Go + 5He       &  4 \\
\bottomrule
\end{tabular}
}
\tablefoot{{\scriptsize Al = Alcaeus, He = Heleus, Go = Gorgophone}}
\end{table}

\begin{table}[ht!]
\caption{The number of NGC1333 members in the literature, in our dataset, and in this work.}
\label{table:ngc1333_literature}
\centering
\resizebox{\columnwidth}{!}{
\begin{tabular}{c|c|c|c|c}
\toprule
Work & Members & In dataset & Recovered & Rejected \\
\midrule
\text{\citet{2016ApJ...827...52L}} & 203 & 93 & 68 + 6Go & 19 \\
\text{\citet{2018ApJ...865...73O}} & 31  & 31 & 28 + 2Go &  1 \\
\text{\citet{2018A&A...618A..93C}} & 50  & 50 & 46 + 3Go &  1 \\
\text{\citet{2020PASP..132j4401A}} & 9   & 3  & 3        &  0 \\ 
\text{\citet{2022AJ....164...57K}} & 34  & 34 & 30 + 3Go &  1 \\
\bottomrule
\end{tabular}
}
\tablefoot{{\scriptsize Go = Gorgophone}}
\end{table}

In addition to the well-known IC348 and NGC1333 groups, the Perseus region contains several substructures that have been recently identified thanks to the high-precision \textit{Gaia} data. For example, \citet{2021MNRAS.503.3232P} identified five Perseus groups, named Alcaeus, Autochthe, Electryon, Heleus, and Mestor. In the rest of this subsection, we discuss the recovery of the members of these groups, with the exception of the Mestor one, whose members are outside the sky region that we analyse. 

Concerning the Electryon group, out of the 329 members of \citet{2021MNRAS.503.3232P}, only 21 are in our dataset. Our methodology recovers 15 of these, although classified as members of the Heleus group. The fact that our methodology classifies these Electryon members into the Heleus group is explained by the small fraction of the former group within our dataset (6\%), its negligible contribution to the Perseus total number of members ($<1$\%), and the entanglement that these two group show in the parallax-proper-motions space. We notice that \citet{2021MNRAS.503.3232P} were able to disentangle Heleus from Electryon due to its relative compactness in the sky coordinates, however, our methodology identifies members of this group spread over a larger sky region.

Concerning the Heleus group, our methodology recovers the 29 members of \citet{2021MNRAS.503.3232P} present in our dataset, and we identify 95 more candidate members. With respect to \citet{2022AJ....164...57K}, out of their 66 candidate members, our dataset contains 22, of which we recover 21, all of them belonging to this group. In addition, we identify 103 more candidate members than these authors.

With respect to Alcaeus, from the 170 members of \citet{2021MNRAS.503.3232P}, 108 are in our dataset, and our methodology recovers 98 of these, with 88 classified into Alcaeus and 10 in Gorgophone. \citet{2022AJ....164...57K} identified 124 candidate members, out of which 83 are in our dataset. We recover 81 of these, although 20 of them are classified into Gorgophone. In addition, our method discovers 64 new candidate members of this group.

Finally, from the 27 members of Autochthe identified by \citet{2021MNRAS.503.3232P}, 25 are in our dataset, and our method recovers 24 of them, with two of them classified into Gorgophone. \citet{2022AJ....164...57K} found 25 candidate members, with 23 of these in our dataset. Our methodology recovers 22 of these, although with one belonging to NGC1333 and 2 to Gorgophone.

As shown in the previous paragraphs, further disentanglement of the Perseus substructures remains a difficult task in which more precise data, particularly radial velocities, and further methodological developments will still be needed.

\subsection{Phase-space structure}
\label{discussion:6D_structure}
We now compare the distribution of positions and velocities of the identified physical groups with those from the literature. The most striking difference with respect to previous pre-\textit{Gaia} works is the number of identified substructures. Classically the region was comprised of IC348 and NGC1333, however, the arrival of \textit{Gaia} unravelled multiple populations. Compared to the latest analyses from the literature \citep{2022ApJ...936...23W,2022AJ....164...57K,2021MNRAS.503.3232P}, we identified one additional physical group that we call Gorgophone, which is composed of a core and halo populations. 

In IC348, our kinematic criterion to identify physical groups (see Assumption \ref{assumption:gaussian} in Appendix \ref{appendix:assumptions}) allow us to conclude that the core and halo of IC348 are distinct physical groups. The existence of these two populations is further supported by the age and mass distribution features of these populations, with the halo population being older and having fewer low-mass stars and more high-mass stars than the core. In addition, the most massive stars of the halo population are located off-centre while those in the core are centrally concentrated. These findings support a distinct halo population that formed earlier than the core population and possibly quenched the formation of high-mass stars in this latter.

\subsubsection{3D distribution}
\label{discussion:3d_structure}
Our results show that the core of IC348 has a ($1\sigma$) radius of $0.65\pm0.20$ pc, with small correlations ($<0.3$) among the X, Y, and Z coordinates. These negligible correlations are in agreement with previous works that found a centrally concentrated distribution (see Sect. \ref{intro:spatial_distribution}). On the contrary, the halo extends over a ($1\sigma$) radius of $6.89\pm0.75$ pc, with non-negligible correlations of 0.3 between X and Y, 0.5 between X and Z, and 0.7 between Y and Z. 

Concerning IC348, our values of the halo radius are larger than the 0.9-1.4 pc radius reported by \citet{1999A&AS..137..305S}, \citet{2000AJ....120.3139C}, and \citet{2003AJ....125.2029M}. On the other hand, our 3D core radius estimate is in agreement with the 2D radius of 0.5~pc reported by \citet{2003AJ....125.2029M}. These latter authors also warned about the possibilities of a larger cluster radius and possible entanglement between the halo sources and the core ones due to projection effects. Evidence of a larger cluster size was also found by \citet{2015ApJ...807...27C}, who identified 63 of their candidate members beyond a radius of 1.8 pc radius (20\arcmin). Our results confirm the following suggestions proposed by \citet{2003AJ....125.2029M}: a larger cluster size extending into larger areas and elongated in the north-south direction, halo sources entangled within the core radius (see Fig. \ref{fig:sky}), and a density profile that is far from trivial. On the other hand, we do not find evidence of substructures within the halo populations, as suggested by the aforementioned authors.

Contrary to IC348, NGC1333 is well described by a single Gaussian. It shows an elongation twice as large in the X coordinate that in the Y and Z, directions, although with negligible correlations and unrelated to the line of sight elongation. Given the cluster's young age and the low number of stars (see Table \ref{table:groups_members}), the observed elongation can be primordial.  

Concerning distances to IC348 and NGC1333, our estimates are in agreement with those of \citet{2018ApJ...865...73O} and \citet{2018ApJ...869...83Z}. The surprising agreement between the early distance estimate of $316\pm22$ pc made by \citet{1974PASP...86..798S} and our $315\pm1$ pc  \textit{Gaia} value for the core of IC348 further supports the hypothesis proposed by those authors about dust grain growth in the dense Perseus clouds (see Sect. \ref{intro:extinction_distribution} and \ref{discussion:extinction}).

We notice that \citet{2018ApJ...865...73O} warned about the impossibility of disentangling the cluster structure along the line of sight due to the fact that the parallax dispersion was dominated by individual uncertainties. However, this is no longer the case, thanks to the high precision of the \textit{Gaia} DR3 and the uncertainty deconvolution applied by \textit{Kalkayotl}.

\subsubsection{Velocity distribution}
\label{discussion:velocity_distribution}

Our results on the velocity distributions of the Perseus groups are shown in Tables \ref{table:groups_sd} and \ref{table:kinematic_indicators}. As can be observed, the total velocity dispersion (third column of Table \ref{table:kinematic_indicators}) of these groups varies from 0.76 $\rm{km\, s^{-1}}$ in Alcaeus to 3.79 $\rm{km\, s^{-1}}$ in the halo of Gorgophone.  

Comparing the halo velocity dispersions of IC348 and Gorgophone with respect to their cores, we observe that in IC348, these values are similar (and compatible within 2$\sigma$ uncertainties), whereas, in Gorgophone, the halo velocity dispersion is significantly more than twice that of its core. The relatively large velocity dispersion and size of Gorgophone's halo to that of IC348, in combination with their similar ages, suggests that different mechanisms have formed these halo populations. We hypothesise that the core and halo of IC348 formed out of the same molecular cloud but at different ages, thus, although physically independent, they inherit similar velocity dispersions as those of the parent molecular cloud. On the contrary, the core and halo of Gorgophone pertain to the same physical group, but the halo-core dichotomy formed as a result of the dynamical interactions, thus the halo is dynamically hotter, and the core is relatively depleted of low-mass stars (see Fig. \ref{fig:mass}).

\citet{2015ApJ...807...27C} fitted the radial velocity profile of IC348 with a GMM with two components (see Sect. \ref{intro:velocity}). They argue that the presence of the second component may have three possible origins: contaminants, a halo population, or lack of relaxation into a Gaussian distribution due to the young age of the cluster. Unfortunately, given the 1.8 pc radius cut that these authors applied to their sample, their results can not be directly compared with those of the IC348 core that we found here. Indeed, at that radius, we find a mixture of both core and halo populations. Our results confirm the hypothesis of the halo population, which we found in the 6D space of positions and velocities rather than only in the 1D radial velocity, as those authors hypothesised. On the other hand, the contamination scenario can be ruled out thanks to our robust membership methodology, in which the expected contamination rate of this group is $\lesssim$ 7\%.

In the case of NGC1333, \citet{2015ApJ...799..136F} used the same methodology as \citet{2015ApJ...807...27C} and found that the radial velocity distribution can be well described by a single Gaussian component. These results agree with our single Gaussian model for the phase-space structure of NGC1333.

\subsubsection{Internal kinematics}
\label{discussion:internal_kinematics}
Concerning IC348, in its core, we find a 3D velocity dispersion of $1.44\pm0.08\, \rm{km\, s^{-1}}$, which is similar than the 3D $1.2 \pm0.1\rm{km\, s^{-1}}$ value ($0.72\pm0.07\,\rm{km\, s^{-1}}$ for the 1D radial velocity dispersion) found by \citet{2015ApJ...807...27C}, and smaller to the 2.28 $\rm{km\, s^{-1}}$ reported by \citet{2018ApJ...865...73O}. On the one hand, the smaller value found by \citet{2015ApJ...807...27C} can be explained by their better treatment of binaries, which in our analysis are not distinguished from single stars. On the other hand, the larger value found by \citet{2018ApJ...865...73O} can be explained by their use of a single Gaussian to model stars in both the core and halo populations. In NGC1333, the 3D velocity dispersion of $1.6\pm0.2\,\rm{km\, s^{-1}}$ ($0.92\pm0.12\,\rm{km\, s^{-1}}$ for the 1D radial velocity dispersion) and $2.0\,\rm{km\, s^{-1}}$ reported by \citet{2015ApJ...799..136F} and \citet{2018ApJ...865...73O}, respectively, are smaller than and compatible with our $2.37\pm0.23\rm{km\, s^{-1}}$ value, respectively (see Table \ref{table:kinematic_indicators}). We notice that the 3D velocity ellipsoid of NGC1333 is 33\% wider in the Y direction than in the Z direction, which makes it far from spherical, as assumed by \citet{2015ApJ...799..136F}. On the other hand, the velocity dispersions measured by us and by \citet{2018ApJ...865...73O} are also probably inflated by unresolved binaries. Future work is warranted upon the arrival of more precise radial velocity measurements.

With respect to expansion and contraction, our results show that the observed precision is not enough to claim any clear trend. Nonetheless, we notice that the largest values of contraction are those of the core of IC348, Alcaeus, and NGC1333. Concerning the signature of internal rotation, our results indicate that it is significant only at the one-sigma level. In IC348, \citet{2015ApJ...807...27C} have already reported a small but significant radial velocity rotation gradient in the plane of the sky: $0.024\pm0.013\,\rm{km\, s^{-1}\, arcmin^{-1}}$ or $0.26\pm0.14\,\rm{km\, s^{-1}\, pc^{-1}}$. Scaling their 1D gradient to 3D values at the cluster core and halo radii, we obtain $0.3\pm0.1\,\rm{km\, s^{-1}}$ and $3.1\pm0.1\,\rm{km\, s^{-1}}$, respectively. Although the rotation values we find for the core and halo of IC348 are larger and smaller, respectively, than those reported by \citet{2015ApJ...807...27C}, they still are not significant enough to rule out the null hypothesis of no internal rotation. Moreover, \citet{2018ApJ...865...73O} also found non-zero values for the rotation vectors of IC348 and NGC1333. However, they concluded that their values were negligible when compared to the internal velocity dispersion.  Similarly, \cite{2019ApJ...870...32K} found non-zero azimuthal velocities in NGC1333 and IC348, but they also judged these as non-significant when compared to the measured uncertainties. In the light of the previous works and current evidence, future efforts are still needed, both on the observational and modelling sides, to test the hypothesis of internal rotation in the Perseus groups.

\subsubsection*{The core and halo of IC348}
As mentioned in Sect. \ref{intro:spatial_distribution} the core and halo populations were first found by \citet{1998ApJ...497..736H} on the base of the spatial distribution of stars with $H_{\alpha}$ emission. He also found an age gradient of 2.4 $\rm{Myr \, pc^{-1}}$ in which older ages are located at larger radii (1.45 Myr at 4' and 2.8 Myr at 10'). Here, we confirm the existence of these two IC348 populations and observe a smaller but still non-negligible age gradient of $0.32\,\rm{Myr\, pc^{-1}}$ computed on the base of the ages and radii of the core and halo.

We notice that \citet{2015ApJ...807...27C} found a correlation between the radial velocity and the reddening in IC348, for which they proposed three different scenarios: a systemic offset in the radial velocities, the contraction of the cluster as a whole, or the convergence of two subclusters aligned in the line of sight. Their third scenario naturally coincides with the core and halo populations proposed by \citet{1998ApJ...497..736H} and confirmed here. These two populations naturally solve the previous issue because the halo population is red-shifted 1.6 $\rm{km\,s^{-1}}$ with respect to the core population and has a slightly larger spread in extinction (see Fig. \ref{fig:av}). 

Furthermore, we observe that the halo contains slightly more massive stars than the core (see Fig. \ref{fig:ages} and \ref{fig:mass}). This effect was already predicted by \citet{1998ApJ...497..736H} when he mentioned that the brightest stars from the halo formed first, and then the molecular cloud retreated behind them, where now the youngest population is formed. 

\subsection{Empirical isochrones and age estimates}
\label{discussion:isochrones}
Our age estimates for NGC1333 and IC348 are compatible with the literature values. In the case of IC348, our 3 Myr estimate for the core is slightly larger than the 2 Myr median value of \citet{2003ApJ...593.1093L}. Similarly, the 5 Myr age of IC348 halo is similar to the 6 Myr reported by \citet{2015MNRAS.454..593B}, and in agreement with the recent 5 Myr age determination of \citet{2022ApJ...931..156P} and \cite{2022AJ....164..125L}. Both our core and halo age estimates fall within the age interval reported by \citet{2003ApJ...593.1093L}. In the case of NGC133, our age estimate of 3 Myr is older but still compatible with the 1-2 Myr derived by \citet{1996AJ....111.1964L}. Furthermore, our age estimates for the halo of IC348 and NGC1333 are in clear agreement with those of  \citet[][see Sect. \ref{intro:ages}]{2022ApJ...936...23W}, who determined isochrone ages of 5.4 Myr and 2.9 Myr for IC348 and NGC1333, respectively. We notice that this agreement is based not only on the theoretical isochrone models but on the similar extinction distribution, as observed when comparing their Fig. 7 to our Fig. \ref{fig:av}.

Comparing the ages estimated by \citet{2021MNRAS.503.3232P} for the newly identified Perseus groups with those obtained here, we observe that the relative internal ages are in agreement. Furthermore, we confirm the coevality of NGC1333 and Authochte. However, the absolute ages differ by a factor of roughly two,  with our estimates being twice older than theirs. In the particular case of the absolute age estimates obtained by \citet{2021MNRAS.503.3232P}, we conclude that their underestimation is a direct consequence of their lack of extinction correction. According to the extinction maps of \citet{2019ApJ...887...93G}, the members of Perseus groups have extinctions as large as 7 mag in $A_v$, with a median value of $A_v\,=\,2.6\pm1.9$ mag. Neglecting these extinction values when estimating isochronal ages results in younger age determinations. The same reasoning applies to the more recent isochrone age determinations by  \citet{2022AJ....164...57K}.

\subsection{Mass distributions}
\label{discussion:mass}

As mentioned in Sect. \ref{results:mass}, the mass distributions obtained with our two methods are compatible within the uncertainties, except in those groups with the largest peaks in extinction, which are the core and halo of IC348, NGC1333, and Autochthe (see Fig. \ref{fig:av}). The differences in the inferred mass distributions arise from the difficulties that the \textit{Miec} method has to recover luminosity distribution of extincted regions based on data sets lacking infrared bands. The \textit{Sakam} method complements \textit{Gaia} data with infrared photometry and thus is not affected by this problem.

Comparing \textit{Sakam} mass distributions of all groups within \textit{Gaia} completeness limits, we observe that there is a general agreement with \citet{2005ASSL..327...41C} initial mass distribution. This comes as no surprise given the young ages of the groups. Nonetheless, we notice a depletion of stars more massive than $\sim$5$\rm{M_{\odot}}$ in Autochte, NGC1333, Heleus and the cores of IC348 and Gorgophone compared to the values expected from \citet{2005ASSL..327...41C} distribution. Unexpectedly, the halo populations of IC348 and Gorgophone seem to have more massive stars than the core populations. In almost all groups, there is a visible peak at masses ranging from 1.5$\rm{M_{\odot}}$ to 2.5$\rm{M_{\odot}}$, which results from abrupt changes in the slope of the mass-luminosity relation of the PARSEC and MIST models at this particular mass interval (see Sect. \ref{methods:mass}).

\begin{figure}[ht!]
    \centering
     \includegraphics[width=\columnwidth]{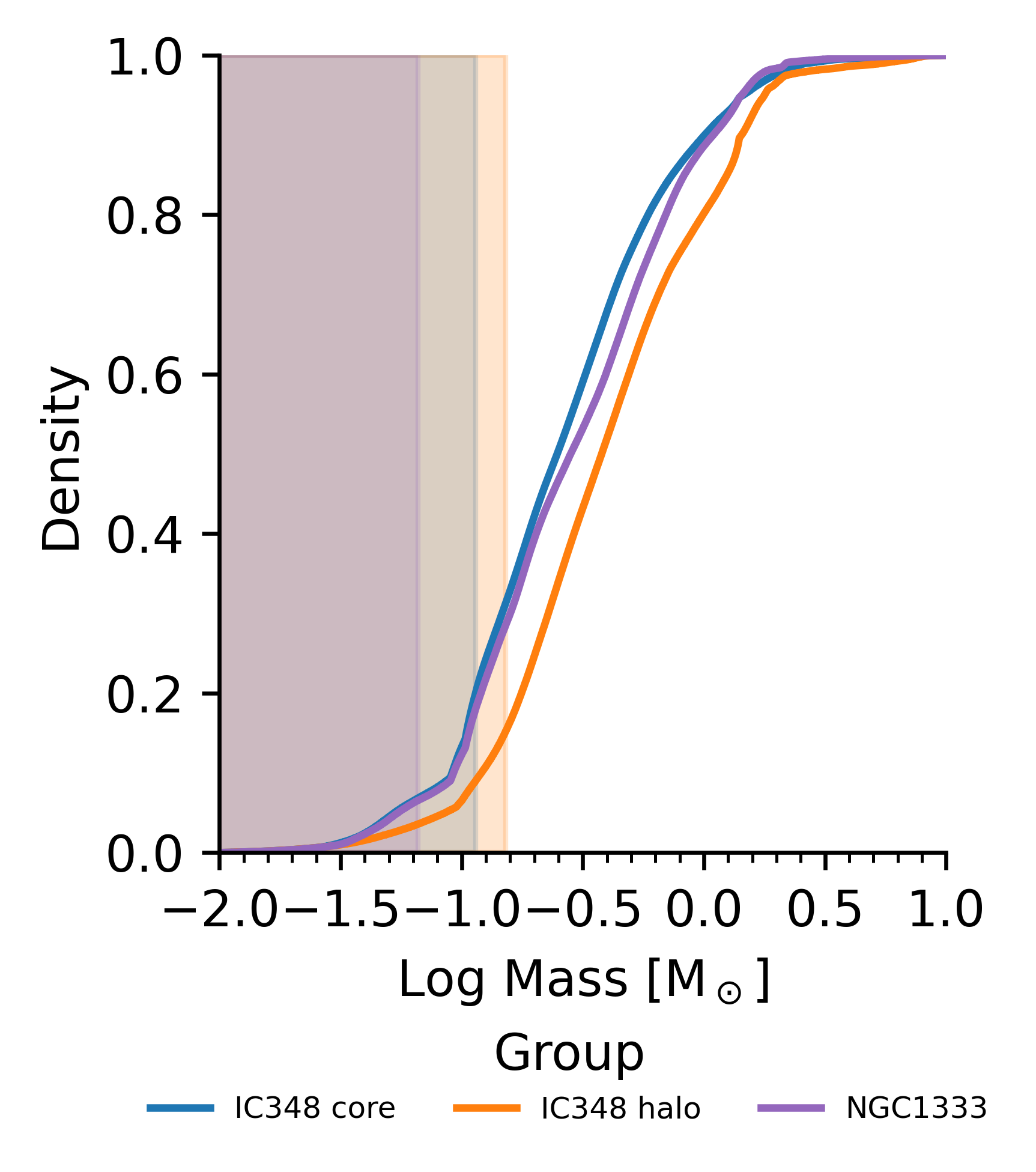}
     \caption{Cumulative mass distribution of IC348 and NGC1333.}
\label{fig:mass_cumulative}
\end{figure}

Figure \ref{fig:mass_cumulative} shows the cumulative mass distribution and incompleteness regions (shaded areas) of the core and halo populations of IC348 as obtained from the unified isochrone model (PMB, see Sect. \ref{methods:mass}). As can be observed in the figure, the core and halo populations are not identical. This observation is confirmed by a two-sample Kolmogorov-Smirnov (KS) test. We applied this test to one hundred samples with bootstrap realisations of the core and halo stars. The mean value of these hundred KS tests results in the rejection (p-value <1\%) of the null hypothesis that both populations come from the same parent distribution.

We observe that the halo region of IC348 has more massive stars than the core region in spite of its lower number density. Furthermore, according to the unified PMB model, the core mass distribution peaks at $0.12\pm0.03M_{\odot}$, whereas the halo peaks at $0.25\pm0.07M_{\odot}$. This result is in disagreement with the observations of \citet{2003AJ....125.2029M}, who found that the core and halo mass distribution peak at $0.56\pm0.18M_{\odot}$ and $0.1\pm0.02M_{\odot}$, respectively. As can be observed in the top right panel of Fig. \ref{fig:mass}, our halo mass distribution starts to decrease at $0.2M_{\odot}$,  within the \textit{Gaia} completeness limits (>$0.11M_{\odot}$). \citet{2003AJ....125.2029M} observed that the alignment of the peaks of their core and halo mass distributions would require a halo age between 5 and 10 Myr. Our age determination for the halo population, 5 Myr, agrees with the lower limit proposed by those authors. However, they mentioned that the hypothesis of two distinct populations with an older halo was difficult to accept. They base their conclusion on the following argument. If the two-population model were correct, then the similar sizes of their core and halo regions will imply a truncation of the halo population below $0.3M_{\odot}$, which was clearly not observed in their mass distributions. Here, we observe that the number of halo stars is half that of the core and that the halo mass distribution peaks at $0.25\pm0.07M_{\odot}$ with a clear cut at $0.2M_{\odot}$. Based on our previous results, we have no compelling evidence to reject Herbig's \citep{1998ApJ...497..736H} hypothesis of distinct core and halo populations. Furthermore, the kinematic evidence further supports the hypothesis of core and halo populations.

Comparing the cumulative distribution of NGC1333 (also shown in Fig. \ref{fig:mass_cumulative}) with the core and halo populations of IC348, we observe the following. On the one hand, comparing the mass distributions of the halo of IC348 and that of NGC1333, we observe that NGC1333 has a general overabundance of stars less massive than $2\rm{M_{\odot}}$. This overabundance of low-mass stars and brown dwarfs was already reported by \citet{2013ApJ...775..138S}, who observed it in several of their mass scenarios (combinations of age, distance and extinction). On the other hand, comparing the mass distributions of NGC1333 and the core of IC348, we observe that although in the mass interval of $0.3-1.0\,\rm{M_{\odot}}$ the core of IC348 appears to have slightly more stars than NGC1333, the two distributions are remarkably similar with a KS p-value of 0.53, which prevents us from rejecting the null hypothesis that the two distributions are random realization of the same underlying distribution. However, due to the \textit{Gaia} completeness limits, we can only claim the previous findings in the domain of masses $>0.16\,\rm{M_{\odot}}$. Therefore, we conclude that, within the completeness limits of the \textit{Gaia} data, we observe no difference in the mass distributions of the core of IC348 and NGC1333, and thus, reject \citet{2013ApJ...775..138S} hypothesis of variation in the initial mass distribution with respect to the crowdedness of the environment.

\subsection{Extinction and $Rv$ distributions}
\label{discussion:extinction}

\begin{figure}[ht!]
    \centering
     \includegraphics[width=\columnwidth]{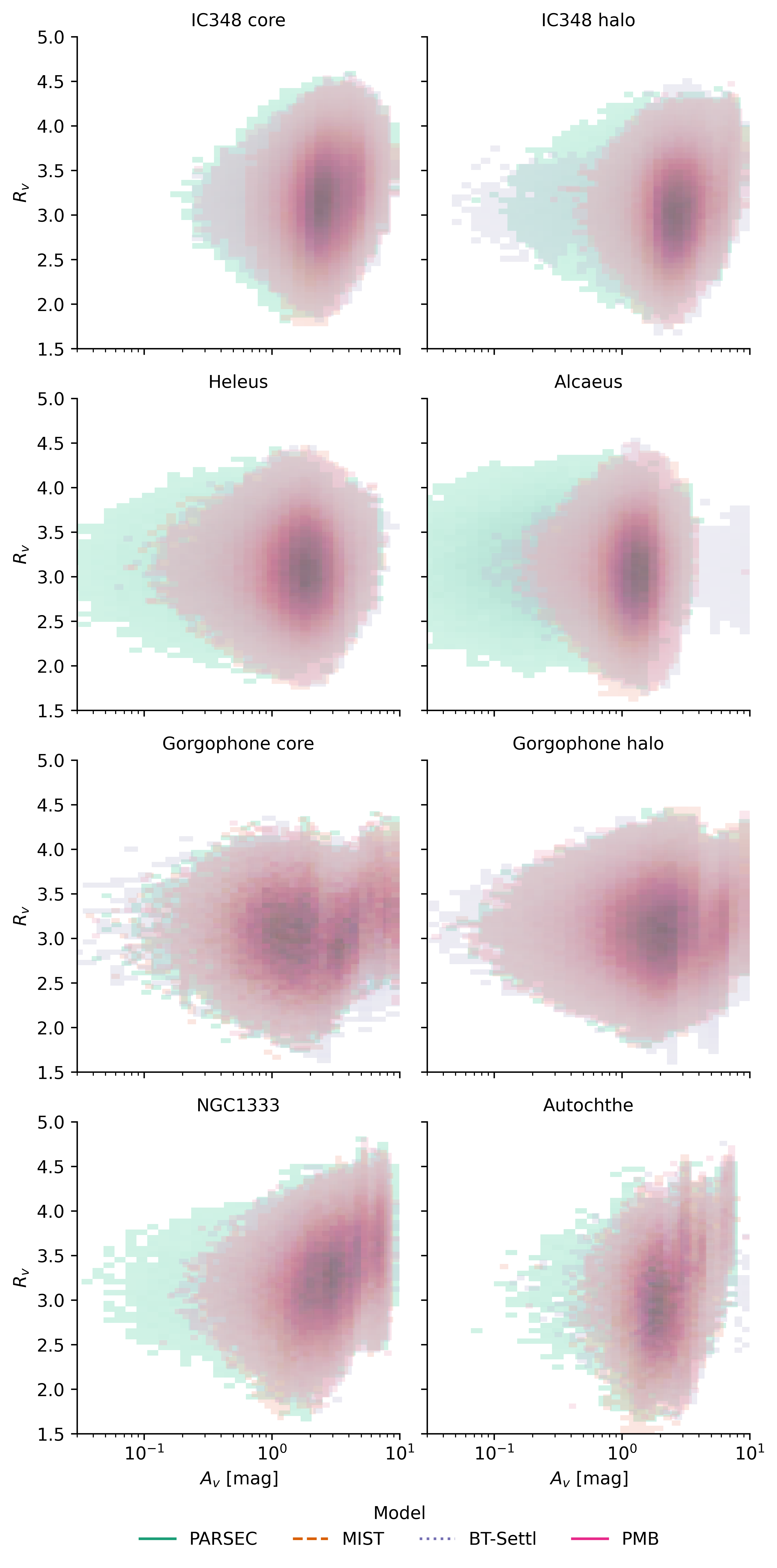}
     \caption{$A_v$ and $R_v$ 2D histograms for stars in the different Perseus groups.}
\label{fig:av_corr}
\end{figure}

\begin{figure}[ht!]
    \centering
     \includegraphics[width=\columnwidth]{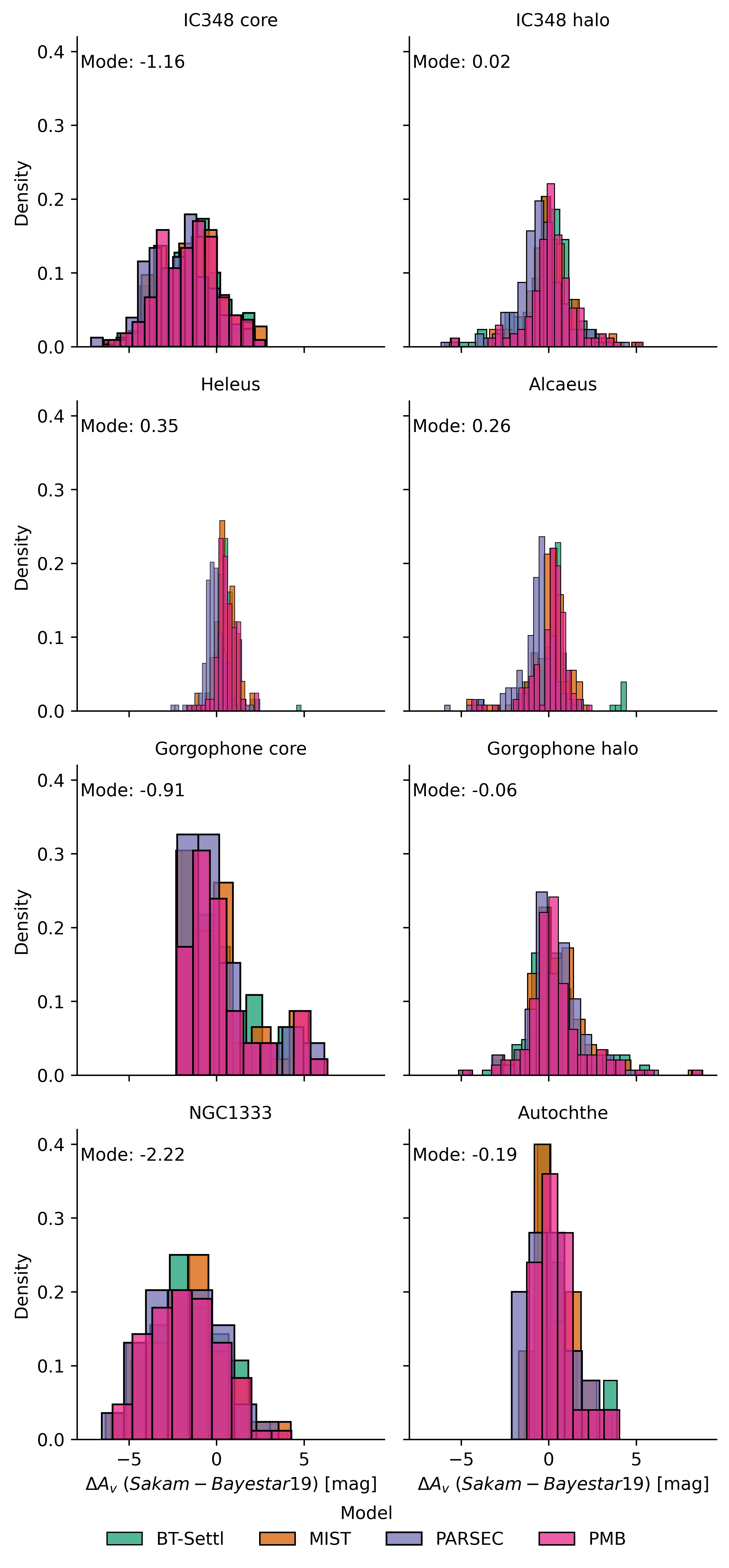}
     \caption{Distributions of $A_v$ difference between the values inferred with \textit{Sakam} and those reported by \textit{Bayestar19} \citep{2019ApJ...887...93G}. Captions as in Fig. \ref{fig:av}.}
\label{fig:av_diff}
\end{figure}

The $A_v$ and $R_v$ values that we derive here confirm previous findings from the literature. We observe cloud-to-cloud and intra-cloud extinction variations, evidence of which was already reported by \citet{2016ApJ...823..102C} for the dust emissivity spectral index. Furthermore, in the extinction interval of the  Perseus group analysed here, we confirm the correlations reported by \citet{2013MNRAS.428.1606F} in the B5 and West-End clouds (see Fig. 9 of those authors). As can be observed from Fig. \ref{fig:av_corr}, there are no significant correlations between $R_v$ and $A_v$ for $A_v<2$ mag in any of the groups. However, for $A_v>3$ mag these correlations are clearly observed in NGC1333 and Authochte, mildly in Gorgophone, almost perceptible in IC348, and clearly absent in Alcaeus and Heleus. Interpreting these correlations as evidence of dust grain growth, and given that Alcaeus and Heleus are off the clouds (see Fig. \ref{fig:sky}), we conclude that there is an east-west gradient in the growth of the dust size across the stars of Perseus ridge.

Comparing the $A_v$ values inferred by \textit{Sakam} with those reported by \textit{Bayestar19} \citep{2019ApJ...887...93G}, see Fig. \ref{fig:av_diff}, we observe that they are, in all cases, compatible (i.e., the zero value is covered by the 95\% credible interval of the distribution). However, NGC1333 and the cores of IC348 and Gorgophone show the largest discrepancies (with modes at -2.2, -1.16, and -0.91 mag, respectively), while Heleus, Alcaeus, and the halos of IC348 and Gorgophone display the smallest differences (with modes at 0.35, 0.26, 0.02, and -0.06 mag, respectively). The largest discrepancies observed in NGC1333 and the cores of IC348 and Gorgophone can be explained by the lack of stars in the 2MASS and PanSTARRS catalogues of these embedded regions that lay in the Perseus ridge (see Fig. \ref{fig:sky}). The lack of stars for anchoring the extinction values causes the \textit{Bayestar19} algorithm to overestimate them.

\subsection{Dynamical analysis}
\label{discussion:dynamical_analysis}

\begin{figure}[ht!]
    \centering
     \includegraphics[width=\columnwidth]{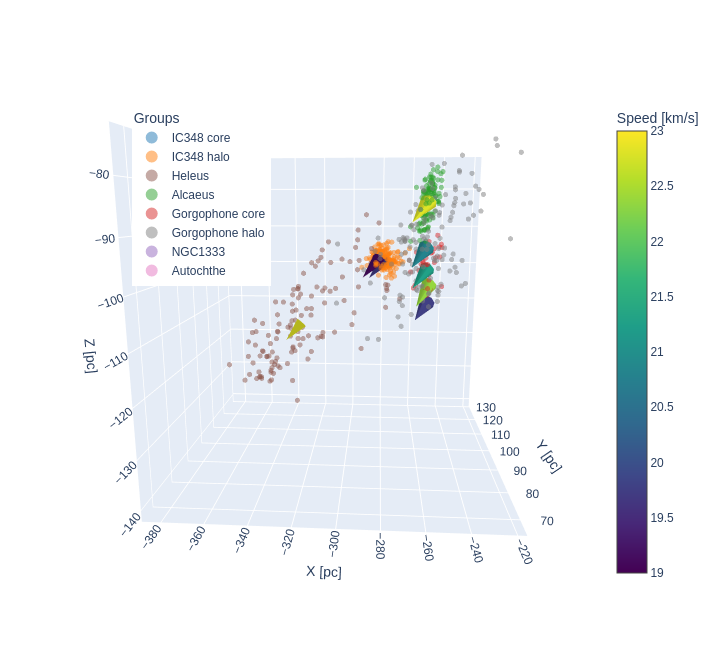}
     \caption{Galactic Cartesian positions and velocities of the Perseus groups (cones) together with the positions of individual members (dots). The interactive version of this figure will be available at \url{www.project-dance.com}.}
\label{fig:3d}
\end{figure}

In Sect. \ref{results:dynamical_analysis}, we presented the CEDFs of each Perseus physical group, and a comparison of its observed velocity dispersion with that expected for virial equilibrium under varying values of the mass and radius parameters. All these results showed that the majority of the stars in the Perseus groups are gravitationally unbound and in a super-virial state. Even in the core of IC348, which is the  most massive ($\rm{146\pm5\,M_{\odot}}$) and compact ($0.65\pm0.20$ pc) group, the upper limit in the fraction of gravitationally bound stars reaches only 40\%. The velocity dispersions at virial equilibrium computed from varying radii and total masses are all similar, although with the more massive clusters assumption resulting in larger virial velocity dispersion (see Sect. \ref{results:dynamical_analysis}). In the case of the core of IC348, the observed velocity dispersion is 1.6 times larger than the most conservative estimate for the virial velocity dispersion ($0.89\pm1.71\,\rm{km\, s^{-1}}$ computed from the core  radius and a non-stellar mass correction of 169\%), whereas in the rest of the groups, the difference is even larger. For the core of IC348 to be in virial equilibrium, a total mass six times larger than the stellar mass will be needed. This value for the dust-and-gas mass exceeds by more than three times our most conservative estimate for this non-stellar mass. Moreover, recent studies have shown that the molecular cloud seen in projection towards IC348 is at 280-300 pc \citep{2022Natur.601..334Z} and thus in front of the cluster. This observation then implies that the core of IC348 has an even larger super-virial state than our conservative estimates. Thus, we conclude that it is highly unlikely that the young Perseus groups formed in virial state.

Our independent estimates of the dynamical state of the Perseus groups confirm \citet{2015ApJ...807...27C} findings about the super-virial state of IC348 and extend it to the rest of the groups. We interpret the super-virial states and large fractions of gravitationally unbound stars in the young ($\lesssim$10 Myr) Perseus groups as evidence that these groups formed through a hierarchical star-formation mechanism \citep[e.g.,][]{2012MNRAS.426.3008K} rather than a monolithic one \citep[e.g.,][]{1991ASPC...13....3L}. For more details about the differences between these star-formation scenarios, see \citet{2018MNRAS.475.5659W} and the references therein. 

Alcaeus is an interesting physical group with 10-30\% of its stars in an energetically bounded state with a steep CEDF (see Fig. \ref{fig:cedfs}). It also has tight velocity dispersion ($0.76\pm0.21\,\rm{km\, s^{-1}}$, see Table  \ref{table:kinematic_indicators}) similar to that of the well-known Pleiades \citep[$0.8\,\rm{km\, s^{-1}}$][]{2017A&A...598A..48G} and Coma Berenices \citep[$0.89\pm0.10\,\rm{km\, s^{-1}}$][]{Olivares2022} open clusters. Therefore, we conclude that this group is a young open cluster in the process of disruption. We notice that its kinematic indicators show hints of contraction. Although this contraction is not significant, if confirmed, it may be explained by either unidentified substructures or contaminants. Future work is warranted.

\subsection{Star-formation history}
\label{discussion:history}

In this section, we use our results to discuss some of the star-formation histories proposed in the literature. Afterwards, we propose our own.

The results that we have collected in this work can be categorised into three major topics: identified physical groups, isochrone ages, and dynamical states. The similar, although not exact same, values of the ages, distances, and kinematics of the Perseus groups suggest that it is unlikely that all these groups randomly formed out of the same star-formation event with this display of properties. The observed spatial (see Fig. \ref{fig:3d}), temporal (see Fig. \ref{fig:relative_ages}), energetic (see Fig. \ref{fig:cedfs}), and mass distribution (see Fig. \ref{fig:mass}) gradients all indicate that a latent (hidden) process links the star-formation history of the Perseus group. Whether this latent process was a supernova explosion, the feedback from nearby OB associations, triggering from within itself, or another unknown phenomenon remains to be explained.

We notice that our results are insufficient to link the star-formation history of the Perseus groups to that of other nearby star-forming regions, as suggested in the literature. Thus, we cannot confirm nor reject the star-formation histories proposed by \citet{1999AJ....117..354D}, \citet{2021ApJ...919L...5B}, \citet{2022Natur.601..334Z}, \citet{2022ApJ...936...23W}, or \citet{2022AJ....164..125L}. On the contrary, the star-formation histories proposed by \cite{1998ApJ...497..736H}, \cite{2002A&A...384..145B}, \cite{2021MNRAS.503.3232P}, \cite{2021ApJ...917...23K}, and \cite{2022AJ....164...57K} are related, up to a certain point, to the intrinsic properties of the region and thus can be scrutinised under the light of the evidence gathered here. We now discuss these works.

\cite{2022AJ....164...57K} discuss two scenarios for the formation of the Perseus regions that are related to the internal properties of the groups (a third scenario associated with the Per-Tau shell is also discussed): a past supernova and cloud-cloud interaction. Those authors concluded that although the region shows evidence of expansion, it is unlikely that it resulted from the triggering of a supernova, given that the observed velocities are not consistent with it. Our results show (see Table \ref{table:groups_mean}) that the velocities of the PerOB2a (i.e., the core and halo of IC348 and Heleus) and PerOB2b (i.e., Alcaeus, Autohochte, NGC1333 and Gorgophone) super-groups have distinct directions but similar magnitudes, with the maximum relative difference amongst the velocities of all the groups being only $4\,\rm{km\,s^{-1}}$. Concerning the cloud-cloud interaction scenario, the aforementioned authors argue that the super-groups PerOB2a and PerOB2b may have originated from two distinct clouds that independently started forming stars when they were in close proximity, a fact that is compatible with the spatial and kinematics derived here. Those authors continue mentioning that after the initial burst, the two super-groups continue forming stars resulting in today's similar ages for all the groups. Although our age estimates are systematically older than those of the previous authors (see Sect. \ref{results:isochrones} and Sect. \ref{discussion:isochrones}), we observe clear age gradients in the groups of PerOBa and PerOB2b. Finally, after highlighting the difficulty of measuring the degree of possible mutual triggering \citep[e.g.,][]{2015MNRAS.450.1199D} between these two super-groups, they stress that there appears to be at least some degree of mutual influence.

\cite{2021ApJ...917...23K} suggested that given the kinematic similarities between Per 1 and Per 2 (which roughly correspond to PerOB2a and PerOB2b of \citealt{2022AJ....164...57K}), these super-groups most likely formed in the same star-forming process (see Sect. \ref{intro:history}). The spatial and kinematic results that we gather here support this conclusion. We consider it unlikely that two unrelated star-forming events produced the observed similar spatial, kinematic, and age properties of the Perseus region. Then, the aforementioned authors argue that given the lack of age gradient and the considerable time lag between the older (Per B) and younger populations (Per A) of the two super-groups (Per 1 or Per 2):$\sim$17 Myr vs $\sim$5 Myr, then they consider unlikely that a continuous star-formation process was at work. Instead, they proposed a scenario in which the formation of the older generation (Per B) dispersed part of the gas but not entirely prevent the formation of the younger generation (Per A). Our results contradict this scenario because we observe age gradients in both super-groups. On the one hand, in Per 2, we have one intermediate age generation with Heleus, and the halo of IC348, both formed at 5 Myr, and a younger generation constituted by the core of IC348, which formed 3 Myr ago. On the other hand, in Per 1, there are three generations with a clear age gradient: Alcaeus at 10 Myr, Gorgophone at 7 Myr, and NGC1333 and Autochthe both at 3 Myr.

\cite{2021MNRAS.503.3232P} proposed that the older groups (Alcaeus and Heleus) formed closer to the Galactic plane while the younger ones did it at higher latitudes. They also point out that NGC1333 and Autochthe are part of the same star-formation event. Both these conclusions are supported by our spatial and age results. These authors also mentioned that only NGC1333, Authochte, and IC348 show evidence of continuing star formation. For this latter conclusion, we have no further evidence.

\cite{2002A&A...384..145B} considered that the Perseus region constitutes an example of propagated star-formation, which started in the edge of Auriga 30 Myr ago, continued to Per OB2a 10 Myr ago, and is now in progress in the southern region where IC348 is located. Our results support these two latter steps, with Alcaeus at 10 Myr and the core of IC348, NGC1333 and Autochthe at 3 Myr.

Finally, \cite{1998ApJ...497..736H} proposed a four steps scenario (see Sect. \ref{intro:history}), which is fully supported by the results we gather here. We confirm the following points: a) continuous star-formation in the region for at least the last 10 Myr, with Autochthe, NGC1333, and the core of IC348 being the youngest examples; b) $\zeta$ Per and $o$ Per are (4$\sigma$ astrometric) members of Alcaeus, a group formed 10 Myr ago; c) there are low-luminosity members of Alcaeus spread over a large region; d) IC348 is composed of two physical groups: an old 5 Myr halo and a young 3 Myr core.

We now propose a star-formation history that closely follows that outlined by \cite{1998ApJ...497..736H}. We highlight the fact that our history is descriptive rather than phenomenological, and it is based on the ages and kinematics of the identified physical group. To arrive at a phenomenological description of the Perseus star-formation history, we still require precise age determinations and the application of the methods similar to the ones used here to a wider sky region encompassing the other actors in the scene: the Taurus star-forming region and its connecting bridge with Perseus. Future steps will be taken to address these points.

\subsubsection{The first generation}
Alcaeus and Gorgophone constitute the first generation of stars in the Perseus region. Alcaeus formed approximately 10 Myr ago, followed by the core and halo of Gorgophone 7 Myr ago. The brightest members of Alcaeus, $\zeta$ Per and $o$ Per are B-type stars with ages of $12.6\pm1.5$ Myr and $11.1\pm0.5$ Myr, respectively, according to \citet{2011MNRAS.410..190T}. These ages are compatible with our 10 Myr age estimate for the group. Most likely, the winds of this first generation sweep up part of the gas and dust from the molecular cloud resulting in the observed lowest extinction values of about $A_v\sim1$ mag (see Fig. \ref{fig:av}). The birth of this generation is an example of high-mass formation in a clustered environment (i.e., Alcaeus) together with low-mass formation in a sparse one (i.e., the core and halo of Gorgophone), which has also been observed in other star-forming regions \citep[e.g.,][]{2022ApJ...937...46K,2022AJ....163..266L} and is predicted by the hierarchical star-formation scenario \citep{2018MNRAS.475.5659W}.

We notice that $\zeta$ Per has been related to the Per OB2 association as a runaway member \citep[e.g.,][]{1999AJ....117..354D,1999Ap.....42..247M}, a member with discrepant radial velocity \citep{2003A&A...402..587S} or a member of the trapezium type system ADS 2843 \citep{2004RMxAC..21..195A,2018MNRAS.481.3953A}, which is located at the central region of Alcaeus.

\subsubsection{The second generation}
Heleus and the halo of IC348 constitute the second generation of Perseus stars. Although these two groups are separated by about  $\sim$54 pc, they have a similar age and direction of their velocity vector (see Fig. \ref{fig:3d} and Table \ref{table:groups_mean}), which indicate that probably these two groups were born from the same star-formation event. The origin of Heleus may also be related to the $\sim$5 Myr, elongated ($\sim$100 pc) but more distant ($\sim$400 pc) Barnard 5 group identified by \citet{2022AJ....164..125L}. However, we currently lack evidence to relate the origin of these two groups.

\subsubsection{The third generation}
Finally, Autochthe, NGC1333, and the core of IC348 make the third and most recent generation of stars. While NGC1333 and IC348's core have some hints that are still contracting (see Table \ref{table:kinematic_indicators}) the value of this contraction is not significant enough to claim this effect. NG1333 and Autochthe have similar space velocities (see Fig. \ref{fig:3d} and Table \ref{table:groups_mean}) and ages. However, the space velocity of IC348's core is different from that of the latter two groups (see Fig. \ref{fig:3d} and Table \ref{table:groups_mean}). Thus, we conclude that the core of IC348 formed in a coeval but rather different, possibly unrelated, or parallel \citep[as suggested by][]{2021ApJ...917...23K} star-formation episode than that giving birth to NGC1333 and Autochthe, with these latter two most likely representing the youngest star-formation episode.

\subsubsection*{Internal triggering?}
Internal triggering, or self-propagated star formation, has been proposed as a mechanism to explain the origin of the Perseus groups \citep[e.g.,][]{2002A&A...384..145B}. Although a clear demonstration of triggered star formation is extremely complex and elusive \citep[e.g.,][]{2015MNRAS.450.1199D}, the Perseus region gathers the following common indicators of internal triggering. Evidence of heating from B stars, embedded protostars, and outflows was found by \citet{2016ApJ...823..102C}. Outflows have also been found by \cite{2010ApJ...715.1170A} and \cite{2013ApJ...774...22P}. These outflows affect the regulation of the star-formation process, given that they can intensify or diminish it, particularly in the environment surrounding the active star formation regions.  Our results show (see Fig. \ref{fig:3d}) that the super-group PerOB2b (i.e., Alcaeus, Gorgophone, NGC1333 and Autochthe) follows a clear spatio-temporal vector gradient that points away from Alcaeus (10 Myr) and towards NGC1333 and Autochthe (3 Myr), passing by Gorgophone (7 Myr). PerOB2a (Heleus and the core and halo of IC348) also follows a clear age gradient but lacks the neat spatial relation observed in PerOB2b. Although the previous evidence is not conclusive enough to claim internal trigger in either PerOB2a, PerOB2b or the entire Perseus region, it nonetheless represents a step forward in understanding the star-formation history of this region.

\section{Conclusions}
\label{conclusions}

We applied Bayesian methodologies to the public \textit{Gaia}, APOGEE, 2MASS, PanSTARRS, and SIMBAD catalogues to identify and characterise the physical groups of the Perseus star-forming region. We found 1052 candidate members distributed into seven physically distinct populations: the core and halo of IC348, Autochthe, Alcaeus, Gorgophone, Heleus, and NGC1333. Gorgophone is a new kinematic group composed of a core and a halo. Our new list of candidate members increases by 31\% with respect to those from the literature in the Perseus region analysed here. The following are our most important conclusions.

\begin{itemize}
\item{Kinematics}. We propose that Alcaeus is an open cluster in a state of disruption. On the other hand, the internal velocities of NGC1333 and the core of IC348 show the largest, although not significant, signals of contraction. In all the Perseus groups, the evidence of internal rotation is not conclusive.

\item{Age}. Our age estimates are compatible with those from the literature. We notice that neglecting or underestimating the contribution of the extinction in this partly embedded star-forming region may result in twice as much younger age estimates, as those determined by \citet{2021MNRAS.503.3232P} and \citet{2022AJ....164...57K}.

\item{Mass}. The mass distribution of NGC1333 shows no over-abundance of low-mass stars compared to that of the core of IC348. This contradicts previous claims about the environmental differences of the initial mass function \citep[e.g.,][]{2013ApJ...775..138S}. On the contrary, the mass distributions of the young Perseus groups are broadly compatible with \citet{2005ASSL..327...41C}, although with distinct features.

\item{Extinction}. We found evidence of dust grain growth in NGC1333 and Autochthe, as previously reported by \citet{2016ApJ...826...95C} for the Perseus clouds.

\item{Dynamical state}. All of the Perseus groups are in a super-virial state, with CEDFs showing large fractions of energetically unbound stars. These findings, together with the lack of clear expansion patterns, support the hierarchical star-formation scenario \citep[e.g.,][]{2018MNRAS.475.5659W} over the monolithic one. 

\item{Star-formation history}. The Perseus region contains stars from at least three generations, which supports the star formation scenario proposed by \citet{1998ApJ...497..736H}.
\end{itemize}

\subsection*{Caveats and future perspectives}

Thanks to the unprecedented quality and abundance of \textit{Gaia} data in combination with our comprehensive Bayesian methodologies, the traditional uncertainty sources in the mass determinations, like distance, extinction, completeness, and membership status, are now minimised. However, the following issues remain.

Age continues to be the largest source of uncertainty for the analysis of the initial mass function and star-formation history of the Perseus region. Precise age determinations, like those provided by dynamical trace-back analyses \citep{2020A&A...642A.179M,2022A&A...667A.163M}, are the next mandatory step before attempting to construct a phenomenological star-formation history of the region.

A phenomenological model for the star-formation history of the Perseus region also demands a joint analysis of its neighbour structures (i.e., Taurus, Auriga, and Pleiades). This analysis will be the next step to linking the origin of the Perseus region to large-scale structures such as the Per-Tau shell. However, the disentanglement of physically coherent structures in large sky regions is a methodological challenge for which our team has made but the first steps \citep[e.g.,][]{Olivares2022}.

The incompleteness limits of the \textit{Gaia} data still prevent sound comparisons of the brown dwarf ratios in the Perseus groups, particularly between IC348 and NGC1333. Deep, wide, and complete astrophotometric surveys as those provided by the COSMIC-DANCe project \citep[e.g.,][]{2013A&A...554A.101B,2015A&A...577A.148B,2022NatAs...6...89M} are still needed. Our team is currently working on collecting and curating these data.

\begin{acknowledgements}
We thank the anonymous referee for the useful comments.
JO acknowledge financial support from “Ayudas para contratos postdoctorales de investigación UNED 2021”.
P.A.B. Galli acknowledges financial support from São Paulo Research Foundation (FAPESP) under grants 2020/12518-8 and 2021/11778-9.

This research has received funding from the European Research Council (ERC) under the European Union’s Horizon 2020 research 
and innovation programme (grant agreement No 682903, P.I. H. Bouy), and from the French State in the framework of the 
”Investments for the future” Program, IdEx Bordeaux, reference ANR-10-IDEX-03-02.

We gratefully acknowledge the support of NVIDIA Corporation with the donation of one of the Titan Xp GPUs used for this research.

We acknowledge Anthony Brown, the Gaia Project Scientist Support Team and the Gaia Data Processing and Analysis Consortium (DPAC) for providing the \textit{PyGaia} code.

This work has made use of data from the European Space Agency (ESA) mission {\it Gaia} (\url{https://www.cosmos.esa.int/gaia}),processed by the {\it Gaia} Data Processing and Analysis Consortium (DPAC, \url{https://www.cosmos.esa.int/web/gaia/dpac/consortium}). Funding for the DPAC has been provided by national institutions, in particular the institutions participating in the {\it Gaia} Multilateral Agreement.

This research has made use of the SIMBAD database,
operated at CDS, Strasbourg, France.

The Pan-STARRS1 Surveys (PS1) and the PS1 public science archive have been made possible through contributions by the Institute for Astronomy, the University of Hawaii, the Pan-STARRS Project Office, the Max-Planck Society and its participating institutes, the Max Planck Institute for Astronomy, Heidelberg and the Max Planck Institute for Extraterrestrial Physics, Garching, The Johns Hopkins University, Durham University, the University of Edinburgh, the Queen's University Belfast, the Harvard-Smithsonian Center for Astrophysics, the Las Cumbres Observatory Global Telescope Network Incorporated, the National Central University of Taiwan, the Space Telescope Science Institute, the National Aeronautics and Space Administration under Grant No. NNX08AR22G issued through the Planetary Science Division of the NASA Science Mission Directorate, the National Science Foundation Grant No. AST-1238877, the University of Maryland, Eotvos Lorand University (ELTE), the Los Alamos National Laboratory, and the Gordon and Betty Moore Foundation.

This publication makes use of data products from the Two Micron All Sky Survey, which is a joint project of the University of Massachusetts and the Infrared Processing and Analysis Center/California Institute of Technology, funded by the National Aeronautics and Space Administration and the National Science Foundation.

Funding for the Sloan Digital Sky 
Survey IV has been provided by the 
Alfred P. Sloan Foundation, the U.S. 
Department of Energy Office of 
Science, and the Participating 
Institutions. 

SDSS-IV acknowledges support and 
resources from the Center for High 
Performance Computing  at the 
University of Utah. The SDSS 
website is www.sdss.org.

SDSS-IV is managed by the 
Astrophysical Research Consortium 
for the Participating Institutions 
of the SDSS Collaboration including 
the Brazilian Participation Group, 
the Carnegie Institution for Science, 
Carnegie Mellon University, Center for 
Astrophysics | Harvard \& 
Smithsonian, the Chilean Participation 
Group, the French Participation Group, 
Instituto de Astrof\'isica de 
Canarias, The Johns Hopkins 
University, Kavli Institute for the 
Physics and Mathematics of the 
Universe (IPMU) / University of 
Tokyo, the Korean Participation Group, 
Lawrence Berkeley National Laboratory, 
Leibniz Institut f\"ur Astrophysik 
Potsdam (AIP),  Max-Planck-Institut 
f\"ur Astronomie (MPIA Heidelberg), 
Max-Planck-Institut f\"ur 
Astrophysik (MPA Garching), 
Max-Planck-Institut f\"ur 
Extraterrestrische Physik (MPE), 
National Astronomical Observatories of 
China, New Mexico State University, 
New York University, University of 
Notre Dame, Observat\'ario 
Nacional / MCTI, The Ohio State 
University, Pennsylvania State 
University, Shanghai 
Astronomical Observatory, United 
Kingdom Participation Group, 
Universidad Nacional Aut\'onoma 
de M\'exico, University of Arizona, 
University of Colorado Boulder, 
University of Oxford, University of 
Portsmouth, University of Utah, 
University of Virginia, University 
of Washington, University of 
Wisconsin, Vanderbilt University, 
and Yale University.
\end{acknowledgements}

\bibliographystyle{aa} 
\bibliography{Perseus.bib}

\begin{appendix}

\section{Assumptions}
\label{appendix:assumptions}

\begin{assumption}
\label{assumption:groups_independency}
In computing the properties of a physical group using its list of candidate members, we assume that this list provides a perfect classification. This assumption is optimistic, given that we have access only to a probabilistic classification in which a certain degree of entanglement is expected.
\end{assumption}

\begin{assumption}
\label{assumption:gaussian}
The positions and velocities of stars belonging to the same physical group are Gaussian distributed or can be modelled as a mixture of Gaussian distributions. Several Gaussian distributions pertain to the same physical group if their medians are mutually contained within one Mahalanobis distance. This latter is the multidimensional extension of the idea of measuring how many standard deviations a point is away from the mean of a distribution.
\end{assumption}

\begin{assumption}
\label{assumption:selection_function}
\textit{Gaia} observes sources with a perfect selection function within its photometric completeness limit \citep[\texttt{G}<19 mag, see Tables 4 and 5 of][]{2021A&A...649A...2L}. This is a simplistic assumption given that \textit{Gaia} misses sources in its expected magnitude coverage according to complex functions of its observables \citep[see, for example,][]{2020MNRAS.497.4246B}.
\end{assumption}

\begin{assumption}
\label{assumption:virial_equilibrium}
Following \citet{2015ApJ...807...27C}, we assume that the velocity dispersion at virial equilibrium, $\sigma_{vir}$ can be computed as:

\begin{equation}
\label{equation:virial}
\sigma_{vir}=\sqrt{3\frac{G\cdot M}{\eta \cdot r_{hm}}},
\end{equation}
where $G$ is the gravitational constant, $\eta$ is a structural parameter, $r_{hm}$ is the half-mass radius, and $M$ is the total mass contained within $r_{hm}$. We added the factor $\sqrt{3}$ to obtain the total velocity dispersion rather than the 1D value computed by \citet{2015ApJ...807...27C}.
\end{assumption}

\begin{assumption}
\label{assumption:EFF_profile}
Following \citet{2015ApJ...807...27C} and \citet{2010ARA&A..48..431P} we assume that $r_{hm}$ and $\eta$ in Eq. \ref{equation:virial} can be estimated by fitting an \citet{1987ApJ...323...54E} profile, and that $\eta\sim10$ for values of the EFF  parameter $\gamma>2.5$.
\end{assumption}

\begin{assumption}
\label{assumption:dust_mass}
We assume that the dust and gas mass in the Perseus groups follows the observed 3D distribution of the stars and that the lower and upper limits to this mass contribution correspond to 65\% and 169\% of the stellar mass, respectively. Although this assumption will be generally correct for the young groups (i.e., IC348 and NGC1333), where the gas and stars are still coupled, it may not be entirely correct  for the oldest groups like Alcaeus, where the stellar winds may have already cleared out some of its gas and dust. Nonetheless, this is the best we can do until the arrival of a 3D dust-and-gas mass map of the Perseus region.
\end{assumption}

\begin{assumption}
\label{assumption:binaries_mass}
We assume that the contribution of unresolved binaries to the total gravitational potential amounts to 20\% of the inferred stellar mass. Generally, an unresolved binary is more luminous than a single star with the same mass as the primary of the binary. Then, it follows that our inferred masses, which are estimated based on the observed photometry of the stars, already partially mitigate the unaccounted mass of unresolved binaries. However, given the lack of information about the mass ratio of possible unresolved binaries within our list of candidate members, we can but make the conservative approach of increasing the inferred stellar mass.
\end{assumption}

\begin{assumption}
\label{assumption:self_gravitating}
When computing the internal energy distribution of the Perseus groups, we assume that the stellar system is self-gravitating. Given the negligible size of the Perseus groups when compared to the Galactic scale, it can be safely assumed that the Gravitational potential of the Galaxy is constant and thus can be removed from Eq. \ref{equation:energy}.
\end{assumption}

\section{3D velocities of the candidate members}
\label{appendix:3D_velocities}
In this Appendix, we show figures with the positions and velocities of our eight statistical Perseus groups. The figures show 2D projections of the Galactic Cartesian positions and velocities of the candidate members in each group. The velocity vectors are shown with respect to the group's central velocity. For visual aid, their magnitude is also shown with the color code.

\begin{figure}[ht!]
    \centering
     \includegraphics[width=\columnwidth]{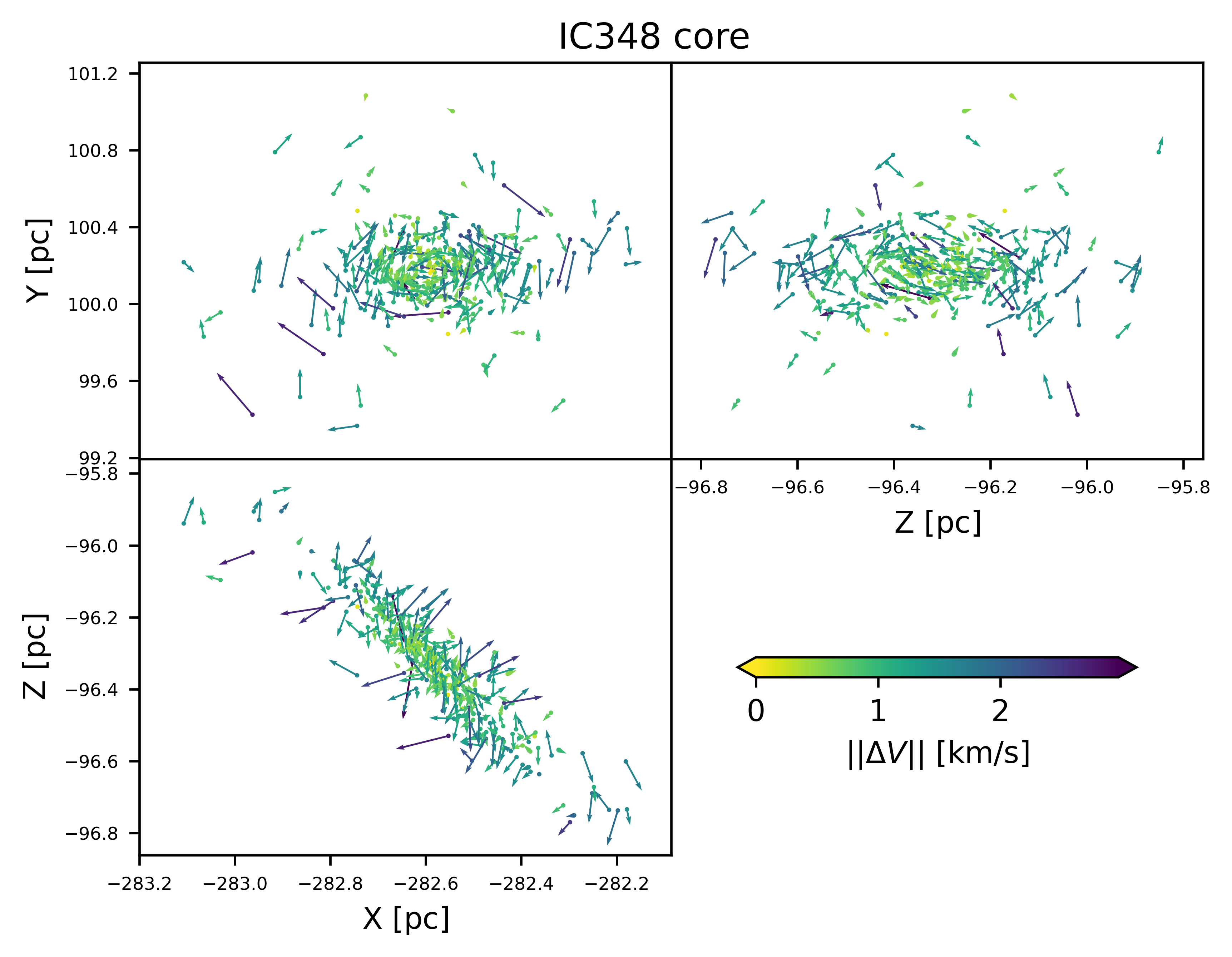}
     \caption{2D projections of the Galactic Cartesian positions (dots) and velocities (arrows) of the candidate members to the core of IC348. The velocities are computed with respect to the group's central velocity. For visual aid, the colour code shows the magnitude of the 3D velocity.}
\label{fig:vel_ic348_core}
\end{figure}

\begin{figure}[ht!]
    \centering
     \includegraphics[width=\columnwidth]{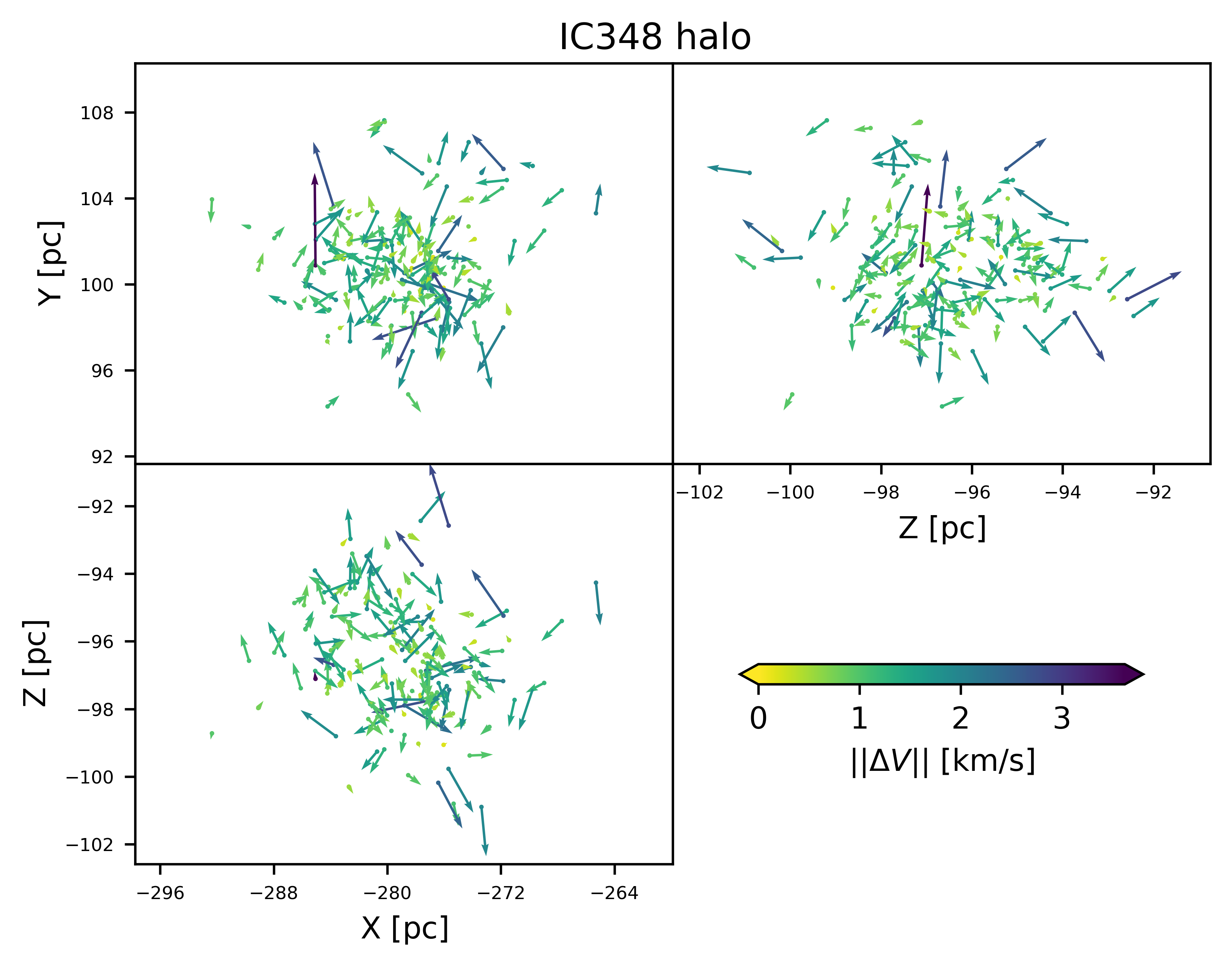}
     \caption{Captions as in Fig. \ref{fig:vel_ic348_core} but for the halo of IC348.}
\label{fig:vel_ic348_halo}
\end{figure}

\begin{figure}[ht!]
    \centering
     \includegraphics[width=\columnwidth]{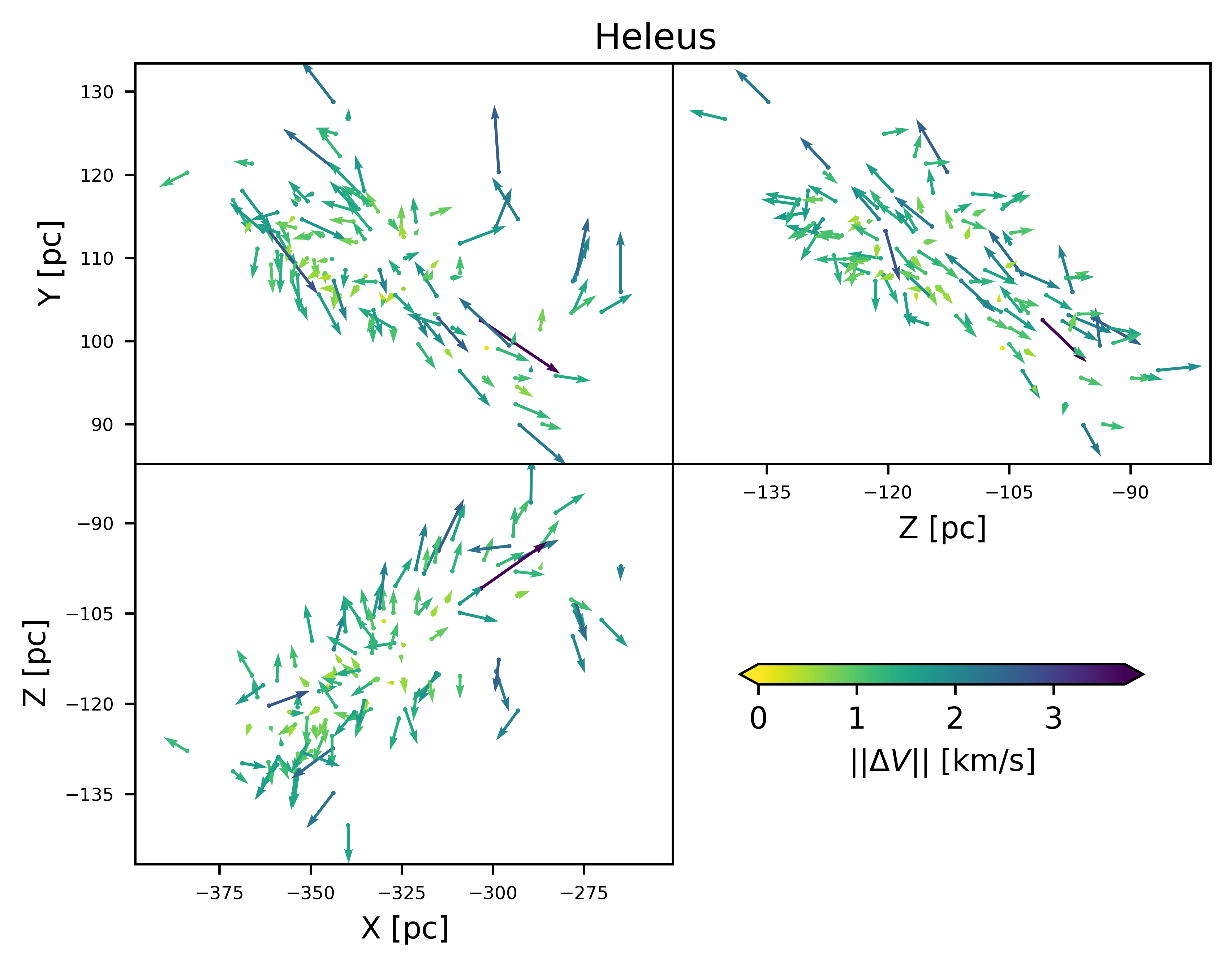}
     \caption{Captions as in Fig. \ref{fig:vel_ic348_core} but for Heleus.}
\label{fig:vel_heleus}
\end{figure}

\begin{figure}[ht!]
    \centering
     \includegraphics[width=\columnwidth]{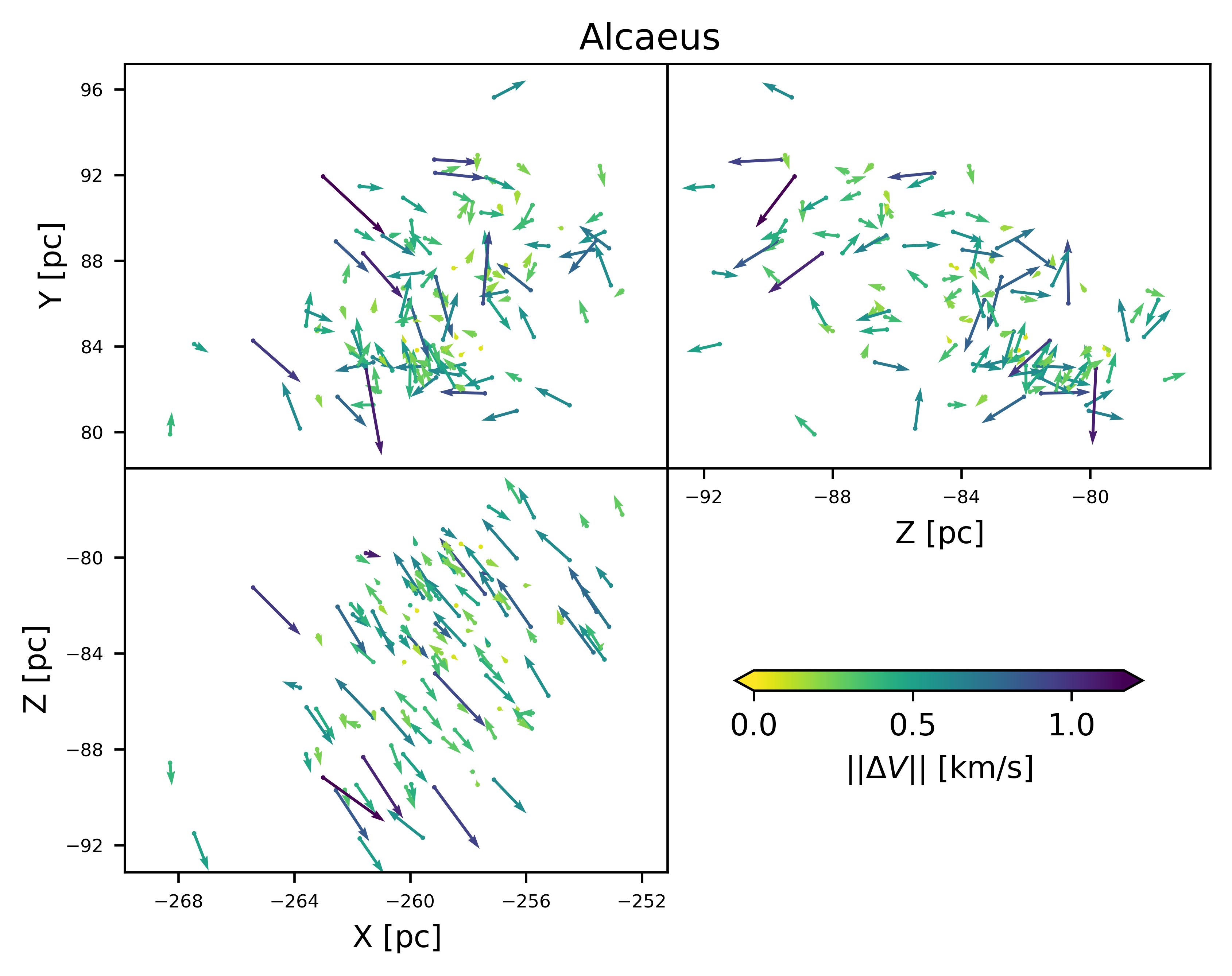}
     \caption{Captions as in Fig. \ref{fig:vel_ic348_core} but for Alcaeus.}
\label{fig:vel_alcaeus}
\end{figure}

\begin{figure}[ht!]
    \centering
     \includegraphics[width=\columnwidth]{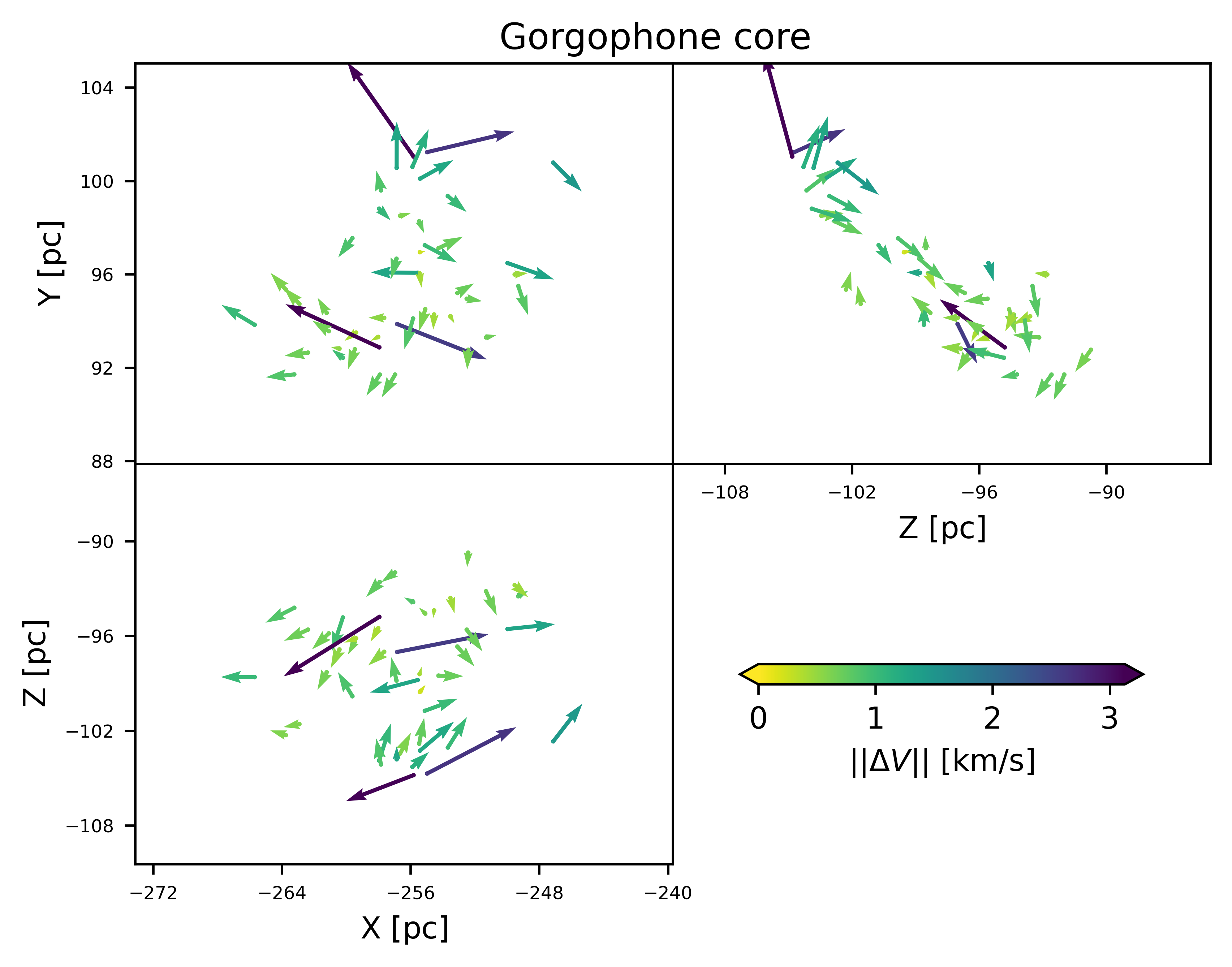}
     \caption{Captions as in Fig. \ref{fig:vel_ic348_core} but for the core of Gorgophone.}
\label{fig:vel_Gorgophone_core}
\end{figure}

\begin{figure}[ht!]
    \centering
     \includegraphics[width=\columnwidth]{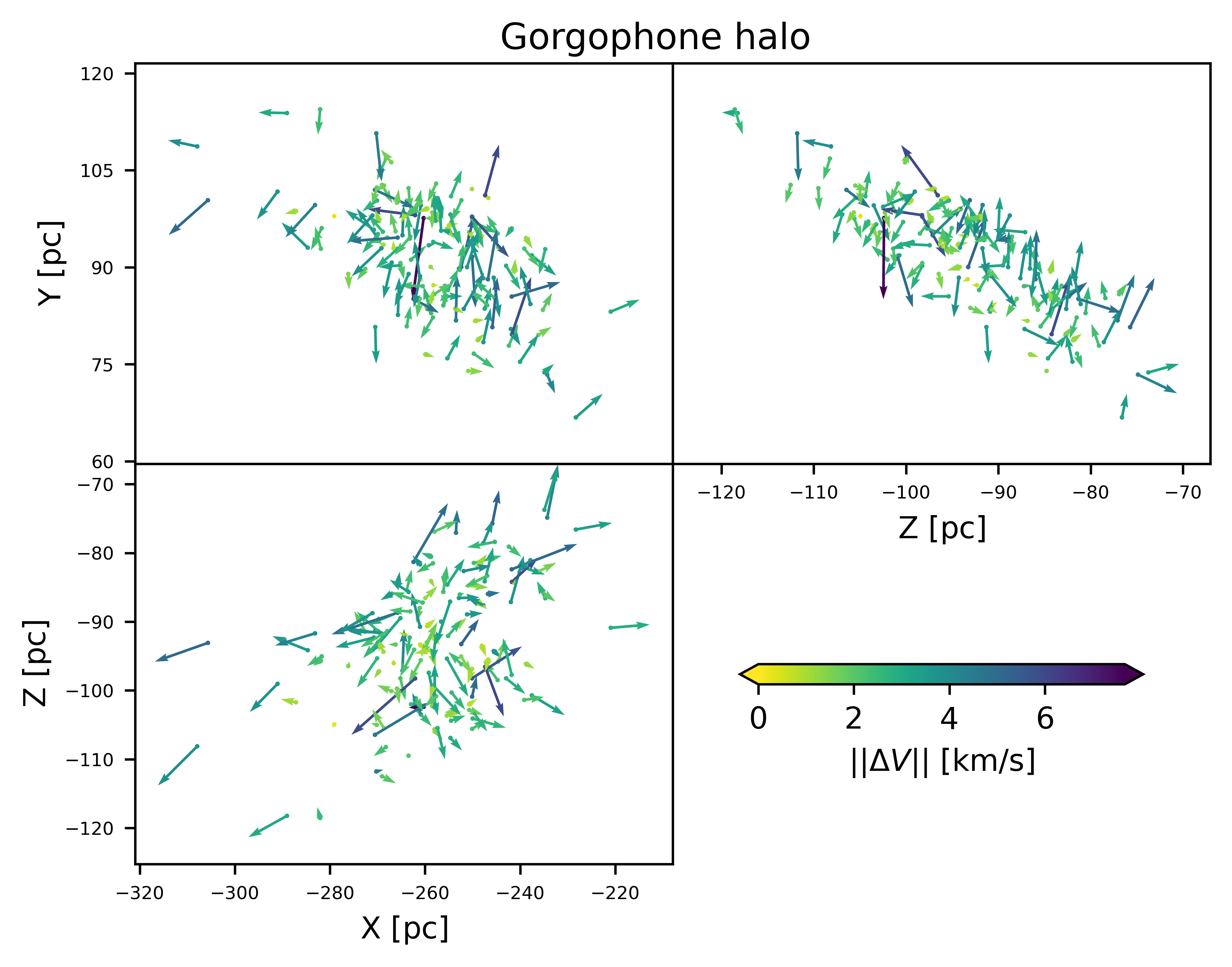}
     \caption{Captions as in Fig. \ref{fig:vel_ic348_core} but for the halo of Gorgophone.}
\label{fig:vel_Gorgophone_halo}
\end{figure}

\begin{figure}[ht!]
    \centering
     \includegraphics[width=\columnwidth]{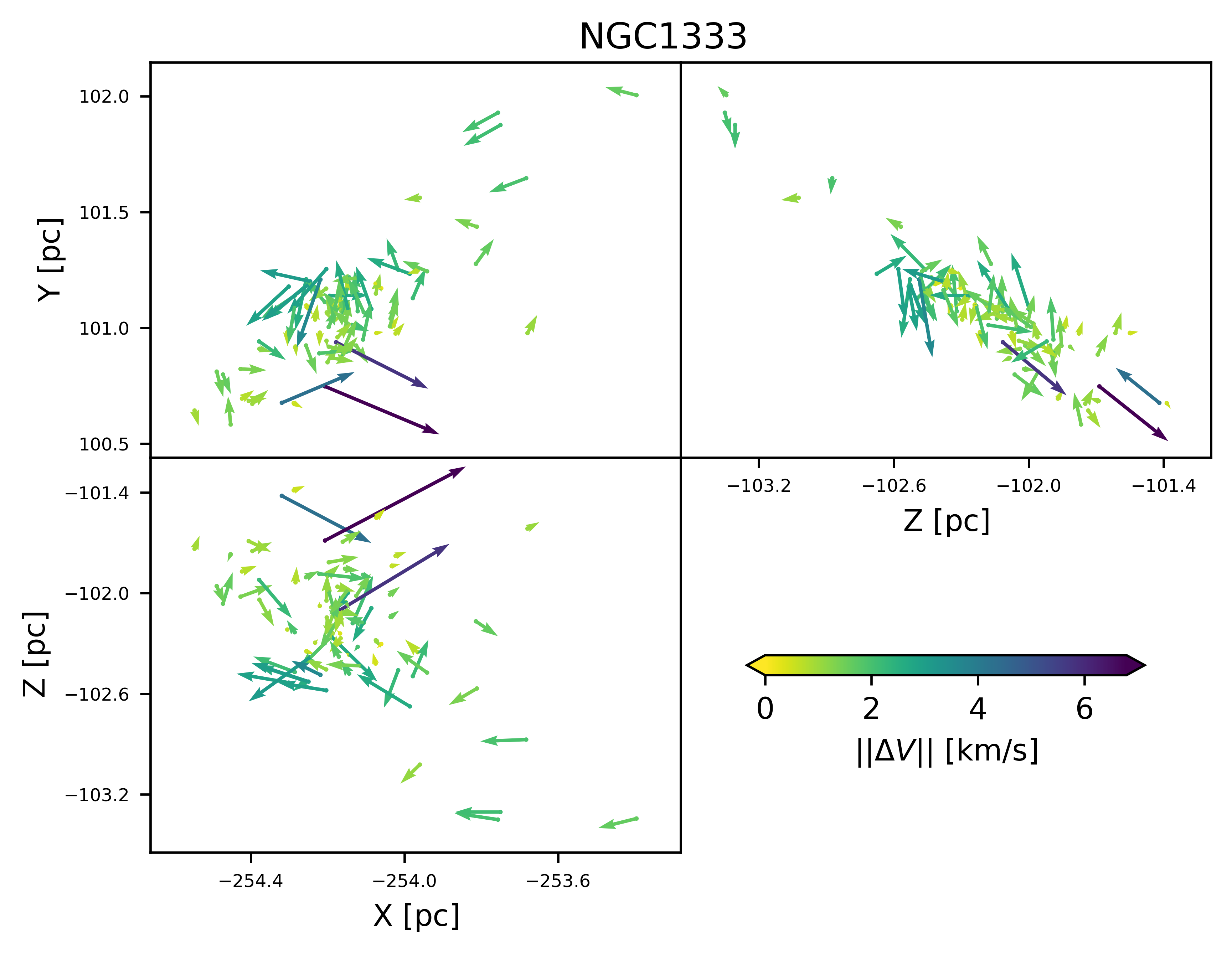}
     \caption{Captions as in Fig. \ref{fig:vel_ic348_core} but for NGC1333.}
\label{fig:vel_ngc1333}
\end{figure}

\begin{figure}[ht!]
    \centering
     \includegraphics[width=\columnwidth]{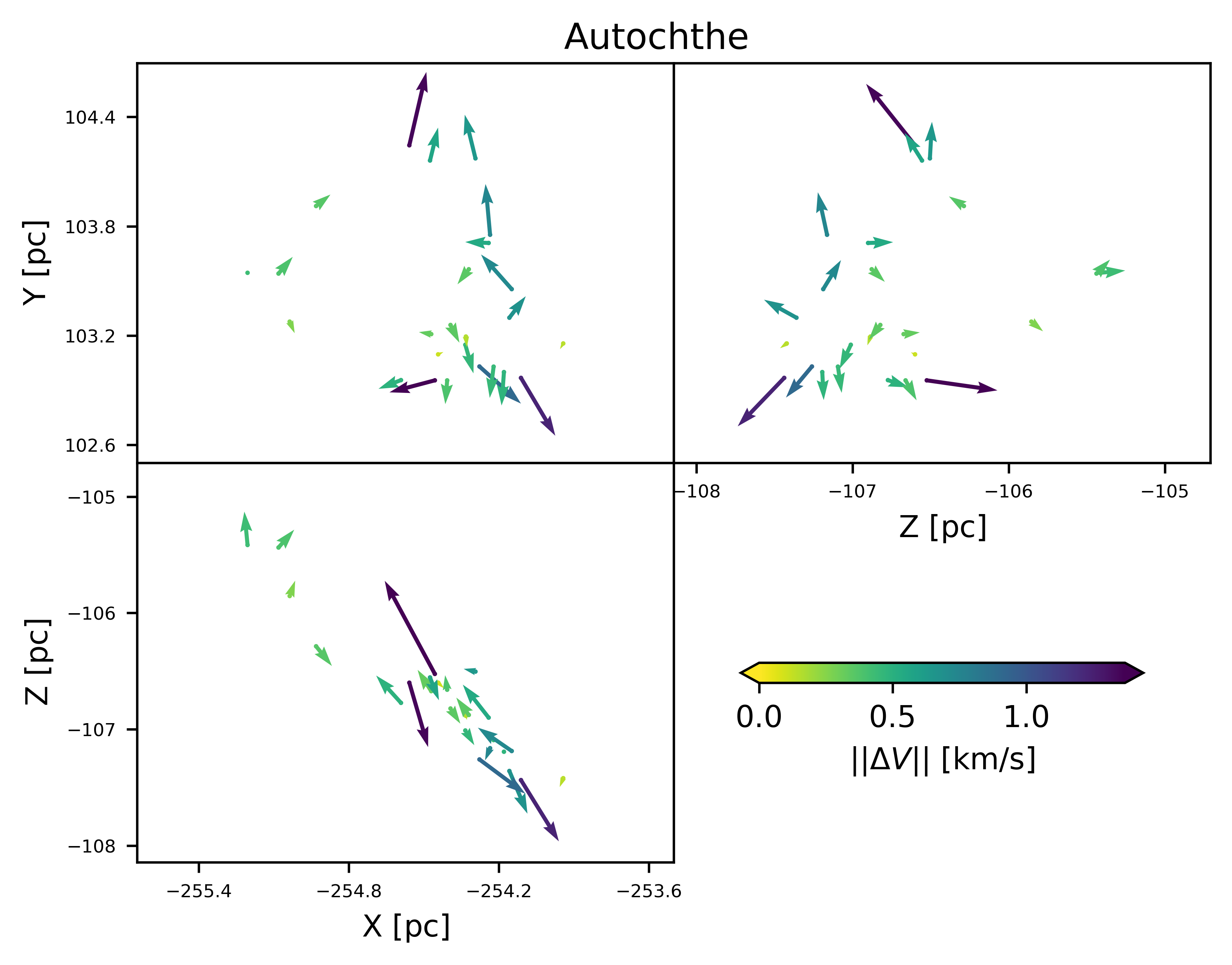}
     \caption{Captions as in Fig. \ref{fig:vel_ic348_core} but for Autochthe.}
\label{fig:vel_autochthe}
\end{figure}

\section{List of candidate members}
\label{appendix:list_of_members}

\begin{table*}

\caption{List of our 1052 candidate members and their properties. This table will be available in its entirety in machine-readable form at the CDS.}
\begin{tabular}{c}
\end{tabular}
\label{table:list_of_members}
\end{table*}

\end{appendix}
\end{document}